\documentclass [12pt]{article}
\usepackage{amsmath,amssymb,cite}
\usepackage{appendix}
\usepackage{feynmp}
\usepackage{float}
\usepackage{subfigure}
\usepackage[vcentermath,enableskew]{youngtab}
\usepackage{tabularx}
\usepackage{booktabs}
\usepackage[english]{babel}
\usepackage{graphicx}
\usepackage{indentfirst}
\usepackage{epsfig}
\usepackage{color}
\usepackage{epstopdf}
\usepackage{tikz}
\usepackage[]{hyperref}
\usepackage[section]{placeins}
\usepackage[stable]{footmisc}
\usepackage{appendix}
\usepackage{tabularx}
\usepackage[english]{babel}
\usepackage{graphicx}
\usepackage{indentfirst}
\usepackage{epsfig}
\usepackage{slashed}
\usepackage{fancyhdr}
\numberwithin{equation}{section}

\fancypagestyle{plain}{\fancyhead{}
}

\begin{document}

\title{Quantum Black Holes and Gauge/Gravity Duality}

\author{
Badis Ydri\\
\textit{Department of Physics, Faculty of Sciences,}\\
\textit{Badji Mokhtar–Annaba University, Annaba, Algeria}
}
\date{November 26, 2024}

\maketitle

\begin{abstract}

  From a conceptual point of view, this chapter may be viewed as an exercise in combining quantum field theory and general relativity in a controlled setting. Despite its apparent simplicity, this exercise is deeply rooted in highly non-trivial developments in superstring theory and holography, and it addresses what is arguably one of the most profound questions in quantum gravity: the fate of information in black hole evaporation. For this reason, one is naturally compelled to examine this exercise as carefully as possible, closely following the original authors.

  This chapter presents thus a detailed overview of recent work by Almheiri et al. and Penington \cite{Almheiri:2019psf,Penington:2019npb} on the ${\bf AdS}^2$ black hole information loss paradox and its proposed resolution within the framework of the AdS/CFT correspondence. The roles of generalized entanglement entropy, quantum extremal surfaces, the island conjecture, holography, and replica wormholes in this resolution are discussed in detail. The distinction between the von Neumann entropy and the Bekenstein–Hawking entropy in black hole physics is carefully clarified. It is shown that a phase transition at the Page time, between the trivial quantum extremal surface at the horizon and a non-vanishing quantum extremal surface located behind the horizon, leads to the correct Page curve.

  A simplified version of the information loss problem for the eternal ${\bf AdS}^2$ black hole, together with its resolution along similar lines, is also presented. The replica trick and its crucial role in computing entanglement entropies for various intervals in ${\bf AdS}^2$ are discussed at some length. However, the use of the replica trick in constructing replica wormholes that dominate the Euclidean path integral—thereby leading to the island rule and the correct quantum extremal surfaces—is only outlined.

  \noindent\textbf{Note.}
This work is published as \textbf{Chapter 7} in
\emph{Lectures on General Relativity, Cosmology and Quantum Black Holes (Second Edition)},
IOP Publishing (2025).
ISBN~978-0-7503-5824-8.
DOI~10.1088/978-0-7503-5824-8.

\medskip
\noindent
The published chapter is available at the publisher’s website:
\href{https://iopscience.iop.org/book/mono/978-0-7503-5824-8/chapter/bk978-0-7503-5824-8ch7}{IOP Science (Chapter 7)}.

\end{abstract}

\newpage
\tableofcontents

\section{Outline}

  \begin{itemize}
  \item The fundamental theory underlying the recent proposed resolution of the black hole information loss paradox is the holographic AdS/CFT correspondence  \cite{Maldacena:1997re,Witten:1998qj,Gubser:1998bc}. See also \cite{Itzhaki:1998dd,Polchinski:1995mt,Witten:1995im}.
  \item More precisely,  we should use, instead of the Bekenstein-Hawking formula \cite{Bekenstein:1972tm,Bekenstein:1973ur} for black hole entropy, the fine-grained entropy formula discovered by Ryu and Takayanagi in \cite{Ryu:2006bv} (See also \cite{Hubeny:2007xt,Engelhardt:2014gca}). This is the correct formula in the case of quantum field theories coupled to gravity.
  \item The black hole information loss paradox consists of the seemingly non-unitary process of evolving a pure state (black hole) into a mixed state (Hawking radiation)  \cite{Hawking:1974rv, Hawking:1974sw,Hawking:1976ra}. Indeed, Hawking radiation is a thermal blackbody radiation. The measurement process in quantum mechanics is the only other known non-unitary process taking a pure state into a mixed state.
    \item The main result of interest to us here is an argument against the  black hole information loss paradox.
  \item This main result goes as follows: Hawking radiation is computed using the Ryu-Takayanagi formula and shown to be consistent with unitarity in \cite{Almheiri:2019psf,Penington:2019npb}. See also \cite{Almheiri:2019qdq,Penington:2019kki,Almheiri:2019hni,Almheiri:2020cfm}.
    
  \item This result  is in fact an evidence for what is termed the "central dogma" in \cite{Almheiri:2020cfm} which can be alternatively stated as follows: From the perspective of an outside observer an evaporating black hole looks like a unitary quantum system with a finite number of degrees of freedom, i.e. a quantum system with a finite-dimensional Hilbert space.
 
  \item Evidence for the "central dogma" comes from the AdS/CFT correspondence \cite{Maldacena:1997re,Witten:1998qj,Gubser:1998bc} but also from the BFSS matrix quantum mechanics \cite{Banks:1996vh}.
  \item Furthermore, it can be shown that the number of degrees of freedom or microstates of the black hole is exactly measured by the Bekenstein-Hawking formula $S=A/4\hbar G_N=A/4l_P^2$ \cite{Strominger:1996sh}.
  \item The  Bekenstein-Hawking entropy is in fact the coarse-grained Boltzmann entropy which is the usual thermodynamic entropy. This thermodynamic entropy is always larger than the fine-grained von Neumann entropy which is the information entropy (density matrix).
    \item The von Neumann entropy of the outgoing Hawking radiation equals the von Neumann entropy of the black hole because the system (black hole $+$ Hawking radiation) must be described by a pure state according to the "central dogma". Thus, the von Neumann entropy of the Hawking radiation must always be smaller than the Bekenstein-Hawking-Boltzmann entropy.
  \item From an operational point of view the unitarity of the Hawking radiation can be characterized by the so-called Page curve \cite{Page:1993wv,Page:2013dx} which gives the von Neumann entropy of the outgoing radiation as a function of time. This curve must have a maximum (turnover point) at the so-called Page time, i.e. the von Neumann fine-grained entropy of the radiation must start to decrease at this time so as to not exceed the Boltzmann coarse-grained entropy of the black hole. Thus, Page time is when the von Neumann entropy of radiation equals the Boltzmann entropy of the black hole. See figure (\ref{fig0}).

    \item Hawking curve, in contrast, increases monotonically and thus the von Neumann entropy of the radiation becomes at a certain time larger than the Boltzmann entropy of the black hole. This is the precise statement of the black hole information loss paradox. 
    %it is stated in \cite{Almheiri:2019qdq} that "The proposal is that Hawking used the wrong formula for computing the entropy. As the theory is coupled to gravity, we should use the proper gravitational formula for entropy: the gravitational fine-grained entropy formula studied by Ryu and Takayanagi \cite{Ryu:2006bv}".
%  \end{itemize}
%  \end{frame}

%\begin{frame}  
%  \frametitle{Outline}
  
  \begin{figure}[htbp]
\begin{center}
  \includegraphics[width=10.0cm,angle=-0]{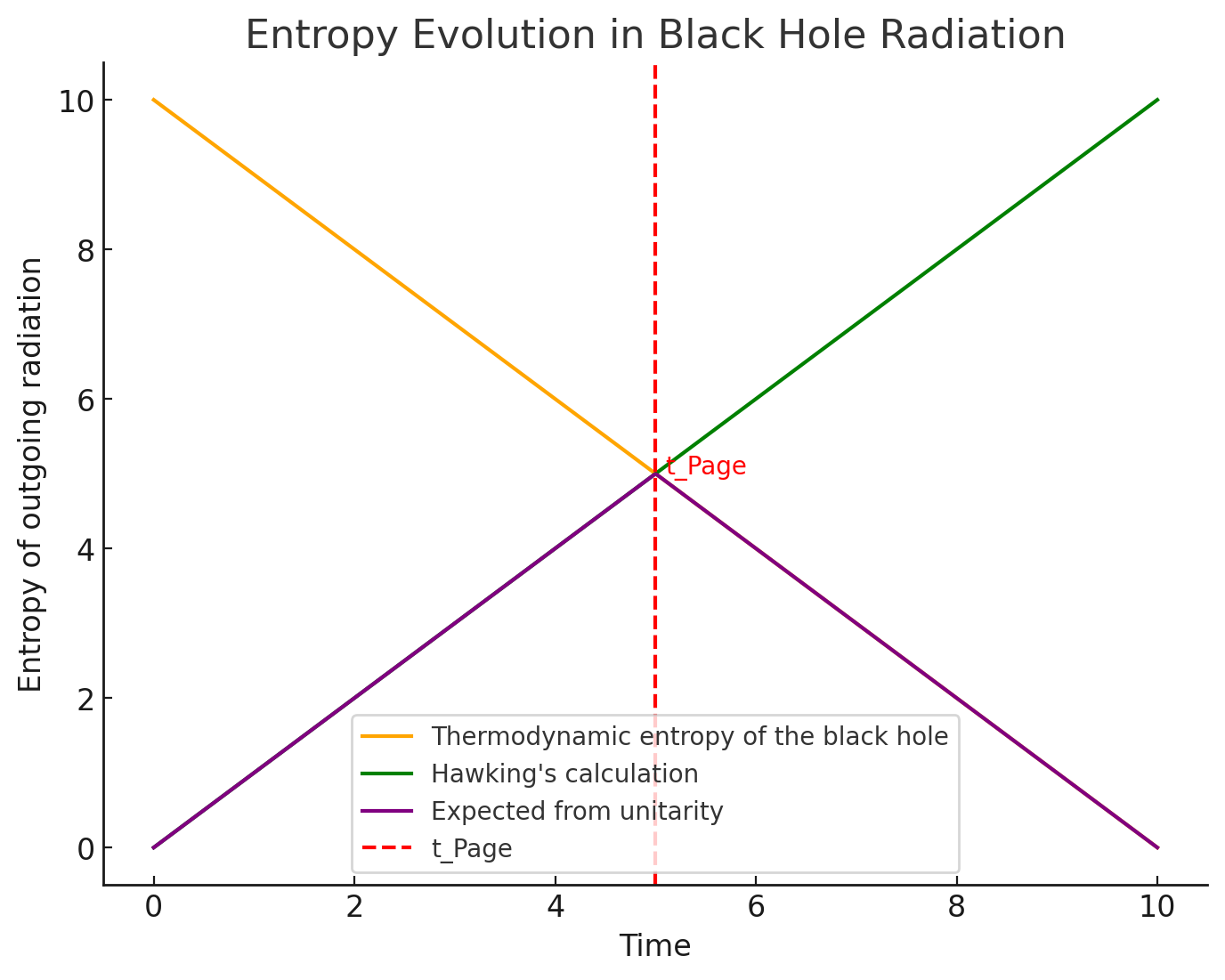}
\end{center}
\caption{Page curve.}\label{fig0}
  \end{figure}
  
%  \begin{itemize}
  \item The main results which underlie the resolution of the black hole information loss paradox proposed in \cite{Almheiri:2019psf,Penington:2019npb,Almheiri:2019qdq,Penington:2019kki,Almheiri:2019hni,Almheiri:2020cfm} are as follows:
%    \begin{enumerate}
    \item[$1)$] The fine-grained von Neumann entropy of an evaporating black hole computed employing the Ryu-Takayanagi formula will follow the Page curve if one uses the so-called quantum extremal surfaces \cite{Engelhardt:2014gca}.
    \item[$2)$] The fine-grained von Neumann  entropy of Hawking radiation is equal to the entropy of the black hole, and thus it too follows the Page curve, if one uses the so-called "island conjecture" in conjunction with the  Ryu-Takayanagi formula \cite{Ryu:2006bv,Hubeny:2007xt}. The boundary of the island is the quantum extremal surface.
    \item[$3)$] The degrees of freedom of the black hole and the Hawking radiation describe the regions of the spacetime given by their respective entanglement wedges where their fine-grained entropies are computed. It is shown that the black hole entanglement wedge describes a portion of the interior whereas the Hawking radiation entanglement wedge is disconnected as it involves an island behind the horizon.
    \item[$4)$] The island  Ryu-Takayanagi formula  for the fine-grained von Neumann entropy can be derived from a  gravitational path integral using the so-called replica trick where it is found that the replica wormhole (zero entropy) dominates over the Hawking saddle point (large entropy) and thus unitarity indeed holds.
      \item[$5)$] For holographic CFT matter it can be shown  that the island is actually connected to the radiation via extra dimensions.
 %     \end{enumerate}
    \end{itemize}

\section{The JT-CFT$_2$ system: A first look}
\subsection{Eternal ${\bf AdS}^2$ black hole}

In this first section we describe the Almheiri-Polchinski (AP) model discussed in \cite{Almheiri:2014cka} which is a particular case of the Jackiw-Teitelboim (JT) model  of dilaton gravity in two dimensions \cite{Jackiw:1984je,Teitelboim:1983ux}. This has striking similarities to the Sachdev–Ye–Kitaev (SYK) model \cite{Kitaev:2015, Sachdev-Ye:1993} (in the sense that everything is encoded in the boundary theory which is given by a Schwarzian action). In here we will mainly follow \cite{Almheiri:2014cka, Almheiri:2019psf} but also \cite{Engelsoy:2016xyb,Maldacena:2016upp}.

The APJT model is a dynamical theory of gravity in two dimensions with no local excitations, i.e. the value of the metric tensor $ds^2=g_{\mu\nu}dx^{\mu}dx^{\nu}$, which is a dynamical variable here (as opposed to pure gravity in two dimensions),  and the value of the dilaton field $\phi$, which plays a crucial role in the existence of stable black hole configurations in two dimensions,  is uniquely fixed in terms of the stress-energy-momentum tensor of the matter field $f$ (through the equations of motion).

We will mostly assume that the matter theory is given by a conformal field theory which can also be holographic. We will also assume that the matter field $f$ interacts with dilaton field $\Phi$ only through the metric tensor and not directly. The action reads explicitly 

  \begin{eqnarray}
    S[g,\Phi,f]&=& S_{\rm JT}[g,\Phi]+D_{\rm CFT}[g,f]\nonumber\\ &=&\frac{1}{16\pi G}\int d^2x\sqrt{-g}(\Phi^2 R-V(\Phi))+S_{\rm CFT}[g,f]. \label{action}\end{eqnarray}
The AP potential $V$ is given explicitly by

  \begin{eqnarray}
    V(\Phi)=2-2\Phi^2.
\end{eqnarray}
This potential guarantees that spacetime has a constant negative scalar curvature (integrate out $\phi=\Phi^2$ to obtain the delta function $\delta (R+2))$, i.e. we must have $R=-2$. In other words, spacetime is precisely  anti-de Sitter spacetime  ${\bf AdS}^2$. In fact this potential also guarantees that the matter action does not depend directly on the dilaton field but depends on it only indirectly through the metric tensor. The dilaton filed itself is the crucial ingredient allowing  the existence of black hole configurations on this two-dimensional ${\bf AdS}^2$ background.

We check these facts more explicitly as follows. First, we write the metric in the conformal gauge as follows

  \begin{eqnarray}
    ds^2=-e^{2\omega(u,v)}du dv.
  \end{eqnarray}
The light-cone coordinates $u$ and $v$ will also be denoted as $u=x^+$ and $v=x^-$ where $x^{\pm}=t\pm z$. The equations of motion (by varying the dilaton field and the metric tensor) will then read

  \begin{eqnarray}
    4\partial_u\partial_v\omega+e^{2\omega}=0.\label{eom1}
  \end{eqnarray}

  \begin{eqnarray}
    -e^{2\omega}\partial_u(e^{-2\omega}\partial_u\Phi^2)=8\pi G T_{uu}.\label{eom2}
  \end{eqnarray}

  \begin{eqnarray}
    -e^{2\omega}\partial_v(e^{-2\omega}\partial_v\Phi^2)=8\pi G T_{vv}.\label{eom3}
  \end{eqnarray}

  \begin{eqnarray}
    2\partial_u \partial_v \Phi^2+e^{2\omega}(\Phi^2-1)=16\pi G T_{uv}.\label{eom4}
  \end{eqnarray}
The stress-energy-momentum tensor of the matter field is of course given by the usual formula, viz

  \begin{eqnarray}
    T_{ab}=-\frac{2}{\sqrt{-g}}\frac{\delta S_{\rm CFT}[g,f]}{\delta g^{ab}}.
  \end{eqnarray}
For conformal matter, such as a massless free scalar field $f$ given by the Klein-Gordon action $S_{\rm CFT}=\frac{1}{32\pi G}\int d^2x \sqrt{-g}(\nabla f)^2$, the off-diagonal component $T_{uv}$ of the stress-energy-momentum tensor vanishes classically.  We also recall that in this case $T_{uu}=T_{++}=\frac{1}{16\pi G}\partial_+f\partial_+f$ and $T_{vv}=T_{--}=\frac{1}{16\pi G}\partial_-f\partial_-f$.

The most general solution of (\ref{eom1}) is the ${\bf AdS}^2$ metric, i.e. $e^{2\omega}=4/(u-v)^2$.  We have explicitly the metric

  \begin{eqnarray}
    ds^2=-\frac{4}{(x^+-x^-)^2}dx^+dx^-=\frac{1}{z^2}(-dt^2+dz^2)~,~x^{\pm}=t\pm z.\label{sol1}
  \end{eqnarray}
The gravitational sector will be treated semi-classically throughout. This means, among many other things, that we will replace the stree-energy-momentum tensor by its expectation value, viz

  \begin{eqnarray}
    T_{ab}=\langle T_{ab}\rangle.
  \end{eqnarray}
For $\langle T_{ab}\rangle=0$ the most general solution of the equations of motion (\ref{eom2}), (\ref{eom3}) and (\ref{eom4}) is given by the dilaton field

  \begin{eqnarray}
    \Phi^2=1+\frac{a-\mu x^+x^-}{x^+-x^-}.\label{sol2}
  \end{eqnarray}
This dilaton field represents an eternal black hole with two asymptotic boundaries. 

The case $a=0$ represents pure ${\bf AdS}^2$ or more precisely the Poincaré patch of ${\bf AdS}^2$ whereas the solution with $a>0$ (and $\mu>0$) represents an ${\bf AdS}^2$ black hole  (it is the Rindler wedge of the Poincaré patch). 

The most general solution of the equations of motion is a conformal transformation $x=f(y)$ of (\ref{sol1}) and (\ref{sol2}), viz
  \begin{eqnarray}
    ds^2=-\frac{4f^{\prime}(y^+)f^{\prime}(y^-)dy^+dy^-}{(f(y^+)-f(y^-))^2}~,~\Phi^2=1+\frac{a-f(y^+)f(y^-)}{f(y^+)-f(y^-)}.
  \end{eqnarray}
A static form of this black hole configuration is obtained by means of the conformal transformation 

  \begin{eqnarray}
    x=f(y)=\frac{1}{\sqrt{\mu}}\tanh \sqrt{\mu} y.\label{diffeo}
  \end{eqnarray}
The black hole solution reads then (we set $a=1$)

  \begin{eqnarray}
    ds^2=-\frac{4\mu dy^{+}dy^{-}}{\sinh^2\sqrt{\mu}(y^{+}-y^{-})}~,~\Phi^2=1+\sqrt{\mu}\coth\sqrt{\mu}(y^{+}-y^{-}).
  \end{eqnarray} 
The $x$ coordinates cover the whole geometry of the spacetime manifold whereas the $y$ coordinates cover only the exterior of the black hole.  At high temperature ($\mu \longrightarrow 0$) we can make the identification 

  \begin{eqnarray}
    z=\frac{y^+-y^-}{2}.
  \end{eqnarray}
Thus, the boundary $z=0$ of ${\bf AdS}^2$ is  located in the $y$ coordinates at $y^+-y^-=0$. On the other hand, the horizon $z\longrightarrow \infty$ of ${\bf AdS}^2$ corresponds in the $y$ coordinates either to the future horizon $y^+\longrightarrow +\infty$ (or equivalently $x^+\longrightarrow 1/\sqrt{\mu}$) or to the past horizon  $y^-\longrightarrow -\infty$ (or equivalently $x^+\longrightarrow -1/\sqrt{\mu}$).

This black hole configuration after Euclidean rotation becomes periodic in the variable $y^+-y^-$ with period $\beta_0=1/T_0$ given by 

  \begin{eqnarray}
    2\sqrt{\mu}=\frac{2\pi}{\beta_0}\Rightarrow T_0=\frac{\sqrt{\mu}}{\pi}.
  \end{eqnarray}
This is Hawking temperature. This can also be checked in the Schwarzschild coordinates defined by

\begin{eqnarray}
  \rho=\sqrt{\mu}\coth \sqrt{\mu}(y^+-y^-)~,~T=\frac{y^++y^-}{2}.
\end{eqnarray}
The metric and the dilaton take then the form

\begin{eqnarray}
  ds^2=-4(\rho^2-\mu)dT^2+\frac{d\rho^2}{\rho^2-\mu}~,~\Phi^2=1+\rho.
\end{eqnarray}
The Hawking temperature is then given by 

\begin{eqnarray}
  T_0&=&\frac{1}{4\pi}\partial_{\rho}\sqrt{-\frac{g_{TT}}{g_{\rho\rho}}}|_{\rho=\sqrt{\mu}}\nonumber\\&=&\frac{\sqrt{\mu}}{\pi}.\label{T_Hawking}
\end{eqnarray}

\subsection{${\bf AdS}^2$ Black hole formed from gravitational collapse}

We look now at the equations of motion more carefully. First, we trivially check that $\exp(2\omega)=4/(x^+-x^-)^2$ solves the equation of motion (\ref{eom1}). Next, in terms of the ansatz $\Phi^2=M/(x^+-x^-)$ we write the constraints (\ref{eom2}) and (\ref{eom3}) in the form

  \begin{eqnarray}
    \partial_+^2M=-8\pi G(x^+-x^-)T_{++}(x^+)~,~ \partial_-^2M=-8\pi G(x^+-x^-)T_{--}(x^-).\label{eom5}
  \end{eqnarray}
The final equation of motion (\ref{eom4}) reads

  \begin{eqnarray}
    \partial_+\partial_-M+(\partial_+M-\partial_-M-2)/(x^+-x^-)=8\pi G(x^+-x^-)T_{+-}.\label{eom6}
  \end{eqnarray} 
The most general solution of (\ref{eom5}) is 

  \begin{eqnarray}
    M=a+bx^++cx^-+dx^{+}x^--I^++I^-.
  \end{eqnarray}

  \begin{eqnarray}
    I^{\pm}=8\pi G\int_{x_0^{\pm}}^{x^{\pm}}dx^{\prime\pm}(x^{\prime \pm}-x^{\mp})(x^{\prime\pm}-x^{\pm})T_{\pm\pm}(x^{\prime\pm}).
  \end{eqnarray}
For conformal theory we have $T_{+-}=0$. The solution of equation (\ref{eom6}) is then given by the requirement

  \begin{eqnarray}
    b-c=2.
  \end{eqnarray}
The solution is then (with $b=b^{\prime}+1$)

  \begin{eqnarray}
    \Phi^2=1+\frac{a+b^{\prime}(x^++x^-)+dx^+x^-}{x^+-x^-}-\frac{I^+-I^-}{x^+-x^-}.\label{res0}
  \end{eqnarray} 
For $T_{uv}=\langle T_{uv}\rangle =0$ we get the solution

  \begin{eqnarray}
    \Phi^2=\frac{a+b^{\prime}(x^++x^-)+dx^+x^-}{x^+-x^-}.\label{eom7}
  \end{eqnarray}
By an ${\bf SL(2,R)}$ symmetry which acts as $t\longrightarrow t^{\prime}=(at+b)/(ct+d)$ with $ad-cb=1$ we can bring this solution to the form 

  \begin{eqnarray}
    \Phi^2=1+\frac{a-\mu x^+x^-}{x^+-x^-}.\label{eom8}
  \end{eqnarray}
This is the dilaton profile corresponding to an eternal ${\bf AdS}^2$ black hole. 

Next, we would like to derive the black hole solution formed from gravitational collapse. We consider then the effect of an infalling matter pulse into the black hole, i.e. the effect of a schokwave of energy $E_S$ traveling on the null curve $x^-=0$ starting from the boundary $z=0$ at time $t=0$. The stress-energy-momentum tensor is given by

  \begin{eqnarray}
    T_{--}=\langle T_{--}\rangle =E_S\delta (x^-).
  \end{eqnarray}
We compute immediately $I^+=0$, $I^-=8\pi G E_S x^+x^-$. The dilaton profile becomes then given by

  \begin{eqnarray}
    \Phi^2=\frac{a+b^{\prime}(x^++x^-)+(d+8\pi GE_S) x^+x^-}{x^+-x^-}.
  \end{eqnarray}
By an ${\bf SL(2,R)}$ symmetry we can bring this solution to the form  

  \begin{eqnarray}
    \Phi^2=\frac{a-(\mu+8\pi GE_S) x^+x^-}{x^+-x^-}.\label{res2}
  \end{eqnarray}
We can deduce from this formula the relationship between the mass of the black hole and its temperature. If we start from a pure ${\bf AdS}^2$ space we can set $\mu=0$ and thus we obtain 

  \begin{eqnarray}
    \Phi^2=\frac{a-8\pi GE_S x^+x^-}{x^+-x^-}.\label{eom9}
  \end{eqnarray}
By comparing (\ref{eom9}) and (\ref{eom8}) and using (\ref{T_Hawking}) we deduce the relationship between the energy and temperature as

  \begin{eqnarray}
    8\pi GE_S=\mu_S=(\pi T_S)^2.
  \end{eqnarray}
Thus, for the eternal ${\bf AdS}^2$ black hole we obtain the relationship 

  \begin{eqnarray}
    8\pi GE_0=\mu_0=(\pi T_0)^2.
  \end{eqnarray}
By considering now the  effect of an infalling matter pulse into the black hole (black hole formed by gravitational collapse) we obtain 

  \begin{eqnarray}
    8\pi GE_1=\mu_1=(\pi T_1)^2.
  \end{eqnarray}
The energy $E_1$ is simply given by the energy $E_0$ of the eternal black hole plus the energy $E_S$ of the infalling matter, viz $E_1=E_0+E_S$. Thus, the relationship between the the new temperature $T_1$, the old temperature $T_0$ and the energy of the pulse $E_S$ is given by 

  \begin{eqnarray}
    (\pi T_1)^2=(\pi T_0)^2+8\pi G E_S.
  \end{eqnarray}

\subsection{Quantum correction and coupling to a heat bath}

The next step is to add quantum corrections, due to matter fields, which in the case of a conformal theory are encoded in the conformal anomaly.  The  Almheiri-Polchinski (AP) model is really characterized by transparent boundary conditions at the boundary as opposed to the reflecting boundary conditions characterizing the usual Jackiw-Teitelboim (JT) model.

In other words, the (right) boundary of ${\bf AdS}^2$ is coupled to a heat bath, at zero temperature, into which Hawking radiation can escape and hence we have a simulated black hole evaporation process with the associated Hawking radiation and the consequent black hole information loss problem.

The matter sector which is independent of the dilaton field (and only interacts with it through the constraints) is treated as a conformal field theory on a fixed ${\bf AdS}^2$ background in the coordinates $x$. The fields are subjected to transparent boundary conditions.

In the external heat bath ${\bf M}^2$ we have the same conformal field theory in the coordinates $y$. The coupling between the ${\bf AdS}^2$ space and the heat bath occurs at $t=0$, i.e. at $x^-=0$ which results in a shokwave of energy $E_S$ infalling from the boundary into the black hole. We have the following metrics

  \begin{eqnarray}
    &&ds^2_{\bf AdS^2}=-\frac{4dx^+dx^-}{(x^+-x^-)^2}=-\Omega^{-2}(y)dy^+dy^-~,~\Omega^{-2}(y)=\frac{4f^{\prime}(y^+)f^{\prime}(y^-)}{(f(y^+)-f(y^-))^2}\nonumber\\&& ds^2_{\bf M^2}=-dy^+dy^.
  \end{eqnarray} 
The boundary conditions of the ${\bf AdS}^2$ metric $ds^2_{\bf AdS^2}=(-dt^2+dz^2)/z^2$ and the dilaton field $\Phi^2=1+(1-\mu(t^2-z^2))/2z$, i.e. their values at the boundary $z=\epsilon$  are given by (where $u$ is the time variable on the boundary)

  \begin{eqnarray}
    g_{tt}|_{\rm boundary}=\frac{1}{\epsilon^2}=\frac{-t^{\prime 2}+z^{\prime 2}}{z^2}~,~\Phi^2|_{\rm boundary}=\frac{1}{2\epsilon}.
  \end{eqnarray}
The diffeomorphism $x=f(y)$ is chosen such that the boundary is simple at constant value in the $y$ coordinates, viz

  \begin{eqnarray}
    \epsilon=\frac{y^+-y^-}{2}.
  \end{eqnarray}
Of course, before the coupling between the ${\bf AdS}^2$ and the heat bath ${\bf M}^2$ is turned on at $t=0$ this diffeomorphsim is given by  (\ref{diffeo}) which corresponds to the static form of the eternal black hole solution.

Approaching the future/past horizon $y^{\pm}\longrightarrow\pm\infty$ on the boundary $y^+-y^-=0$ means that $u=T|_{\rm boundary}=\frac{y^++y^-}{2}|_{\rm boundary}\longrightarrow \pm\infty$ which in the $x$ coordinates is equivalent to spending the times $\pm t_{\infty}={\rm lim}_{u\longrightarrow\pm\infty}f(u)$.

Before the coupling between ${\bf AdS}^2$ and the heat bath is switched on these times $\pm t_{\infty}$ are precisely the future/past horizon times $\pm t_{\infty}=f(\pm \infty)=x^{\pm}=\pm 1/\sqrt{\mu_0}=\pm 1/\pi T_0$.

After the coupling the temperature changes to $T_1$ and in order for the wormhole to remain not traversable  the new event horizon is required to lie outside the original horizon and hence it can be reached in less time, i.e. $t_{\infty}$ after the coupling must satisfy $t_{\infty}< 1/\pi T_0$. The idea is that the so-called Averaged Null Energy Condition (ANEC) on the horizon must always be satisfied in order to maintain boundary causality \cite{Maldacena:2018lmt,Galloway:2018dak}.

Let us now introduce the Euclidean time $\tau=i t$ and the Euclidean (complex) coordinates $x$ and $\bar{x}$ by 

  \begin{eqnarray}
    x^+=\bar{x}=t+z~,~x^-=-x=t-z.
  \end{eqnarray}
Thus, $t=(x^++x^-)/2=(\bar{x}-x)/2$ and $z=(x^+-x^-)/2=(\bar{x}+x)/2$. The ${\bf AdS}^2$ boundary is at $z=0$ whereas the bulk is $z>0$. The initial state at $t=0$ is therefore the Hartle-Hawking state on ${\bf AdS}^2$ which is given by the vacuum on the half-line $z>0$.  The heat bath is another half-line $z<0$ with the same CFT prepared in the same vacuum state. The physical time (time on the boundary and in the heat bath) is $u$ and not $t$ (which is the bulk time). They are related by a diffeomorphism $t=f(u)$ (we choose $0=f(0)$).

We must therefore go from the coordinates $x$ (defined on ${\bf AdS}^2$) to the coordinates $y$ (defined on the heat bath) by the diffeomorphism $f$, viz $x=f(y)$ and $\bar{x}=f(\bar{y})$. The boundary is simple in the $y$ coordinates located at $y+\bar{y}=0$ while the physical time is $T=(\bar{y}-y)/2$ (we choose $f(y)=-f(-y)$). 

At the initial time $t=u=0$ we have $y=\bar{y}=f^{-1}(z)$ and ${\bf AdS}^2$ corresponds to the right half-line $y>0$ whereas the heat bath corresponds to the left half-line $y<0$ . In general, ${\bf AdS}^2$ corresponds to the right half-plane $y+\bar{y}>0$ whereas the heat bath corresponds to the left half-plane $y+\bar{y}<0$. 

The initial quantum state of the heat bath is therefore given by the Euclidean path integral on the left lower half-plane (the half-line vacuum). On the other hand, the initial quantum state of ${\bf AdS}^2$ is given by an Euclidean path integral on the right lower half-plane with a deformed boundary (a Virasoro descendant of the usual CFT vacuum on the half-line). In conclusion, we have in the $y$ coordinates a combined coupled system evolving in the physical time by the usual Hamiltonian of conformal field theory on the line.

The initial state is then time-reflection symmetric given by an Euclidean path integral over a simply connected space with a single boundary. Therefore it is a descendent of the half-line vacuum. The Cauchy surface at $t=0$ can thus be mapped by a means of an appropriate diffeomorphism to a half-line parameterized by a coordinate $w\in [0,\infty[$, i.e. we can map our initial quantum state to the half-line vacuum.

The goal is to derive the diffeomorphism $w=w(x)$ (in the ${\bf AdS}^2$ region $x>0$), the diffeomorphism $w=w(y)$ (in the heat bath ${\bf M}^2$ region $y<0$), the stress-energy-momentum tensor of the conformal matter in both regions and the energy $E_S$ of the initial shockwave due to the turning on of the coupling between ${\bf AdS}^2$ and ${\bf M}^2$ at time $t=0$. This will allow us to determine the quantum corrections to the Hawking temperature.

First, at $t=0$ the stress-energy-momentum tensor of ${\bf AdS}^2$ is zero in the physical coordinates $y$, and also zero in the Poincare coordinates $x$ since the Weyl anomaly between $x$ and $y$ is zero ($x=x(y)$ is an ${\bf SL(2,R)}$ transformation). Similarly, the heat bath ${\bf M}^2$ is at zero tempertaure and thus the corresponding stress-energy-momentum tensor is also zero. We have then

  \begin{eqnarray}
    \langle T_{xx}(x)\rangle=0~(x>0)~,~\langle T_{yy}(y)\rangle=0~ (y<0)~,~t=0.
  \end{eqnarray}
  At later times we use the transformation law of the stress-energy-momentum tensor under the diffeomorphism $w$, viz 

  \begin{eqnarray}
    (\frac{dw}{dx})^2\langle T_{ww}(w)\rangle=\langle T_{xx}(x)\rangle-\frac{c}{24\pi}S(w,x).
  \end{eqnarray}

  \begin{eqnarray}
    (\frac{dw}{dy})^2\langle T_{ww}(w)\rangle=\langle T_{yy}(y)\rangle+\frac{c}{24\pi }S(w,y).
  \end{eqnarray}
The number $c$ is the central charge of the conformal field theory and $S$ is the so-called Schwarzian which is defined by 

  \begin{eqnarray}
    S(w,x)=\{w,x\}&=&\frac{w^{\prime\prime\prime}(x)}{w^{\prime}(x)}-\frac{3}{2}\frac{w^{\prime\prime 2}(x)}{w^{\prime 2}(x)}\nonumber\\&=&(\frac{w^{\prime\prime}}{w^{\prime}})^{\prime}-\frac{1}{2}(\frac{w^{\prime\prime}}{w^{\prime}})^2.
  \end{eqnarray}
We take the diffeomorphism $w$ to be an ${\bf SL(2,R)}$ transformation of the relevant coordinates, i.e.  a Mobius map which guarantees the vanishing of the stress-energy-momentum tensor  in the $w$ coordinates. Hence, we obtain the energy-momentum tensor 

  \begin{eqnarray}
    \langle T_{xx}(x)\rangle=-\frac{c}{24\pi}S(w,x).
  \end{eqnarray}

  \begin{eqnarray}
    \langle T_{yy}(y)\rangle=-\frac{c}{24\pi }S(w,y).
  \end{eqnarray}
We go back to the initial time $t=0$. The diffeomorphism (or conformal transformation) $w$ will map the ${\bf AdS}^2$ region $x>0$ to the interval $[0,w_0]$ while it will map the heat bath ${\bf M}^2$ region $y<0$ to the interval $[w_0,\infty[$ and it is given explicitly by \cite{Almheiri:2019psf}

    \begin{eqnarray}
      w(x)=\frac{w_0^2}{w_0+x}~,~x>0.
    \end{eqnarray}

    \begin{eqnarray}
      w(y)=w_0+f^{-1}(-x)~,~x<0.
    \end{eqnarray}
We write this as

  \begin{eqnarray}
    w(x)=\frac{w_0^2}{w_0+x}\theta(x)+(w_0-y)\theta(-x).
  \end{eqnarray}
We compute (using $f(0)=0$)

  \begin{eqnarray}
    w^{\prime}(x)=-\frac{w_0^2}{(w_0+x)^2}\theta(x)-y^{\prime}(x)\theta(-x).
  \end{eqnarray}
Then, we compute (using $y^{\prime}(x)=1/f^{\prime}(y)$ and $f^{\prime}(0)=1$ where primes denote derivatives with respect to the appropriate variable) 

  \begin{eqnarray}
    w^{\prime\prime}(x)=\frac{2w_0^2}{(w_0+x)^3}\theta(x)-y^{\prime\prime}(x)\theta(-x).
  \end{eqnarray}
Hence, we obtain

  \begin{eqnarray}
    \frac{w^{\prime\prime}(x)}{w^{\prime}(x)}=-\frac{2}{w_0+x}\theta(x)+\frac{y^{\prime\prime}(x)}{y^{\prime}(x)}\theta(-x).
  \end{eqnarray}
As a consequence, we have (using $y^{\prime\prime}/y^{\prime}=-f^{\prime\prime}/f^{\prime 2}$ and hence $\frac{y^{\prime\prime}}{y^{\prime}}|_{x=0}=\frac{y^{\prime\prime}}{y^{\prime}}|_{y=0}=-f^{\prime\prime}(0)$) 

  \begin{eqnarray}
    (\frac{w^{\prime\prime}(x)}{w^{\prime}(x)})^{\prime}=\frac{2}{(w_0+x)^2}\theta(x)+(\frac{y^{\prime\prime}(x)}{y^{\prime}(x)})^{\prime}\theta(-x)-\frac{2}{w_0}\delta(x)+f^{\prime\prime}(0)\delta(x).
  \end{eqnarray}
We get then the Schwarzian and the energy-momentum tensor which are given explicitly by

  \begin{eqnarray}
    S(w,x)=\{w,x\}=\{y,x\}\theta(-x)-(\frac{2}{w_0}-f^{\prime\prime}(0))\delta(x).
  \end{eqnarray}

  \begin{eqnarray}
    \langle T_{xx}(x)\rangle=-\frac{c}{24\pi}\{y,x\}\theta(-x)+\frac{c}{24\pi}(\frac{2}{w_0}-f^{\prime\prime}(0))\delta(x).\label{res1}
  \end{eqnarray}
Thus, we can make the identification:

  \begin{eqnarray}
    E_S= \frac{c}{24\pi}(\frac{2}{w_0}-f^{\prime\prime}(0)).
  \end{eqnarray}
However,  ${\bf AdS}^2$ is dual to a conformal quantum mechanics at the boundary and thus it is more natural to map the ${\bf AdS}^2$ region to a single point. In other words, we must take the limit $w_0\longrightarrow 0$ and hence $E_S\longrightarrow \infty$. As it turns out, this is indeed the physically sensible limit in order to avoid acausal correlations \cite{Almheiri:2019psf}. The diffeomorphism $w$ becomes a mapping to the upper-half plane given by 

  \begin{eqnarray}
    w(x)=(\frac{12\pi E_S}{c})^{-1}\frac{1}{x}\theta(x)+f^{-1}(-x)\theta(-x).
  \end{eqnarray}
We also write the result (\ref{res1}) in the form 

  \begin{eqnarray}
    \langle T_{x^{-}x^{-}}(x^{-})\rangle=-\frac{c}{24\pi}\{y^{-},x^{-}\}\theta(x^{-})+E_S\delta(x^{-}).
  \end{eqnarray}
We use this result in (\ref{res0}). We compute $I^+=0$ as before but now we have

  \begin{eqnarray}
    I^-=8\pi GE_S x^+x^--\frac{k}{2}\int_0^{x^-}dt(x^+-t)(x^--t)\{u,t\}~,~t=f(u)~,~k=\frac{c.G}{3}.
  \end{eqnarray}
The dilaton field, including quantum corrections, becomes (compare with equation (\ref{res2}) and subsequent equations)

  \begin{eqnarray}
    \Phi^2&=&\frac{1-(\mu_0+8\pi GE_S)x^+x^-+\frac{k}{2}\int_0^{x^-}dt(x^+-t)(x^--t)\{u,t\}}{x^+-x^-}\nonumber\\&=&\frac{1-(8\pi T_1)^2x^+x^-+\frac{k}{2}I(x^+,x^-)}{x^+-x^-}.
  \end{eqnarray}

\subsection{More on the boundary theory and  the Schwarzian}

The  Schwarzian plays a crucial role in this problem since the underlying dynamics is one-dimensional on the boundary. Indeed, the space ${\bf AdS}^2$ is characterized by a boundary and the action should be enhanced by a boundary term, viz

  \begin{eqnarray}
    S[g,\Phi,f]\longrightarrow S[g,\Phi,f]&=& S_{\rm JT}[g,\Phi]+D_{\rm CFT}[g,f]+S_{b}[g,\Phi]\nonumber\\
    &=&\frac{1}{16\pi G}\int_{\cal M} d^2x\sqrt{-g}(\Phi^2 R-V(\Phi))+S_{\rm CFT}[g,f]+S_{b}[g,\Phi].\nonumber\\
  \end{eqnarray} 
The boundary term is given by 

  \begin{eqnarray}
    S_{b}[g,\Phi]=\frac{1}{8\pi G}\int_{\partial\cal M} du\sqrt{-\gamma}\Phi^2K.
  \end{eqnarray}
  Here, $K$ is the scalar extrinsic curvature. More precisely, $K=g^{\mu\nu}K_{\mu\nu}=\gamma^{\mu\nu}K_{\mu\nu}$ where $\gamma_{\mu\nu}$ is the induced metric at the boundary and the extrinsic curvature tensor $K_{\mu\nu}$ describes how the boundary ${\partial \cal M}$ is curved with respect to the manifold ${\cal M}={\bf AdS}^2$ in which it is embedded.

  Since the equation of motion of the dilaton enforces the ${\bf AdS}^2$ geometry, the bulk action is zero and we only need to focus on the boundary term.

  Let us consider the Euclidean metric $ds^2=(dt^2+dz^2)/z^2$. The boundary conditions become then

  \begin{eqnarray}
    g_{tt}|_{\rm boundary}=\frac{1}{\epsilon^2}=\frac{t^{\prime 2}+z^{\prime 2}}{z^2}~,~\Phi^2|_{\rm boundary}=\frac{1}{2\epsilon}\Rightarrow \epsilon=\frac{z}{\sqrt{t^{\prime 2}+z^{\prime 2}}}.
  \end{eqnarray}
We can solve this equation explicitly in powers of the cutoff $\epsilon$ to find

  \begin{eqnarray}
    z=\epsilon t^{\prime}+O(\epsilon^2).
  \end{eqnarray}
We compute then

  \begin{eqnarray}
    z^{\prime}=\epsilon t^{\prime\prime}+O(\epsilon^2)~,~z^{\prime\prime}=\epsilon t^{\prime\prime\prime}+O(\epsilon^2).
  \end{eqnarray}

  \begin{eqnarray}
    \epsilon^{\prime}=\frac{z^{\prime}}{\sqrt{t^{\prime 2}+z^{\prime 2}}}-\frac{z(t^{\prime}t^{\prime\prime}+z^{\prime}z^{\prime\prime})}{(t^{\prime 2}+z^{\prime 2})^{3/2}}=0+O(\epsilon^2).
  \end{eqnarray}
The primes are derivatives with respect to $u$ which is the time parameter on the boundary. The tangent vector at the boundary is $e^{\mu}=\partial_{u}x^{\mu}=(t^{\prime},z^{\prime})$ whereas the normal vector is $n^{\mu}=\epsilon(z^{\prime},-t^{\prime})$. By construction these two vectors are orthogonal, i.e. $n_{\mu}e^{\mu}=0$ and furthermore $n^{\mu}$ is normalized, i.e. $n_{\mu}n^{\mu}=1$.

We compute the scalar curvature by the formula

  \begin{eqnarray}
    K=\nabla_{\mu}n^{\mu}&=&\partial_{\mu}n^{\mu}+\Gamma_{\mu\nu}^{\mu}n^{\nu}\nonumber\\
    &=&\partial_{\mu}n^{\mu}+\frac{1}{2}g^{\mu\beta}\partial_{\nu}g_{\beta\mu} n^{\nu}\nonumber\\
    &=&-\frac{t^{\prime 2}-z^{\prime 2}}{t^{\prime}\sqrt{t^{\prime 2}+z^{\prime 2}}}+\frac{2z}{(t^{\prime 2}+z^{\prime 2})^{3/2}}(z^{\prime\prime}t^{\prime}-t^{\prime\prime}z^{\prime})+2\frac{t^{\prime 2}-z^{\prime 2}}{t^{\prime}\sqrt{t^{\prime 2}+z^{\prime 2}}}\nonumber\\
    &=&-\frac{z}{\epsilon t^{\prime}}+\frac{2}{(t^{\prime 2}+z^{\prime 2})^{3/2}}(t^{\prime}(zz^{\prime\prime}+t^{\prime 2}+z^{\prime 2})-z z^{\prime}t^{\prime\prime})\nonumber\\
    &=&1+2\epsilon^2\{t,u\}.
  \end{eqnarray}
So, the extrinsic curvature to leading order in $\epsilon^2$ is equal to the Schwarzian. Note, that in going from the second line to the third line we have replaced in both terms the derivatives $\partial_t$ and $\partial_z$ with $\partial_u/t^{\prime}$ and $\partial_u/z^{\prime}$ respectively. The boundary term becomes given by 

  \begin{eqnarray}
    S_{b}[g,\Phi]&=&(-1)(\frac{1}{2})\frac{1}{8\pi G}\int_{\partial\cal M} du\frac{1}{\epsilon}\frac{1}{2\epsilon}. 2\epsilon^2\{t,u\}\nonumber\\
    &=&-\frac{1}{16\pi G}\int du \{t,u\}.
  \end{eqnarray} 
The minus sign is due to the Euclidean signature whereas the factor $1/2$ is due to the fact that the boundary of ${\bf AdS}^2$ is constituted of two identical disconnected segments.

Thus, we have spontaneous symmetry breaking of conformal symmetry along the boundary down to the ${\bf SL(2,R)}$ Mobius transformations $t\longrightarrow (at+b)/(ct+d)$ with $ad-cb=1$. Indeed, the Schwarzian is only invariant under these transformations, viz

  \begin{eqnarray}
    \{\frac{at+b}{ct+d},u\}=\{t,u\}.
  \end{eqnarray}
The field $t=t(u)$ acts as the corresponding pseudo Nambu Goldstone modes associated with this spontaneous breaking \cite{Maldacena:2016upp}.

The ADM energy associated with boundary translations $u\longrightarrow u+\delta u$ is immediately given from the above action by the Schwarzian, viz

  \begin{eqnarray}
    E(u)=-\frac{1}{16\pi G} \{t,u\}.\label{ADM}
  \end{eqnarray}
This can be obatined by varying the boundary metric and computing the corresponding stress-energy-momentum tensor \cite{Almheiri:2014cka}. 

We can check that for $u<0$, where the diffeomorphism $t=f(u)=\frac{1}{\pi T_0}\tanh (\pi T_0 u)$, this ADM energy $E(u)$ is precisely equal to the energy of the eternal black hole $E_0=\pi T_0^2/8G$.

%\section{Exercises}

%Exercise $1$: Calculate the equations of motion (\ref{eom1}), (\ref{eom2}), (\ref{eom3}) and (\ref{eom4}). Determine the constraints.

%Exercise $2$: Show by using an ${\rm SL}(2,R)$ symmetry that we can bring the solution (\ref{eom7}) to the form (\ref{eom8}).

%Exercise $3$: By varying the boundary metric compute the boundary stress-energy-momentum tensor and show that the ADM energy is given by equation (\ref{ADM}).

%Exercise $4$: Show that for $u<0$ the ADM energy $E(u)$ is precisely equal to the energy of the eternal black hole $E_0=\pi T_0^2/8G$.

\section{More on JT gravity coupled to conformal matter}
\subsection{The eternal ${\bf AdS}^2$ black hole}

    In this section we follow  the presentation of \cite{Almheiri:2019psf}. We consider  Jackiw-Teitelboim (JT) gravity coupled to conformal matter given by the action ($\phi$ being the dilaton field)
    \begin{eqnarray}
      &&S=S_{0}+S_G+S_M\nonumber\\
      S_0&=&\frac{1}{16\pi G_N}\int_{\cal M}d^2x\sqrt{-g}\phi_0 (R+2)+\frac{1}{8\pi G_N}\int_{\partial\cal M}\phi_0K\nonumber\\
      S_G&=&\frac{1}{16\pi G_N}\int_{\cal M}d^2x\sqrt{-g}\phi (R+2)+\frac{1}{8\pi G_N}\int_{\partial\cal M}\phi|_b K\nonumber\\
      S_M&=&S_{\rm CFT}[g].
    \end{eqnarray}
This action  describes the Almheiri-Polchinski (AP) model \cite{Almheiri:2014cka}.  As we will show, this theory describes an eternal ${\bf AdS}^2$ black hole.
    
    By varying the action with respect to the dilaton field $\phi$ we obtain the constraint $R+2\equiv 0$, i.e. spacetime has a constant negative scalar curvature. In fact,  the corresponding spacetime is locally ${\bf AdS}^2$ given by the Poincare metric
      \begin{eqnarray}
     ds^2=g_{\mu\nu}dx^{\mu}dx^{\nu}=\frac{1}{z^2}(-dt^2+dz^2)=-\frac{4}{(x^+-x^-)^2}dx^+dx^-~,~x^{\pm}=t\pm z.
      \end{eqnarray}
      Furthermore, by varying the action with respect to the metric $g_{\mu\nu}$ we obtain equations of motion coupling the dilaton field to the bulk CFT matter (stress-energy-momentum tensor).
      
    The gravitational sector will be treated semi-classically, i.e. we replace the stress-energy-momentum tensor by its expectation value, viz $T_{ab}=\langle T_{ab}\rangle$. For $\langle T_{ab}\rangle=0$ the dilaton field is found to be given by
       \begin{eqnarray}
     \phi=2\bar{\phi}_r\frac{1-(\pi T_0)^2x^+x^-}{x^+-x^-}.
       \end{eqnarray}
       This dilaton field $\phi$ represents an eternal black hole with two asymptotic boundaries and a Hawking temperature $T_0$. This black hole can be rewritten into a static form by means of the following conformal transformation 
        \begin{eqnarray}
     x^{\pm}=f(y^{\pm})=\frac{1}{{\pi T_0}}\tanh {\pi T_0} y^{\pm}.
        \end{eqnarray}
        The coordinates $x$ cover the whole spacetime whereas $y$ cover the exterior of the black hole. The physical boundary $z=0$ of ${\bf AdS}^2$ is located in the $y$ coordinates at $y^+-y^-=0$. The diffeomorphism $x=f(y)$ is in fact such that the boundary is simple at constant value in the $y$ coordinates, viz
    \begin{eqnarray}
      \frac{y^+-y^-}{2}=\epsilon.
      \end{eqnarray}
    The physical boundary proper time $u$ is different from the bulk Poincare time $t$ near the boundary  (boundary particle formulation of JT gravity \cite{Maldacena:2016upp,Engelsoy:2016xyb}). The physical time $u$ corresponds to $(y^++y^-)/2$. The physical boundary is defined by the boundary conditions (with $\epsilon\longrightarrow 0$)
      \begin{eqnarray}
     g_{uu}|_{b}=\frac{1}{z^2}(-t^{'2}+z^{'2})=-\frac{1}{\epsilon^2}~,~\phi|_{b}=\frac{\bar{\phi}_r}{\epsilon}.
    \end{eqnarray}
    With these boundary conditions the JT action reduces to a boundary term given by the so-called Schwarzian action, viz
       \begin{eqnarray}
            S_G=\frac{1}{8\pi G_N}\int_{\partial\cal M}\phi|_b K=\frac{\bar{\phi}_r}{8\pi G_N}\int du \{f(u),u\}.
    \end{eqnarray}
  The diffeomorphism relating the boundary proper time $u$ to the Poincare time $t$ is precisely given in terms of the function $f$ which converted the black hole into its static form  and made the boundary simple located at constant value in the $y$ coordinates. This function $f$ is also what appears in the  Schwarzian action.
    
  The Noether charge under physical time translations $u\longrightarrow u+\delta u$ is precisely the ADM energy of the ${\bf AdS}^2$ black hole. This is given explicitly by \cite{Maldacena:2016upp,Engelsoy:2016xyb} 
  \begin{eqnarray}
    E(u)=-\frac{\bar{\phi}_r}{8\pi G_N}\{f(u),u\}=\frac{\pi\bar{\phi}_rT_0^2}{4G_N}\equiv E_0.
  \end{eqnarray}

        \subsection{The coupling to conformal matter}

This ${\bf AdS}^2$ black hole is static and does not radiate and evaporate (black hole is in thermal equilibrium at temperature $T_0$). In order for Hawking radiation to escape to infinity we couple the right boundary of this ${\bf AdS}^2$ black hole to a heat bath $B$ at zero temperature. This heat bath $B$ is given by an identical copy of the bulk conformal field theory. The coupling will produce a transient effect given by an infalling shock of positive energy $E_S$. After this initial transient effect, which is due to the coupling between the eternal black hole and the heat bath, Hawking evaporation of the black hole begins. The ${\bf AdS}^2$ black hole becomes of energy $E_1$ and temperature $T_1$ given by
   \begin{eqnarray}
           E_1=E_S+E_0~,~ E_1=\frac{\pi\bar{\phi}_rT_1^2}{4G_N}.
   \end{eqnarray}
 The initial transient shock is required to satisfy the ANEC (Averaged Null Energy Condition) on the horizon in order to maintain boundary causality and prevent the formation of a traversable wormhole. In particular, there should be  no interaction between the left and right boundaries and the new event horizon must lie outside the original event horizon. For more detail see \cite{Maldacena:2018lmt,Galloway:2018dak,Faulkner:2016mzt}.

 The stress-energy-momentum tensor determines the dynamics of the dilaton field. The computation of the stress-energy-momentum tensor is greatly simplified by assuming conformally invariant matter (the heat bath is an identical copy of the bulk CFT) and conformally invariant boundary conditions. In fact, the expectation value of the stress-energy-momentum tensor is completely determined by the conformal anomaly $c$.

 As we have already shown, the  ingoing stress-energy-momentum tensor is explicitly given by the expectation value \cite{Engelsoy:2016xyb}
     \begin{eqnarray}
          \langle T_{x^{-}x^{-}}(x^{-})\rangle=E_S\delta(x^{-})-\frac{c}{24\pi}\{y^{-},x^{-}\}\theta(x^{-}).
   \end{eqnarray}
   The  energy of the black hole $E(u)$ can be obtained by varying the action $S_G+S_{CFT}[g]$ with respect to the boundary time $u$ giving the  change in energy $\partial_uE(u)$ as the difference between the ingoing and incoming fluxes, viz
     \begin{eqnarray}
          \partial_uE(u)=f^{'2}(u)\bigg(T_{x^{-}x^{-}}(x^{-})-T_{x^{+}x^{+}}(x^{+})\bigg).
   \end{eqnarray}
    Explicitly, the energy of the black hole is found to be given by 
\begin{eqnarray}
          E(u)=E_0\theta(-u)+E_1\theta(u)e^{-ku}~,~k=\frac{G_Nc}{3\bar{\phi}_r}\ll 1.
\end{eqnarray}
For $u<0$ the diffeomorphism $f$ is given explicitly by 
  
  \begin{eqnarray}
     t=f(u)=\frac{1}{{\pi T_0}}\tanh {\pi T_0} u~,~u<0\Rightarrow  \{u,t\}=2(\pi T_0)^2.
  \end{eqnarray}
  The horizon $z\longrightarrow\infty$ of ${\bf AdS}^2$ corresponds in the $y$ coordinates either to the future horizon $y^+\longrightarrow +\infty$ or to the past horizon  $y^{-}\longrightarrow -\infty$. In the $x$ coordinates this horizon corresponds to  
    \begin{eqnarray}
     x^{\pm}=\pm \frac{1}{\pi T_0}.
    \end{eqnarray}
    This is precisely the Poincare time $t_{\infty}$ at which the boundary particle reaches the horizon $u\longrightarrow\pm \infty$, viz
    \begin{eqnarray}
     t_{\infty}={\rm lim}_{u\longrightarrow\pm \infty}f(u)=\pm \frac{1}{\pi T_0}.
    \end{eqnarray}
For $u>0$ this diffeomorphism $f$ is given explicitly by a complicated expression in terms of Bessel functions. This expression is crucial at very late times $u\sim k^{-1}\ln k$ when the black hole is almost extremal and the semi-classical treatment is no longer trusted. The Poincare time $t_{\infty}$ at which the boundary particle reaches the horizon $u\longrightarrow+\infty$ is given now by
    \begin{eqnarray}
     t_{\infty}={\rm lim}_{u\longrightarrow + \infty}f(u)=\frac{1}{\pi T_1}+\frac{k}{4\pi^2T_1^2}+O(k^2).
    \end{eqnarray}
    The causality requirement $t_{\infty}<1/\pi T_0$ (the new event horizon is required to lie outside the original horizon and hence it can be reached in less time) gives the lower bound $E_S>cT_0/24$.

         At early times $u\longrightarrow 0$ the black hole starts giving up Hawking radiation through the right boundary. At these times the diffeomorphism $f$ is given by 
       \begin{eqnarray}
     f(u)=\frac{1}{{\pi T_1}}\tanh {\pi T_1} u~,~u<0\Rightarrow \{u,t\}=\frac{2(\pi T_1)^2}{(1-(\pi T_1t)^2)^2}.
       \end{eqnarray}
At late times $u\sim k^{-1}$ the black hole is still far from being extremal despite the continued Hawking radiation through the right boundary. At these times the diffeomorphism $f$ is given through the equation
     \begin{eqnarray}
     \{u,t\}=\frac{1}{2(t_{\infty}-t)^2}\big(1+O(k^2e^{-ku})\big).
     \end{eqnarray}
     The double pole is shifted to $t_{\infty}$ which is crucial for the behavior of the quantum extremal surfaces.

     The ${\bf AdS}^2$ black hole solution after the shock involves a different dilaton field configuration which is given explicitly by \cite{Almheiri:2014cka}
     \begin{eqnarray}
     \phi=2\bar{\phi}_r\frac{1-(\pi T_1)^2x^+x^-+\frac{1}{2}kI(x^+,x^-)}{x^+-x^-}~,~I(x^+,x^-)=\int_0^{x^-} dt(x^+-t)(x^--t)\{u,t\}.
     \end{eqnarray}

        \subsection{The Hartle-Hawking state}

    Starting in this section, we revert to the notation of \cite{Almheiri:2019psf}.

    The matter sector is given by a two-dimensional conformal field theory which couples to the dilaton field only through the constraints. Thus, this CFT$_2$ will be treated as a quantum field theory on a fixed ${\bf AdS}^2$ background.

    The boundary conditions at the boundary of ${\bf AdS}^2$ are also chosen to be conformally invariant, e.g. reflecting boundary conditions.

    The initial state is prepared by a path integral in Euclidean signature (Wick rotation to imaginary time: $t\longrightarrow \tau=i t$). The Euclidean metric can be given by
    \begin{eqnarray}
      ds^2=\frac{4dxd\bar{x}}{(x+\bar{x})^2}~,~-x\equiv x^-=-z-i\tau~,~\bar{x}\equiv x^+=z-i\tau. 
      \end{eqnarray}
  The initial state on ${\bf AdS}^2$ (after a conformal scale transformation to the metric $dxd\bar{x}$) is  therefore given by the vacuum on the half-line $z>0$ with conformally invariant boundary conditions at $z=0$. This is the Hartle-Hawking state on the ${\bf AdS}^2$ black hole at $t=0$.

  The construction of the initial state of the total system formed by the ${\bf AdS}^2$ black hole and the heat bath is much more involved. 

  First, recall that the heat bath contains an identical copy of the bulk CFT$_2$ with identical boundary conditions. The time parameter in the bath is precisely the physical time $u$ at the boundary whereas the time parameter in the bulk is the Poincare time $t=f(u)$. We can use the diffeomorphism $f$ to define the coordinates $x=f(y)$ and $\bar{x}=f(\bar{y})$ in which the metric is given by 
     \begin{eqnarray}
       ds^2=\Omega^{-2}(y,\bar{y})dyd\bar{y}~,~\Omega^{-2}=\frac{4f^{\prime}(y)f^{\prime}(\bar{y})}{(f(y)+f(\bar{y}))^2}.
     \end{eqnarray}
     Thus, by a Weyl transformation $\Omega_y$ we can go to the flat metric $dyd\bar{y}$.

     Recall also that in the $y$ coordinates the  ${\bf AdS}^2$ boundary is at $(y+\bar{y})/2=0$ whereas $(y-\bar{y})/2$ corresponds to the physical time $u$.
       The left half-line $y+\bar{y}<0$ corresponds to the bath while the right half-line $y+\bar{y}>0$ corresponds to the  bulk or ${\bf AdS}^2$ black hole.

       Obviously, the evaporation process of the black hole corresponds to $u>0$. We extend $f$ to $u<0$ by imposing $f(-u)=-f(u)$. Thus, we obtain a time-reversal invariant Hamiltonian of the total system formed by the ${\bf AdS}^2$ black hole and the heat bath. This choice makes also the dynamics highly non-trivial, i.e. it makes the Hamiltonian of the total system  genuinely coupled.
%          \end{itemize}
%\end{frame}

%\begin{frame}
%  \frametitle{The Hartle-Hawking state}
  
%  \begin{itemize}
     The initial state on ${\bf AdS}^2$ is  given by the vacuum on the right half-line $y+\bar{y}>0$ with conformally invariant boundary conditions at $y+\bar{y}=0$.  In general, the state on ${\bf AdS}^2$ is given by the vacuum on the right half-cylinder ${\rm Re}~y>0$ (since the matter sector, after Wick rotation, is in fact given by a thermal field theory which is periodic with period $T^{-1}$). 

     The initial state in the heat bath  is  given by the vacuum on the left half-line $y+\bar{y}<0$ with conformally invariant boundary conditions at $y+\bar{y}=0$. In general, the state in the heat bath is given by the vacuum on the left half-plane ${\rm Re}~y<0$.

     The total Hamiltonian of the  coupled system formed by the ${\bf AdS}^2$ black hole and the heat bath is then given by the standard CFT$_2$ Hamiltonian on the line with the times identified in such a way that the state in the right half-line is a descendant of the half-line vacuum with some complicated boundary.

     This state is then mapped to the half-space vacuum by means of the diffeomorphism $x\longrightarrow w=\omega(x)$, $\bar{x}\longrightarrow \bar{w}=\bar{\omega}(\bar{x})$, i.e. this state is defined in the half-space $w+\bar{w}>0$. The metric in the coordinate $w$, $\bar{w}$ is defined by
       \begin{eqnarray}
       ds^2=\Omega^{-2}(w,\bar{w})dwd\bar{w}~,~\Omega=\frac{x+\bar{x}}{2}\sqrt{w^{\prime}(x)\bar{w}^{\prime}(\bar{x})}.
       \end{eqnarray}
     At $t=0$,  the diffeomorphism or conformal transformation $w(x)$ to the coordinates $w$ is determined by means of conformal invariance to be a Mobius map of $x=f(y)$. The ${\bf AdS}^2$ region $x>0$ is mapped to $w\in[0,w_0]$ and the bath region $y<0$ is mapped to $[w_0,\infty[$ under the following transformation to the upper half-plane:
          \begin{eqnarray}
       w(x)=\frac{w_0^2}{w_0+x}~,~x>0.
          \end{eqnarray}
               \begin{eqnarray}
       w(x)=w_0+f^{-1}(-x)~,~x<0.
          \end{eqnarray}
             By using the conformal anomaly we can show that the scale $w_0$ is related to the energy $E_S$ of the shock by the relation
                \begin{eqnarray}
E_S=\frac{c}{24\pi}\bigg(\frac{2}{w_0}-f^{\prime\prime}(0)\bigg).
                \end{eqnarray}
              In order to avoid acausal influences we must take $E_S\longrightarrow\infty$ or $w_0\longrightarrow 0$. Indeed, ${\bf AdS}^2$ is dual to a conformal quantum mechanics at the boundary and thus it is more natural to map the ${\bf AdS}^2$ region to a single point. In this limit the above conformal transformation becomes
          \begin{eqnarray}
       w(x)=\bigg(\frac{c}{12 \pi} E_S\bigg)^2\frac{1}{x}~,~x>0.
          \end{eqnarray}
               \begin{eqnarray}
       w(x)=f^{-1}(-x)~,~x<0.
               \end{eqnarray}
             In summary, the ${\bf AdS}^2$ black hole is described by the Poincare coordinates $x$. We first work in the conformally flat coordinates $y$ (in the right half-line $y>0$) and then transform back to $x$ at the end of the calculation. The initial state here is a Virasoro descendant of the half-line vacuum (Euclidean path integral with a deformed boundary). In fact, the ${\bf AdS}^2$ state on the half-line $y>0$ is given by an Euclidean path integral on a half-cylinder (since we are dealing with a thermal field theory with a Hawking temperature $T_1$, i.e. the Euclidean time is periodic with period $T_1^{-1}$).

             The bath is described by the coordinates $y$ (in the left half-line $y<0$).  The initial state here is the Hartle-Hawking state which is given by the Euclidean path integral on the left half-line $y<0$.

             The state of the combined system is then described by the standard CFT Hamiltonian on a line (with a complicated shape).

             The state of the combined system is described, in the coordinates $w$, by an Euclidean path integral on the half-line. In other words, the state in the coordinate $w$ is the half-line vacuum (in the region $w+\bar{w}>0$). Everything reduces to calculation in the half-space vacuum in the metric $dwd\bar{w}$ followed by a Weyl transformation to the physical metric $\Omega_w^{-2}dwd\bar{w}$.

             The diffeomorphism $w=w(x)$ is derived from the fact that the stress-energy tensor vanishes in the half-line vacuum. The stress-energy tensor vanishes also in the coordinates $x$ (${\bf AdS}^2$) and the coordinates $y$ (heat bath). The stress-energy tensor picks up an anomaly under the diffeomorphisms $w=w(x)$ and $w=w(y)$. This allows us to conclude that $w$ is Mobius map of $x$ and $y$.

\begin{figure}[htbp]
                        \begin{center}
  \includegraphics[width=10cm,angle=-0]{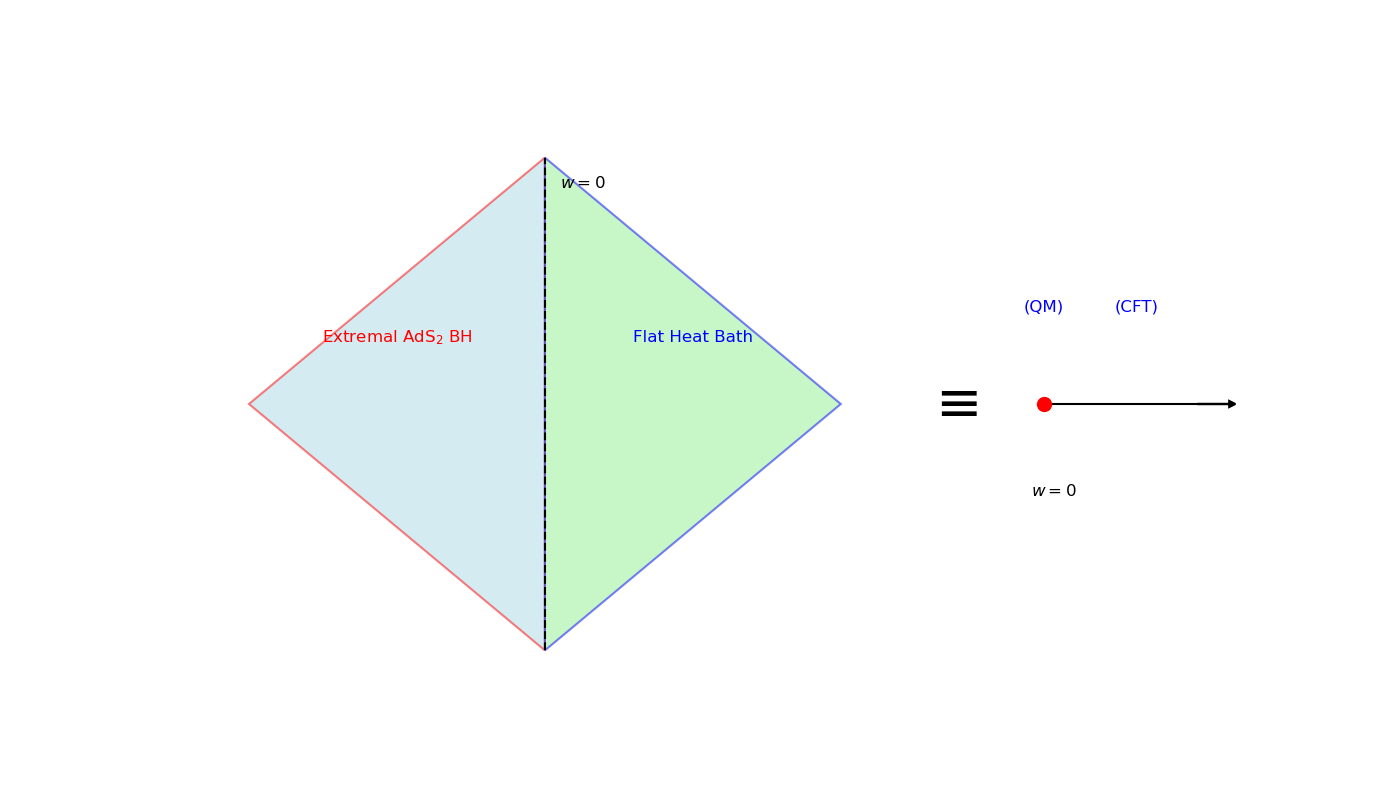}
\end{center}
                        \caption{
                              The whole ${\bf AdS}^2$ theory maps to the point $\sigma=0$ where the boundary is located and where the dual quantum mechanics lives (holographic correspondence).}\label{fig1}
     \end{figure}

             The ${\bf AdS}^2$ is mapped to the region $[0,w_0]$ while the heat bath is mapped to the region $[w_0,\infty[$. In order to avoid acausal action we take $w_0\longrightarrow 0$. Hence, the whole ${\bf AdS}^2$ theory maps to the point $w=0$ where the boundary is located and where the dual quantum mechanics lives (holographic correspondence). See figure (\ref{fig1}).
                      
\subsection{An evaporating ${\bf AdS}^2$ black hole}

 The dilaton configuration including the perturbation ($E_S=E_1-E_0\neq 0$ and $T_1\ne T_0$) and the backreaction ($I\ne 0$) to the future of the shock $x^{-}>0$ is given by
    \begin{eqnarray}
      \phi=2\phi_r\frac{1-(\pi T_1)^2x^+x^{-}+\frac{k}{2}I(x^+,x^-)}{x^+-x^-}.\label{dilaton}
    \end{eqnarray}
    The backreaction integral $I$ is given in terms of the Schwarzian $\{u,t\}$, where $u=f^{-1}(t)$, by the relation
    \begin{eqnarray}
    I(x^+,x^-)=\int_0^{\infty}dt (x^{+}-t)(x^{-}-t)\{u,t\}.
    \end{eqnarray}
  The expansion parameter $k$ is given in terms of the central charge $c$ and Newton's constant $G_N$ by
     \begin{eqnarray}
    k=\frac{c}{12}\frac{4G_N}{\bar{\phi}_r}.
    \end{eqnarray}
Recall that the diffeomorphism $t=f(u)$ (which relates the boundary proper time $u$ to the Poincare time $t=(x^++x^-)/2$) is given for $u<0$ (bath) by the simple expression
     \begin{eqnarray}
   t=f(u)=\frac{1}{\pi T_0}\tanh(\pi T_0u).
     \end{eqnarray}
     In the region $u>0$ this diffeomorphism is more complicated (it is required to solve the condition $\{t,u\}=2(\pi T_1)^2\exp(-ku)$). For the unperturbed black hole ($E_S=0$ and $k=0$) we get the same reparameterization. But for the perturbed black hole with backreaction the reparameterization $f(u)$ is a complicated expression given in terms of Bessel functions \cite{Almheiri:2019psf}.

     The new future horizon in the $x-$coordinates is located at
     \begin{eqnarray}
       x^{+}=t_{\infty}={\rm lim}_{u\longrightarrow \infty}f(u).
     \end{eqnarray}
     This is the required time for a boundary particle to reach the horizon. Indeed,  the physical time $u$, which is  given by $u=\frac{1}{2}(y^++y^-)|_{\rm boundary}$, approaches $\infty$ when we approach the future horizon $y^{+}\longrightarrow +\infty$.

     Recall also that the boundary in the $y-$coordinates is located at the surface $\frac{1}{2}(y^+-y^-)=\epsilon\longrightarrow 0$.
     
     By using the expression of $f(u)$ in the region $u>0$ and expanding in $k$ we obtain the horizon 
     \begin{eqnarray}
       x^{+}=t_{\infty}=\frac{1}{\pi T_1}+\frac{k}{4(\pi T_1)^2}+...
     \end{eqnarray}
     In order to maintain causality we must have $t_{\infty}<1/\pi T_0$ which is the old horizon.

     The relevant Bessel functions found are $K_{0,1}(z)$ and $I_{0,1}(z)$ where the variable $z$ is either $z=2\pi T_1/k$ or $z=2\pi T_1\exp(-ku/2)/k$. This means that the regime of very late times $u\gg 1$ corresponds to $e^{-ku/2}/k\ll 1$ or equivalently $u\gg -\log k/k$. At these times the ${\bf AdS}^2$ black hole becomes extremal and the semi-classical description becomes unreliable.

     We are more interested in late times $u\gg 1$ corresponding to $u\gg 1/k$ where the ${\bf AdS}^2$ black hole has evaporated a sufficient fraction of its mass but remains non-extremal. In this regime of interest the diffeomorphism $t=f(u)$ is given by the Schwarzian equation
     \begin{eqnarray}
      \log\frac{t_{\infty}-f(u)}{2t_{\infty}}=-\frac{4\pi T_1}{k}\big(1-e^{-\frac{ku}{2}}\big)\Rightarrow  \frac{f^{'}(u)}{t_{\infty}-t}=2\pi T_1e^{-\frac{ku}{2}}.\label{diffeo}
     \end{eqnarray}
     This expression is actually obtained at small $k$ keeping $ku$ fixed.

     At early times the diffeomorphism $t=f(u)$ is simply given by
     \begin{eqnarray}
   t=f(u)=\frac{1}{\pi T_1}\tanh(\pi T_1u).
     \end{eqnarray}

\section{Entropy and quantum extremal surface in JT gravity}
\subsection{Entropies, Ryu-Takayanagi formula and the island conjecture}

  In this section we follow the beautiful review   \cite{Almheiri:2020cfm}.

The most fundamental result concerning classical black holes is the Bekenstein-Hawking entropy formula \cite{Bekenstein:1972tm,Bekenstein:1973ur}. This states that the entropy $S_{B-H}$ of a classical black hole is proportional to the area $A$ of the event horizon, viz

\begin{eqnarray}
S_{\rm B-H}=\frac{A}{4\hbar G_N}.
\end{eqnarray}
By including the entropy $S_{\rm outside}$ of matter and gravitons in the outside region of the black hole we obtain the Bekenstein-Hawking total entropy

\begin{eqnarray}
S_{\rm gen}=\frac{A}{4\hbar G_N}+S_{\rm outside}.
\end{eqnarray}
For classical black holes the area always increases in time and hence we obtain the second law of thermodynamics. In fact, we obtain the second law of thermodynamics even if we include the outside entropy, viz

\begin{eqnarray}
\Delta S_{\rm gen}\geq 0.
\end{eqnarray}
However, it was shown by Hawking that a quantum black holes evaporate at a temperature $T$ proportional to the surface gravity $\kappa$ of the black hole \cite{Hawking:1974rv,Hawking:1974sw}. More precisely, we have

  \begin{eqnarray}
    T=\frac{\hbar \kappa}{2\pi}.
  \end{eqnarray}
In other words, surface gravity and the area of the horizon are conjugate variables in general relativity in the same way that temperature and entropy are conjugate variables in thermodynamic. 

Thus, an evaporating quantum black hole should be described instead by the  quantum von Neumann-Landau entropy defined in terms of the density matrix $\rho$ by the standard formula 

  \begin{eqnarray}
    S_{vN}=-{\rm Tr}\rho ln\rho.
  \end{eqnarray}
This vanishes for a pure state, i.e. for $\rho=|\psi\rangle\langle \psi|$ we have $S_{vN}=0$. Thus, $S_{vN}$ measures the degree of mixing of the quantum state, i.e. our ignorance about the quantum state of the system. This entropy is precisely Shannon's entropy of information.   

The Bekenstein-Hawking entropy should be thought of as the coarse-grained Gibbs thermodynamic entropy (measures Boltzmann's number of microscopic states of the black hole). In contrast von Neumann entropy is thought of as the fine-grained entropy of the black hole (measures Bell's quantum entanglement characterizing the quantum state of the black hole). The coarse-grained entropy is obtained from the fine-grained entropy by a maximization procedure over all possible choices of density matrices. Thus, we must have

  \begin{eqnarray}
    S_{vN}\leq S_{\rm B-H}.
  \end{eqnarray}
The generalized entropy should also be thought of as a coarse-grained thermodynamic entropy.

In contrast to the coarse-grained Bekenstein-Hawking entropy which can only increase in time, if there is no black hole evaporation, the fine-grained von Neumann entropy can both increase or decrease in time after the start of the evaporation process of the black hole.

An evaporating black hole can thus be viewed, by an outside observer located at infinity, as a unitary quantum system with a total number of degrees of freedom given by the area term $A/4\hbar G_N$ of the Bekenstein-Hawking entropy. This area term is precisely the logarithm of the dimension of the Hilbert space of quantum states of the black hole. 

The most important discovery regarding von Neumann entropy in the past 20 years, which was originally motivated by the gauge/gravity duality and quantum entanglement, is the result that the fine-grained von Neumann  entropy  of quantum systems coupled to gravitational theories can be computed using the so-called Ryu-Takayanagi formula \cite{Ryu:2006bv,Hubeny:2007xt} which is a quantum generalization of the Bekenstein-Hawking fromula where the horizon is replaced with the so-called quantum extremal surfaces. 

\begin{figure}[htbp]
   \begin{center}
     \includegraphics[width=10cm,angle=-0]{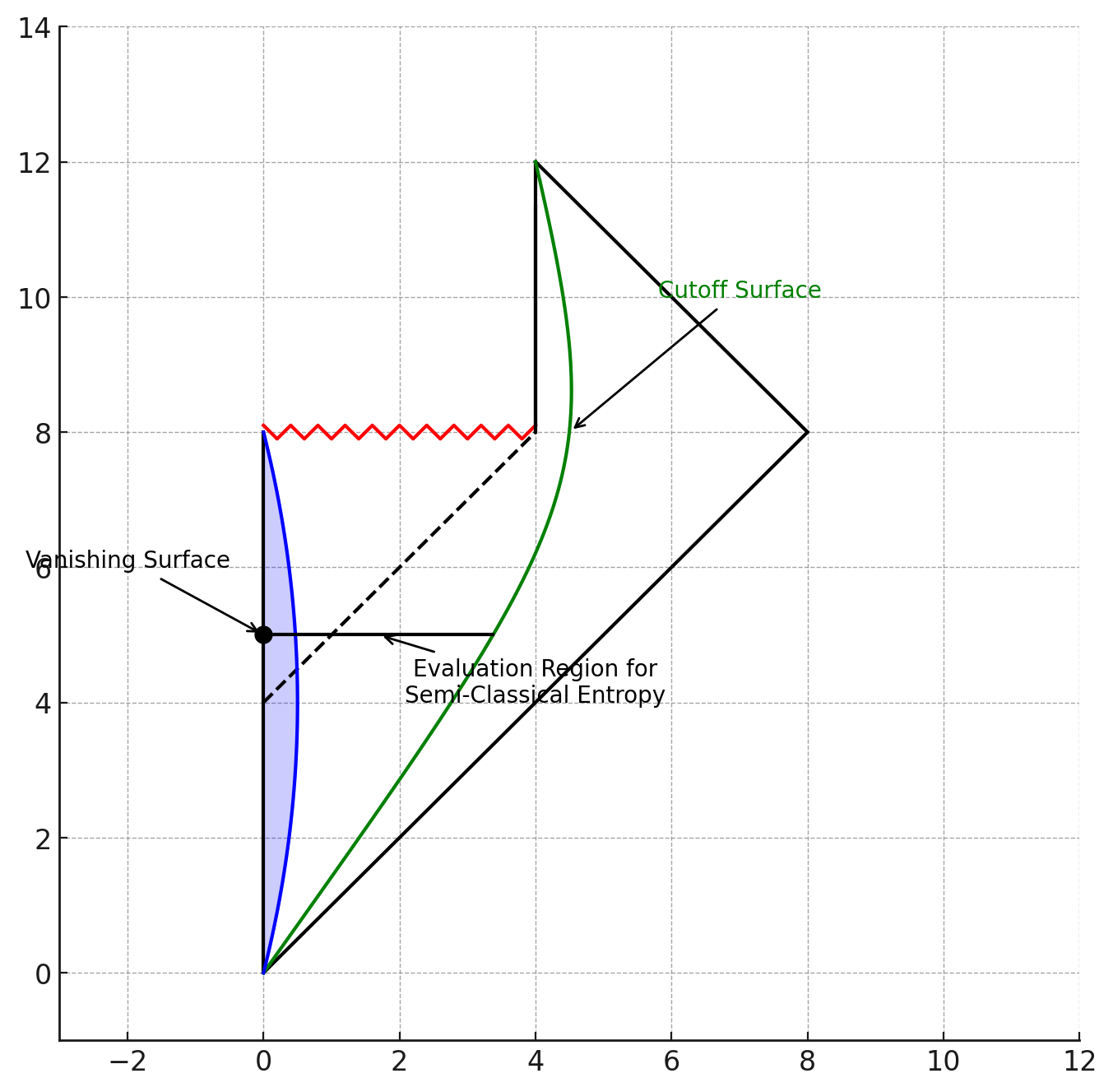}
     \includegraphics[width=10cm,angle=-0]{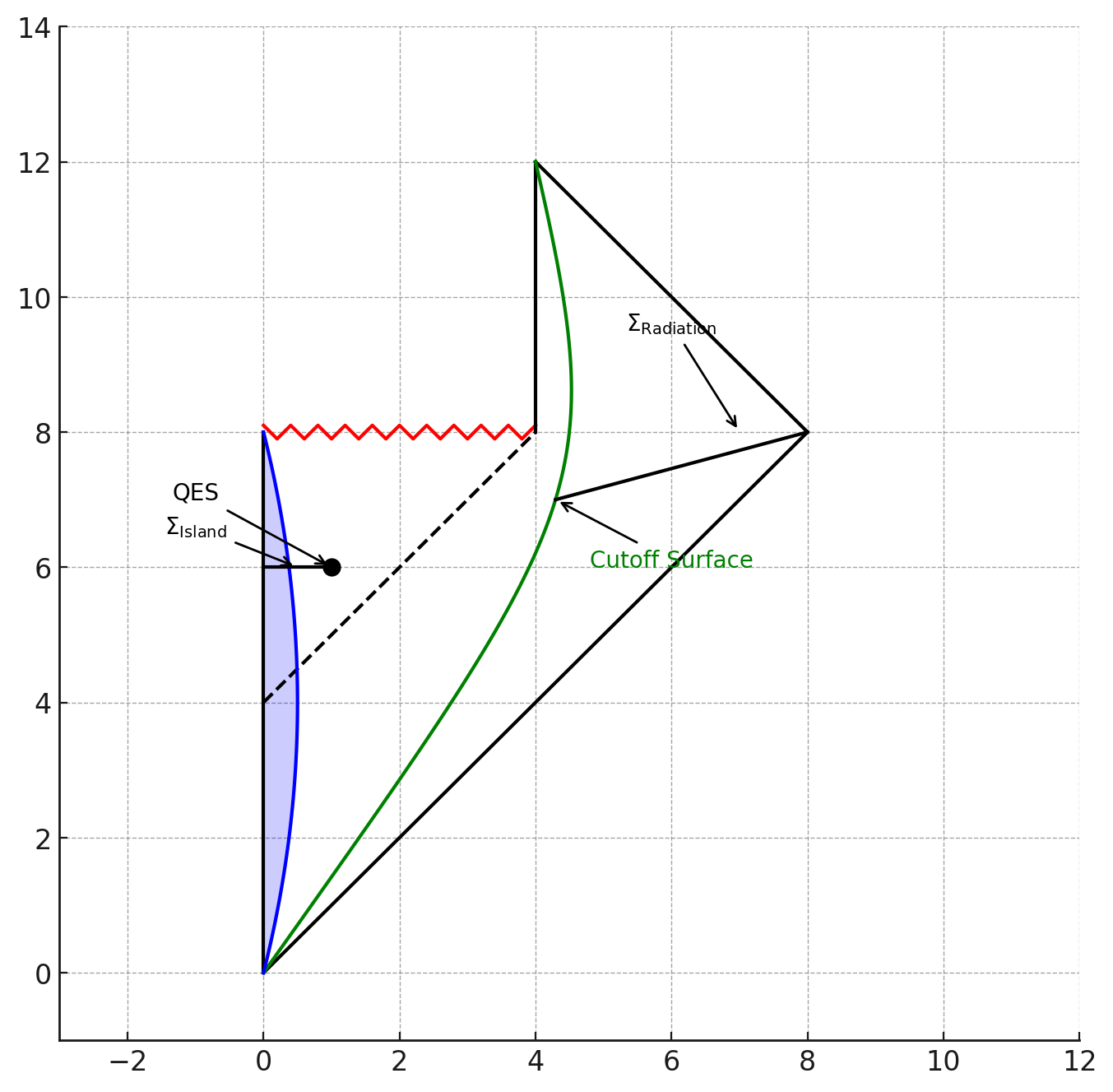}
\end{center}
\caption{The vanishing surface (before Page time) and quantum extremal surface or QES (after Page time).}\label{fig2}
  \end{figure}

In this formulation a codimension-2 surface $X$ of area $A(X)$ is used to compute the generalized entropy given by the formula 

  \begin{eqnarray}
    S_{\rm gen}(X)&=&\frac{A(X)}{4\hbar G_N}+S_{\rm semi-cl}(\Sigma_X).
  \end{eqnarray}
Here, $\Sigma_X$ is the region bounded by a cutoff surface (an arbitrarily chosen surface demarcating the black hole system and separating it from its Hawking radiation) and the codimension-two surface $X$. The entropy $S_{\rm semi-cl}(\Sigma_X)$ is then the von Neumann entropy of the quantum fields of matter and gravitons which are propagating on the classical geometry of the region $\Sigma_X$ (semi-classical approximation). This fine-grained quantum entropy is therefore computed using the density matrix $\rho_{\Sigma_X}$ of the region $\Sigma_X$, viz 

\begin{eqnarray}
  S_{\rm semi-cl}(\Sigma_X)&=&S_{\rm vN}(\rho_{\Sigma_X}).
\end{eqnarray}
The codimension-2 surface $X$ is a surface which has two dimensions less than the embedding spacetime and it is chosen in such a way that the generalized entropy is minimized in the spatial direction but maximized in the temporal direction. In other words, this surface is in fact an extremal surface, i.e. the generalized entropy takes an extremal value on this so-called ''quantum extremal surface''. This extremal value of the generalized entropy is precisely the fine-grained von Neumann entropy of the black hole system, viz 

  \begin{eqnarray}
    S_{b-h}&=&{\rm min}_X\{{\rm ext}_X\big[S_{\rm gen}[X]\big]\}\nonumber\\&=&{\rm min}_X\{{\rm ext}_X\bigg[\frac{A(X)}{4\hbar G_N}+S_{\rm semi-cl}(\Sigma_X)\bigg]\}.
  \end{eqnarray}
The quantum extremal surface can be shown to lie behind the event horizon. There are two different surfaces which dominate respectively the early and late times of the evaporation process. At early times the quantum extremal surface is found to be the vanishing surface, i.e. a trivial surface of zero size. At late times the quantum extremal surface is found to be a non-vanishing surface which lies just behind the event horizon. See figure (\ref{fig2}).

The so-called ''entanglement wedge'' of this fine-grained entropy is the region bounded between the cutoff surface and the quantum extremal surface, i.e. it includes only a portion of the interior of the black hole. The degrees of freedom of the black hole in this region describe the geometry up to the extremal surface.

In more detail, at very early times there is only the vanishing quantum extremal surface.   Thus, at these very early times, the entropy of the area term is zero. Furthermore, since no Hawking modes has enough time to escape the black hole region at these very early times, the semi-classical von Neumann entropy of the matter enclosed by the cutoff and the vanishing surfaces is zero because there is no quantum entanglement. 

At early times when some Hawking radiation has the chance to escape the black hole region the semi-classical fine-grained von Neumann entropy of the matter modes enclosed by the cutoff and the vanishing surfaces becomes non-zero due to the quantum entanglement between these inner modes and the outer modes of the Hawking radiation. This entropy increases as the black hole evaporates and thus as more Hawking modes escape or more matter modes accumulate. This increasing semi-classical entropy dominates the generalized entropy as the entropy of the area term always vanishes for the vanishing surface. In other words, the generalized entropy of the vanishing quantum extremal surface is increasing in time.

However, at some early time another quantum extremal surface appears. This surface is non-vanishing and time-dependent and lies close to the event horizon.  This surface can be found as follows. At time $t$ on the cutoff surface (which determines how much Hawking radiation has escaped) we must go backward an amount of time of the order of the so-called srcambling time $r_s\ln S_{B-H}$ and then shoot a light ray towards the event horizon. It is near the intersection point behind the event horizon where the non-vanishing surface is found. The entropy of the area term is now non-zero given precisely by the entropy of the black hole which is decreasing in time as the area of the black hole is constantly shrinking due to the evaporation process. The semi-classical von Neumann entropy is clearly negligible  compared to the Bekenstein-Hawking entropy . In other words, the generalized entropy of the non-vanishing quantum extremal surface is decreasing in time.

In summary, the generalized entropy of the vanishing surface is dominated by the area term and is increasing while the generalized entropy of the non-vanishing surface is dominated by the semi-classical entropy and is decreasing. Thus, the vanishing surface gives the minimum at early times whereas the non-vanishing surface gives the minimum at late times and as a consequence the generalized entropy goes through a maximum at some time called the Page time. In other words, the generalized entropy follows the so-called Page curve where a phase transition at the Page time occurs between the vanishing and non-vanishing surfaces. See figure (\ref{fig0}).

This behavior can be seen more explicitly as follows. For a generic surface $X$ (which starts on the horizon) the area term is always decreasing as we move the surface inward along a null direction whereas the semi-classical entropy starts by decreasing (since the included interior modes purify the exterior modes inside the black hole region) and then becomes increasing (since at this stage the interior modes are  entangled with the Hawking modes outside the black hole region). It is in this regime where the area term is decreasing while the semi-classical entropy is increasing and thus their derivatives can be balanced, i.e. the switch or phase transition  from the vanishing surface to the non-vanishing surface occurs.

By employing the assumption of unitarity we know that the black hole system  and the Hawking radiation system should be described by a pure state. This means that the fine-grained entropy of the Hawking radiation should be equal to the fine-grained entropy of the black hole which must always be less than the Bekenstein-Hawking entropy, viz  

  \begin{eqnarray}
    S_{\rm rad}={\rm S}_{\rm b-h}\leq S_{B-H}
  \end{eqnarray}
As we have discussed, the fine-grained entropy region (entanglement wedge) of the black hole system is bounded between the cutoff surface and the quantum extremal surface $X$. 

\begin{figure}[htbp]
   \begin{center}
     \includegraphics[width=10cm,angle=-0]{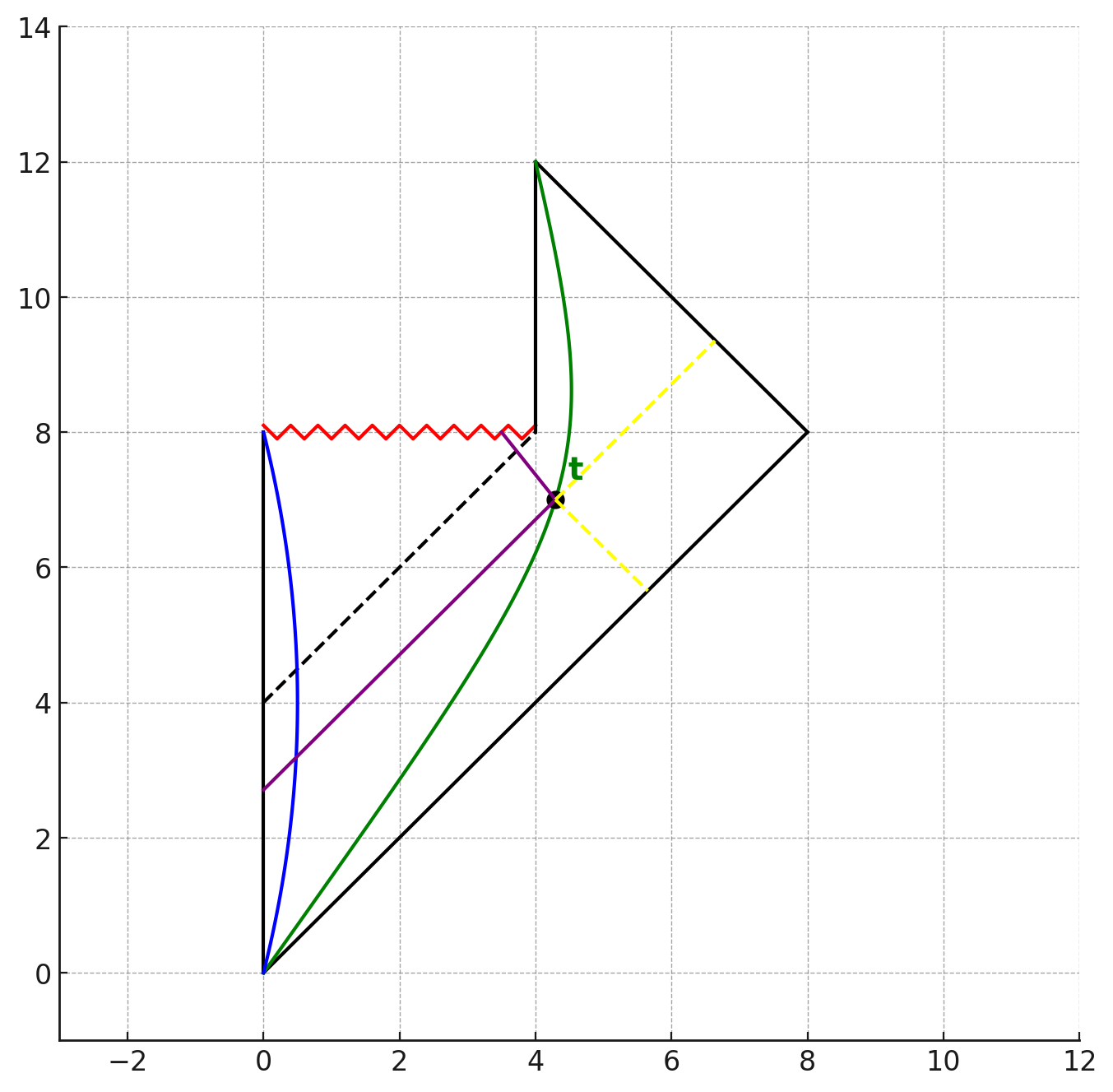}
      \includegraphics[width=10cm,angle=-0]{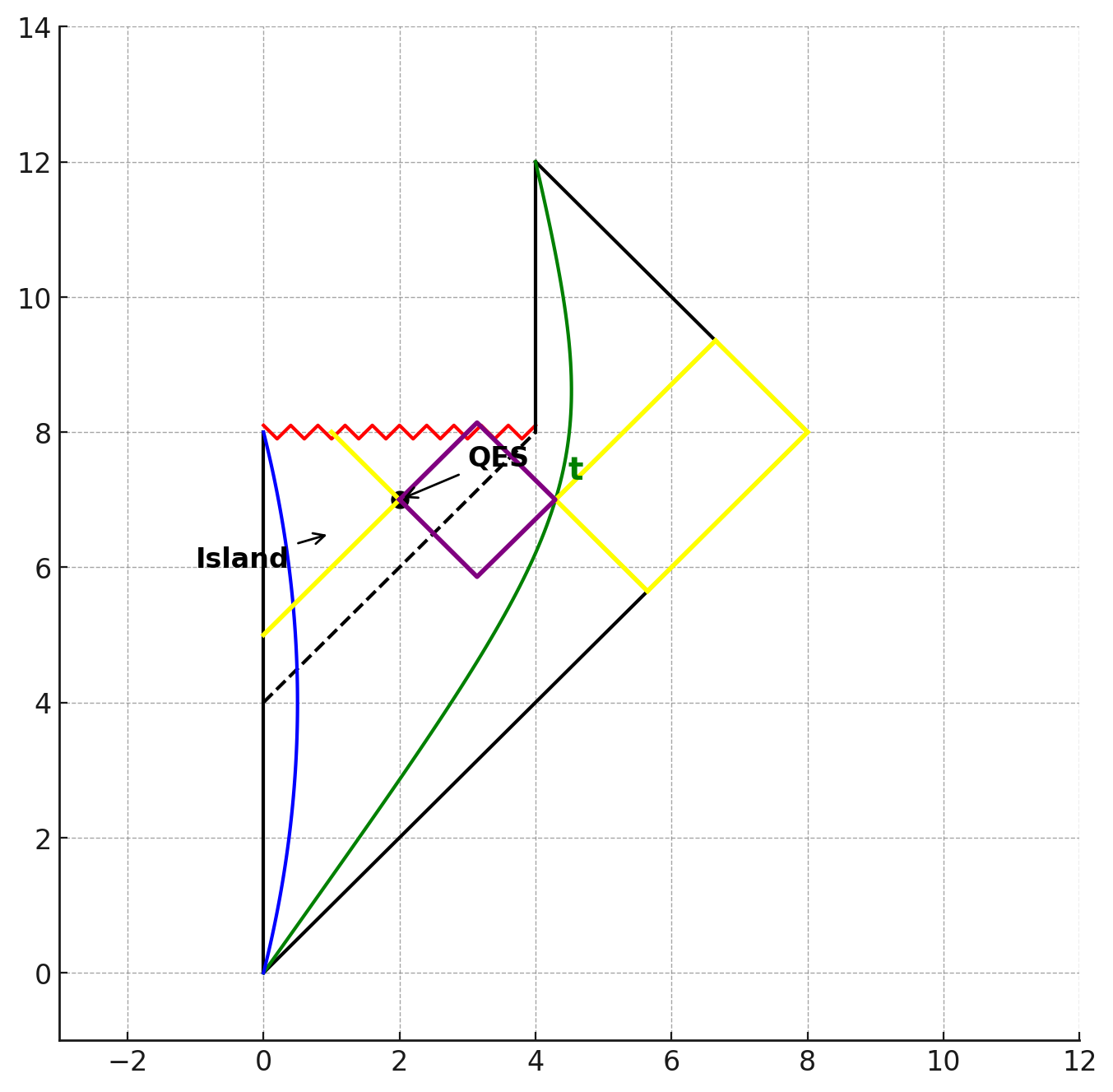}
\end{center}
\caption{Entanglement wedges before (first diagram) and after (second diagram) the Page time. The radiation wedge is drawn in yellow whereas the black hole wedge is drawn in purple. The island is counted in the radiation wedge after the Page time. }\label{fig3}
  \end{figure}

Naively, the fine-grained entropy region (entanglement wedge) of the Hawking radiation system should be located beyond the cutoff surface where Hawking radiation has escaped and where gravity can be neglected and spacetime is flat. However, and as it turns out, the entanglement wedge of the Hawking radiation is a disconnected surface which contains, in addition to the region beyond the cutoff surface, the region inside the black hole behind the quantum extremal surface $X$. See figure (\ref{fig3}).

The degrees of freedom in the Hawking radiation are obviously entangled with the degrees of  freedom in the interior of the black hole. Thus, the fine-grained von Neumann entropy of the emitted Hawking radiation which has escaped beyond the cutoff surface increases steadily in the early stages of the evaporation. As we accumulate more radiation the entropy keeps rising until it reaches in value the Bekenstein-Hawking entropy which defines the maximum number of degrees of freedom contained originally in the black hole, i.e. it defines the maximum number of degrees of freedom which the radiation can be entangled with. The time at which the fine-grained entropy of the radiation ceases increasing and starts decreasing is precisely the Page time and it is the time when the quantum extremal surface $X$ changes or jumps from the trivial or vanishing surface to the non-trivial quantum extremal surface which lies just behind the horizon.

The fine-grained von Neumann entropy of the radiation contains no area term and the semi-classical entropy of the region $\Sigma_{\rm rad}$ beyond the cutoff surface is always increasing. The area can be increased and the semi-classical entropy $S_{\rm semi-cl}(\Sigma_{\rm rad})$ can be decreased if we modify the region where we compute the entropy, i.e. the entanglement wedge of the radiation in such a way that it becomes a disconnected region (the area term becomes thus increasing) which contains entangled matter in far away parts (the semi-classical entropy becomes increasing). 

The only obvious solution to achieve this outcome is to modify the entanglement wedge of the radiation in such a way that it includes a portion of the interior of the black hole. More precisely,  the region $\Sigma_{\rm rad}$ is replaced with the disconnected region  $\Sigma_{X}^{'}=\Sigma_{\rm rad}+\Sigma_{\rm island}$ where $\Sigma_{\rm island}$ represents the portion of the black hole interior which must be included in the entanglement wedge of the radiation. This region $\Sigma_{\rm island}$  is also known as an ''island" and it is centered around the origin. It appears a time $r_s\log S_{B-H}$ (scrambling time) after the formation of the black hole. The boundary of the island, i.e. the boundary of the disconnected region $\Sigma_{X}^{'}$  is precisely given by the quantum extremal surface $X$ and thus the area term is given as before by $A(X)/4\hbar G_N$. 

The fine-grained von Neumann entropy of the radiation is then given explicitly by

  \begin{eqnarray}
    S_{\rm rad}&=&{\rm min}_X\{{\rm ext}_X\big[S_{\rm gen}[X]\big]\}\nonumber\\&=&{\rm min}_X\{{\rm ext}_X\bigg[\frac{A(X)}{4\hbar G_N}+S_{\rm semi-cl}(\Sigma_{\rm rad}+\Sigma_{\rm island})\bigg]\}.
  \end{eqnarray}   
This is the von Neumann entropy of the exact quantum state of the radiation computed using only the state of the radiation in the semi-classical approximation. This island formula can be derived from the gravitational path integral. Also, it is not difficult to show that the von Neumann entropies of the Hawing radiation and of the black hole system are identically equal. Indeed, we have $S_{\rm rad}=S_{\rm b-h}$ since the area term is the same in both cases and by using the properties of quantum entanglement we have $S_{\rm semi-cl}(\Sigma_{\rm rad}+\Sigma_{\rm island})=S_{\rm semi-cl}(\Sigma_X)$. 

We can check that the von Neumann entropy of the radiation as defined  in terms of the island by the above formula will decrease at late times. In fact, this entropy follows exactly the Page curve. This von Neumann entropy is the minimum of the vanishing and non-vanishing islands contributions. The increasing segment for early times corresponds to the no-island contribution whereas the decreasing segment for late times corresponds to the with-island contribution.


\subsection{Entanglement entropy of an interval in CFT$_2$ and ${\bf AdS}^2$}

    In this section we follow \cite{Almheiri:2019psf}.

    The computation of the fine-grained von Neumann entropy in ${\bf AdS}^2$ is equivalent to the use of conformal field theory techniques to calculate entropies in curved spacetime settings. The basic technique is the replica trick \cite{Faulkner:2013ana,Lewkowycz:2013nqa,Dong:2016hjy,Dong:2017xht}. The von Neumann entropy $S$ is obtained as the limit $n\longrightarrow 1$ of the so-called  Renyi entropy $S_n$, viz
    \begin{eqnarray}
      S&=&-\rho\log \rho=S_n~,~n\longrightarrow 1\nonumber\\
      S_n&=&\frac{1}{1-n}Tr\log\rho^n.
    \end{eqnarray}
    To find the von Neumann entropy $S$, we use the replica trick. The idea is to first compute the Renyi entropy $S_n$ for integer $n$, and then analytically continue the expression to non-integer $n$, eventually taking the limit as $n\longrightarrow 1$. We can apply L'Hopital's rule to the limit $n\longrightarrow 1$ as follows
     \begin{eqnarray}
      S=\frac{f(n)}{1-n}=\frac{f^{\prime}(n)}{-1}=-\frac{d}{dn}\log Tr\rho^n=-\frac{Tr\rho\log\rho}{Tr\rho}=-Tr\rho\log\rho~,~Tr\rho=1.
     \end{eqnarray}
     The Renyi entropy can also be given by
     \begin{eqnarray}
      S_n=\frac{1}{1-n}Tr\rho^n=-\frac{\partial}{\partial n}Tr\rho^n~,~n\longrightarrow 1.
     \end{eqnarray}
   The most important thing in the replica trick is the fact that the calculation of the Renyi entropy involves considering $n$ copies (called replicas) of the system and then cyclically gluing them together along the interval whose entropy is being calculated. This procedure creates a new manifold $\tilde{\cal M}_n$ with a specific structure where the field configurations on the different replicas are connected. The replicated geometry will be discussed further when we discuss the replica wormhole picture.
   
   The Renyi entropy is given in terms of the effective action $I_n=-\ln Z_n$  by the relation 
     \begin{eqnarray}
       S_n={\partial}_n\big(\frac{I_n}{n}\big)=(1-n\frac{\partial}{\partial n})\ln Z_n=-\frac{1}{n-1}\ln Z_n~,~n\longrightarrow 1.
     \end{eqnarray}
     Here, $Z_n$ is the partition function on the replicated $n$ geometry $\tilde{\cal M}_n$.

     The replicated geometry  $\tilde{\cal M}_n$ is symmetric under cyclic permutation. This replica symmetry reduces the replicated geometry $\tilde{\cal M}_n$ to a single manifold given by the orbifold ${\cal M}_n=\tilde{\cal M}_n/Z_n$. In other words, correlation functions are computed on this orbifold.

     We have also to insert twist operators in the replicated geometry at the endpoints $(x_1,\bar{x}_1)$ and $(x_2,\bar{x}_2)$ of the interval of interest. These twist fields are primary operators in the orbifold theory  ${\cal M}_n=\tilde{\cal M}_n/Z_n$ with scaling dimensions 
 \begin{eqnarray}
\Delta_n=\frac{c}{12}\frac{n^2-1}{n}.
     \end{eqnarray}
     The partition function and the entropy of an interval on the half-plane are then given by
     \begin{eqnarray}
     Z_n=\langle\sigma(x_1,\bar{x}_1)\tilde{\sigma}(x_2,\bar{x}_2)\rangle_{{\cal M}_n}~,~S_n=-\frac{1}{n-1}\ln\langle\sigma(x_1,\bar{x}_1)\sigma(x_2,\bar{x}_2)\rangle_{{\cal M}_n}.
     \end{eqnarray}
      The replicated geometry is singular at the endpoints. The entropies are renormalized by normalizing the operator product expansion (OPE) in such a way that the identity operator appears with coefficient $1$, i.e. $\sigma(x_1,\bar{x}_1)\tilde{\sigma}(x_2,\bar{x}_2)\sim |x_1-x_2|^{-2\Delta_n}$ when $x_1-x_2\longrightarrow 0$. This normalization gives the standard result $S=\frac{c}{3}\log l$  for the entropy of an interval of length $l$ in two-dimensional conformal field theory \cite{Holzhey:1994we,Calabrese:2004eu}. 

      As we have shown, the half-plane and ${\bf AdS}^2$ are essentially related by a Weyl transformation. The partition function and the entropy of an interval in ${\bf AdS}^2$ are then obtained by using the transformation law under Weyl transformations  $g\longrightarrow \Omega^{-2}g$ given by
     \begin{eqnarray}
     \langle\sigma(x_1,\bar{x}_1)\sigma(x_2,\bar{x}_2)\rangle_{\Omega^{-2}g}=\Omega(x_1,\bar{x}_1)^{\Delta_n}\Omega(x_2,\bar{x}_2)^{\Delta_n}\langle\sigma(x_1,\bar{x}_1)\sigma(x_2,\bar{x}_2)\rangle_{g}.
     \end{eqnarray}
\begin{eqnarray}
  S_{\Omega^{-2}g}=S_{g}-\frac{c}{6}\sum_{\rm endpoints}\log \Omega.
\end{eqnarray}
     We are mostly interested in an interval for which both endpoints are in the bulk and they are spacelike separated. The entropy of such interval on the half-plane is given by \cite{Almheiri:2019psf}
    \begin{eqnarray}
      S=\frac{c}{6}\log\big((w_1-w_2)(\bar{w}_1-\bar{w}_2)\eta\big)+\log G(\eta).
    \end{eqnarray}
    Here, $\eta$ is the cross-ratio and $G(\eta)$ is defined in terms of the  function $G_n(\eta)$ by the relation $\log G(\eta)=-\partial_n\log G_n(\eta)$~,~$n\longrightarrow 1$. The function  $G_n(\eta)$, which depends on the theory and the boundary conditions, is such that $G(1)=1$. More importantly, the function $G_n(\eta)$ determines the two-point function of twist operators. Explicitly, we have
    
     \begin{eqnarray}
\eta=\frac{(w_1+\bar{w}_1)(w_2+\bar{w}_2)}{(w_1+\bar{w}_2)(w_2+\bar{w}_1)}.
     \end{eqnarray}
     \begin{eqnarray}
\langle \sigma(w_1,\bar{w}_1)\sigma(w_2,\bar{w}_2)\rangle=\frac{G_n(\eta)}{(w_1-w_2)^{\Delta_n}(\bar{w}_1-\bar{w}_2)^{\Delta_n}\eta^{\Delta_n}}.
     \end{eqnarray}
    We can now compute the entropy of an interval in ${\bf AdS}^2$ by using the transformation law of the entropy under Weyl transformations $g\longrightarrow \Omega^{-2}g$.

We have two cases to consider here. We can have an interval with one endpoint to the future of the shock $x_1^{\pm}>0$ and the other endpoint to the past of the shock $x_2^+>0$, $x_2^-<0$. Recall that $x=f(y)$. The entropy in this case is given by \cite{Almheiri:2019psf}
  \begin{eqnarray}
 S=\frac{c}{6}\log\bigg[\frac{48\pi E_S}{c}\frac{-y_1^-x_1^+x_2^-(x_2^+-x_1^+)\sqrt{f^{\prime}(y_1^-)}}{x_2^+(x_1^+-x_1^-)(x_1^+-x_2^-)}\bigg]+\log G(\eta).
  \end{eqnarray}
In the second case, we can have an interval where both endpoints are to the future of the shock $x_{i}^{\pm}>0$. The entropy in this case is given by \cite{Almheiri:2019psf}
  \begin{eqnarray}
 S=\frac{c}{6}\log\bigg[\frac{4(y_1^--y_2^-)(x_2^+-x_1^+)\sqrt{f^{\prime}(y_1^-)f^{\prime}(y_2^-)}}{(x_1^+-x_1^-)(x_2^+-x_2^-)}\bigg].
  \end{eqnarray}
Finally, we take the limit in which one of the endpoints is located on the cutoff surface (regularized boundary) $z=\epsilon f^{'}(u)$ at the physical time $u=f^{-1}(t)$. We obtain the two equations \cite{Almheiri:2019psf}
   \begin{eqnarray}
 S=\frac{c}{6}\log\bigg[\frac{24\pi E_S}{\epsilon c}\frac{-u t x^-(x^+-t)}{x^+(t-x^-)\sqrt{f^{\prime}(u)}}\bigg]+\log G(\frac{t(x^+-x^-)}{x^+(t-x^-)})~,~x^-<0<t<x^+.\label{early}
  \end{eqnarray}
  \begin{eqnarray}
 S=\frac{c}{6}\log\bigg[\frac{2(u-y^-)(x^+-t)}{\epsilon(x^+-x^-)}\sqrt{\frac{f^{\prime}(y^-)}{f^{\prime}(u)}}\bigg]~,~0<x^-<t<x^+.\label{late}
  \end{eqnarray}

\subsection{The late time quantum extremal surfaces}

  First, we are interested in the behavior at late times to the future of the shock $x^+>t>x^->0$. The central quantity is the  generalized entropy defined by 
    \begin{eqnarray}
      S_{\rm gen}=\frac{A}{4G_N}+S_{\rm bulk}=\frac{\phi+\phi_0}{4G_N}+S_{\rm CFT}.
    \end{eqnarray}
    In two dimensions, the area of a point is given by the coefficient of the Ricci scalar in the action.

    The bulk term $S_{\rm bulk}$ is the entanglement entropy $S_{\rm CFT}$  of conformal matter in the region between the cutoff surface  (near the boundary of ${\bf AdS}^2$)   and the quantum extremal surface determined by the coordinates $(x^+,x^-)$.  This matter entropy $S_{\rm CFT}$ is computed using the formula (\ref{late}).

    The area term $A$ is the surface area of the quantum extremal surface which is given by the dilaton field.

    The entanglement region corresponding to the correct value of the entropy, i.e. the value which is consistent with unitarity (Page curve) and thus resolving the information loss problem, is called the entanglement wedge of the black hole. This entanglement region is found by extremizing the generalized entropy over $(x^+,x^-)$. In other words, the  quantum extremal surfaces are given by the conditions
     \begin{eqnarray}
      \partial_{\pm}S_{\rm gen}=0\Rightarrow \partial_{\pm}\phi=-4G_N \partial_{\pm}S_{\rm CFT}.\label{gen}
     \end{eqnarray}
     In the late time regime the generalized entropy is dominated by the area term yet the derivative of the bulk term plays an essential role in shifting the extremal surface from the location of the trivial vanishing surface which dominates at early times.
\begin{itemize}
    
   \item {\bf The variations $\partial_{\pm}\phi$:} We start by computing the derivatives of the dilaton field $\partial_{\pm}\phi$.
     \begin{itemize}
       \item We will assume that the sought after quantum extremal surface is near the horizon $x^+=t_{\infty}$.  

     \item We will start on the boundary $y^+-y^-=0$ or equivalently $x^+-x^-=0$ where the dilaton field is infinite and then move inward.

     \item We expand then the dilaton as follows
     \begin{eqnarray}
       \phi|_{x^+}=\phi|_{x^+=x^-}+(x^+-x^-)\partial_+\phi+...
     \end{eqnarray}
   \item The {\bf variation of the dilaton field (\ref{dilaton}) in the $x^{+}$ direction} is given by
     \begin{eqnarray}
       (x^+-x^-)^2\partial_+\phi=\bar{\phi}_r\bigg(-2+2(\pi T_1x^{-})^2-k\int_0^{x^{-}}dt(x^{-}-t)^2\{u,t\}\bigg)\equiv F(x^-).\nonumber\\
     \end{eqnarray}
     The right-hand side is a function $F$ of only $x^-$. In other words, we have
     \begin{eqnarray}
       \phi|_{x^+}=\phi|_{x^+=x^-}+\frac{F(x^{-})}{x^+-x^-}+...
       \end{eqnarray}     
     This shows that the function $F(x^-)$ is always negative for $x^-<t<x^+<t_{\infty}$. Clearly, for $x^-=x^+=t_{\infty}$ this function must necessarily vanish. We obtain then the integral
     \begin{eqnarray}
       2(\pi T_1t_{\infty})^2-2=k\int_0^{t_{\infty}}dt(t_{\infty}-t)^2\{u,t\}\equiv kI_{\infty}.\label{Iin}
     \end{eqnarray}

     \item At late times $u  \gg 1/k$ the diffeomorphism $t=f(u)$ is given by equation (\ref{diffeo}) from which we can derive the crucial double pole at $t=t_{\infty}$ in the Schwarzian at late times. Indeed, we compute
     \begin{eqnarray}
       \{u,t\}=\frac{1}{2(t_{\infty}-t)^2}.
     \end{eqnarray}
     \item We use this formula to compute the following derivative
      \begin{eqnarray}
        \partial_{-}\int_0^{x^{-}}dt(x^{-}-t)^2\{u,t\}=\frac{t_{\infty}-x^-}{t_{\infty}}-1-\log\big(\frac{t_{\infty}-x^-}{t_{\infty}}\big).
      \end{eqnarray}
      \item Then, by integrating both sides of this equation and using the integral (\ref{Iin}), we get 
      \begin{eqnarray}
        \int_0^{x^{-}}dt(x^{-}-t)^2\{u,t\}&=&-\frac{(t_{\infty}-x^-)^2}{2t_{\infty}}+(t_{\infty}-x^-)\log\big(\frac{t_{\infty}-x^-}{t_{\infty}}\big)+\frac{2}{k}\big((\pi T_1 t_{\infty})^2-1\big).\nonumber\\
      \end{eqnarray}
      \item Next, by substituting this result back in the derivative $\partial_{+}\phi$ and use $x^+\simeq t_{\infty}$ we get the desired result 
      \begin{eqnarray}
        \Rightarrow\partial_+\phi%&=&\bar{\phi}_r\bigg[-\frac{1}{2}(2\pi T_1)^2\frac{t_{\infty}+x^-}{t_{\infty}-x^-}-\frac{k}{t_{\infty}-x^-}\log \frac{t_{\infty}-x^-}{t_{\infty}}+\frac{k}{2t_{\infty}}\bigg]\nonumber\\
        &=&\frac{\bar{\phi}_r}{t_{\infty}-x^-}\bigg[-2(\pi T_1)^2(t_{\infty}+x^-)-k\log \frac{t_{\infty}-x^-}{t_{\infty}}+\frac{k}{2t_{\infty}}(t_{\infty}-x^-)\bigg]\nonumber\\
         &=&\frac{\bar{\phi}_r}{t_{\infty}-x^-}\bigg[-(2\pi T_1)^2t_{\infty}-k\log \frac{t_{\infty}-x^-}{t_{\infty}}+(2\pi^2T_1^2+\frac{k}{2t_{\infty}})(t_{\infty}-x^-)\bigg]\nonumber\\
        &=&\frac{\bar{\phi}_r}{t_{\infty}-x^-}\bigg[-4\pi  T_1e^{-\frac{k}{2}y^-}-k (1+\log 2)+(2\pi^2T_1^2+\frac{k}{2t_{\infty}})(t_{\infty}-x^-)\bigg].\nonumber\\
     \end{eqnarray}
      In the last line we have used $t_{\infty}=1/(\pi T_1)+k/(2\pi T_1)^2...$ and also used equation (\ref{diffeo}) in the form
      \begin{eqnarray}
      \log\frac{t_{\infty}-x^{-}}{2t_{\infty}}=-\frac{4\pi T_1}{k}\big(1-e^{-\frac{k y^{-}}{2}}\big).\label{diffeo1}
      \end{eqnarray}
      \item In the above variation we can neglect the second term compared to the first one since $k$ is small with fixed  $ky^-$. And also neglect the third term since $t_{\infty}-x^-$ is sufficiently small. We get the final variation
      \begin{eqnarray}
        \Rightarrow\partial_+\phi
        &=&-4\pi T_1\bar{\phi}_r\frac{e^{-\frac{k}{2}y^-}}{t_{\infty}-x^-}.\label{xplus}
     \end{eqnarray}
      \item   We consider now the {\bf variation of the dilaton field (\ref{dilaton}) in the $x^{-}$ direction} given by
     \begin{eqnarray}
       (x^+-x^-)^2\partial_-\phi=\bar{\phi}_r\bigg(2-2(\pi T_1x^{+})^2+k\int_0^{x^{-}}dt(x^{+}-t)^2\{u,t\}\bigg).
     \end{eqnarray}
     \item The integral is now computed by expanding in powers of $t_{\infty}-x^+$ (since $x^+\simeq t_{\infty}$) and neglecting quadratic powers in this variable as follows 
      \begin{eqnarray}
        \int_0^{x^{-}}dt(x^{+}-t)^2\{u,t\}-I_{\infty}&=& \int_0^{x^{-}}dt(x^{+}-t)^2\{u,t\}- \int_0^{t_{\infty}}dt(t_{\infty}-t)^2\{u,t\}\nonumber\\
        &=&\int_0^{x^{-}}dt\bigg[(x^{+}-t_{\infty})^2+(t_{\infty}-t)^2+2(x^{+}-t_{\infty})(t_{\infty}-t)\bigg]\{u,t\}\nonumber\\
        &-& \int_0^{x^-}dt(t_{\infty}-t)^2\{u,t\}- \int_{x^-}^{t_{\infty}}dt(t_{\infty}-t)^2\{u,t\}\nonumber\\
        &=&\frac{1}{2}\frac{(t_{\infty}-x^+)^2}{t_{\infty}-x^-}+(t_{\infty}-x^+)\log\frac{t_{\infty}-x^-}{t_{\infty}}-\frac{(t_{\infty}-x^+)^2}{2t_{\infty}}\nonumber\\
        &-&\frac{1}{2}(t_{\infty}-x^-)\nonumber\\
          &=&(t_{\infty}-x^+)\log\frac{t_{\infty}-x^-}{t_{\infty}}-\frac{1}{2}(t_{\infty}-x^-).
      \end{eqnarray}
      \item The derivative $\partial_-\phi$ becomes
      \begin{eqnarray}
        \partial_-\phi&=&\frac{\bar{\phi}_r}{(t_{\infty}-x^-)^2}\bigg(4 \pi T_1(t_{\infty}-x^{+})+k(t_{\infty}-x^+)\log\frac{t_{\infty}-x^-}{t_{\infty}}-\frac{k}{2}(t_{\infty}-x^-)\bigg)\nonumber\\
        &=&\frac{\bar{\phi}_r}{(t_{\infty}-x^-)^2}\bigg(4 \pi T_1 (t_{\infty}-x^+)e^{-\frac{k y^{-}}{2}} +k(t_{\infty}-x^+)\log 2-\frac{k}{2}(t_{\infty}-x^-)\bigg)\nonumber\\
         &=&\frac{\bar{\phi}_r}{(t_{\infty}-x^-)^2}\bigg(4 \pi T_1 (t_{\infty}-x^+) e^{-\frac{k y^{-}}{2}} -\frac{k}{2}(t_{\infty}-x^-)\bigg).\label{xminus}
      \end{eqnarray}
      In the second line we can neglect the second term compared to the first one since $k$ is small with fixed  $ky^-$. The first term is more important since $t_{\infty}-x^-$ is sufficiently small.
    \item From this equation we can determine the apparent horizon to be given by
      \begin{eqnarray}
        \partial_-\phi=0\Rightarrow x^+=t_{\infty}-\frac{1}{3}(t_{\infty}-t).
      \end{eqnarray}
      In other words, the new horizon is shifted outside the old horizon $x^+=t_{\infty}$.
      \end{itemize}
   \item {\bf The variations $\partial_{\pm}S_{\rm CFT}$ and $\partial_{\pm}S_{\rm gen}$:} Now we compute the derivatives of the bulk entropy  $\partial_{\pm}S_{\rm CFT}$ and the derivatives of the generalized entropy $\partial_{\pm}S_{\rm gen}$. Next, we extremize by setting   $\partial_{\pm}S_{\rm gen}=0$ or equivalently (\ref{gen}) and then using equations (\ref{xplus}) and (\ref{xminus}).
     \begin{itemize}
      \item First, the bulk entropy of an interval to the future of the shock between the point $t=f(u)$ on the boundary and the point $(x^+,x^-)$ in the bulk is given explicitly by
     \begin{eqnarray}
       S_{\rm CFT}=\frac{c}{6}\log \bigg[\frac{2(u-y^-)(x^+-t)}{\epsilon(x^+-x^-)}\sqrt{\frac{f^{'}(y^-)}{f^{'}(u)}}\bigg].
     \end{eqnarray}
   \item The {\bf derivative in the $x^+$ direction of the bulk entropy} is dominated by the lightcone singularity at $x^+-t\longrightarrow 0$. We compute then
     \begin{eqnarray}
       &&\partial_+S_{\rm CFT}=\frac{c}{6}\partial_+\log \bigg[\frac{(x^+-t)}{(x^+-x^-)}\bigg]=\frac{c}{6}\bigg[\frac{1}{x^+-t}+\frac{1}{x^+-x^-}\bigg]\nonumber\\
       &&\partial_+\phi=-4G_N\partial_+S_{\rm CFT}=-2\bar{\phi}_rk\bigg[\frac{1}{x^+-t}-\frac{1}{t_{\infty}-x^-}\bigg]\nonumber\\
       &&2\pi T_1\frac{e^{-\frac{k}{2}y^-}}{t_{\infty}-x^-}+\frac{k}{t_{\infty}-x^-}=\frac{k}{x^+-t}.
     \end{eqnarray}
     \item We can neglect the second term compared to the first one since $k$ is small with fixed  $ky^-$. We get then the result
     \begin{eqnarray}
             2\pi T_1\frac{e^{-\frac{k}{2}y^-}}{t_{\infty}-x^-}=\frac{k}{x^+-t}.\label{cond1}
     \end{eqnarray}
   \item The {\bf derivative in the $x^-$ direction of the bulk entropy} will involve the second derivative of the diffeomorphism $f(y^-)$. By taking the derivative with respect of $x^-$ twice  of equation (\ref{diffeo1}) we obtain
     \begin{eqnarray}
       &&\frac{1}{t_{\infty}-x^-}=2\pi T_1\frac{1}{f^{'}(y^-)}e^{-\frac{k}{2}y^-}\nonumber\\
       &&\partial_-\log f^{'}(y^-)=-\frac{1}{t_{\infty}-x^-}-\frac{k}{2}\frac{1}{f^{'}(y)}.
       \end{eqnarray}
     \item We can now compute the derivative in the $x^-$ direction of the bulk entropy and then solve  the extremum condition $\partial_{-}S_{\rm gen}=0$ by  using equation (\ref{xminus}) as follows
     
     \begin{eqnarray}
       \partial_-S_{\rm SFT}&=&\frac{c}{6}\partial_-\log \bigg[\frac{u-y^-}{x^+-x^-}\sqrt{f^{'}(y^-)}\bigg]\nonumber\\
       &=&\frac{c}{6} \bigg[-\frac{1}{f^{'}(y^-)}\frac{1}{u-y^-}+\frac{1}{x^+-x^-}+\frac{1}{2}\partial_-\log f^{'}(y^-)\bigg]\nonumber\\
  &&     \partial_{-}\phi=-4G_N\partial_-S_{\rm CFT}
         =-2\bar{\phi}_rk\bigg[-\frac{1}{f^{'}(y^-)}\frac{1}{u-y^-}+\frac{1}{2}\frac{1}{t_{\infty}-x^-}-\frac{k}{4}\frac{1}{f^{'}(y^-)}\bigg]\nonumber\\
       &&4\pi T_1\frac{(t_{\infty}-x^+)e^{-\frac{k}{2}y^-}}{(t_{\infty}-x^-)^2}+\frac{k}{2(t_{\infty}-x^-)}=\frac{2k}{f^{'}(y^-)}\frac{1}{u-y^-}+\frac{k^2}{2}\frac{1}{f^{'}(y^-)}.
     \end{eqnarray}
     \item Both terms in the right-hand side are neglected since $k$ is small with fixed  $ky^-$. We get then the result
      \begin{eqnarray}
 4\pi T_1\frac{(t_{\infty}-x^+)e^{-\frac{k}{2}y^-}}{(t_{\infty}-x^-)^2}+\frac{k}{2(t_{\infty}-x^-)}=0. \label{cond2}       
      \end{eqnarray}
    \item The two conditions (\ref{cond1}) and (\ref{cond2}) lead immediately to the location of the quantum extremal surface which is given by
      \begin{eqnarray}
        x^+=t_{\infty}+\frac{1}{3}(t_{\infty}-t).
      \end{eqnarray}
      In other words, the quantum extremal surface is located inside the old horizon $x^+=t_{\infty}$.

    \item This surface can be rewritten in terms of the proper time $u=f^{-1}(t)$ and the corresponding coordinate $y^-=f^{-1}(x^-)$ as follows. First, the two condition (\ref{cond1}) and (\ref{cond2}) lead to the equation
      \begin{eqnarray}
        t_{\infty}-x^-=\frac{8\pi T_1}{3k}(t_{\infty}-t)e^{-\frac{ky^-}{2}}\Rightarrow \log \frac{t_{\infty}-x^-}{t_{\infty}-t}=\log\big(\frac{8\pi T_1}{3k}e^{-\frac{ky^-}{2}}\big).
        \end{eqnarray}
      But from equations (\ref{diffeo}) and  (\ref{diffeo1}) we have
      \begin{eqnarray}
        \frac{e^{\frac{ky^-}{2}}}{2\pi T_1}\log \frac{t_{\infty}-x^-}{t_{\infty}-t}=\frac{2}{k}\big(1-\exp(\frac{k}{2}(y^--u))\big).
      \end{eqnarray}
      Hence, we obtain
      \begin{eqnarray}
        u=y^-+\frac{e^{\frac{ky^-}{2}}}{2\pi T_1}\log\big(\frac{8\pi T_1}{3k}e^{-\frac{ky^-}{2}}\big).\label{uy}
      \end{eqnarray}
      \end{itemize}
   %   It is immediately behind $x^+=t_{\infty}$.
    \item The generalized entropy of this quantum extremal surface is dominated by the area term. Indeed, the value of the dilaton field at the quantum extremal surface is equal to the thermodynamic entropy at the corresponding coordinate $y^-$, viz
      %and thus to the effective temperature $T_1^{\rm eff}=T_1e^{-ky^-/2}$, viz
      
 \begin{eqnarray}
A=\phi_0+\phi=\phi_0+\bar{\phi}_r 2\pi T_1e^{-\frac{ky^-}{2}}\Rightarrow S_{\rm gen}-S_0=\frac{\bar{\phi}_r}{4G_N}2\pi T_1e^{-\frac{ky^-}{2}}.\label{Sent}
        \end{eqnarray}
 This entropy is a linearly decreasing function of time which starts at the Page time (which is to be defined shortly) from the entropy of the perturbed black hole at temperature $T_1$ and then goes to zero at infinite time.
\item This late time behavior should be contrasted with the early time behavior which, as we will see in the next section, gives a linearly increasing function of time which starts at time zero from the entropy of the unperturbed black hole at temperature $T_0$ then reaches after the Page time the entropy  corresponding to temperature $T_1$.
\item By using (\ref{uy}) and (\ref{Sent}) we obtain the delay time
                    \begin{eqnarray}
                      t_{\rm HP}=u-y^-=\frac{\beta}{2\pi}\log \frac{16}{c}(S_{\rm gen}-S_0)~,~\beta=\frac{1}{T_1(y^-)}=\frac{1}{T_1e^{-\frac{ky^-}{2}}}.
                      \end{eqnarray}
                    \item This delay is precisely the scrambling time. The quantum extremal surface at the boundary proper time $u$ corresponds actually to the intersection with the horizon of the ingoing null geodesic which was emitted from the boundary at the earlier time $u-t_{\rm HP}=y^{-}$. This quantum extremal surface, as we have seen, lies immediately behind the horizon. The scrambling time is also related to the Hayden-Preskill effect as it demarcates the region in spacetime beyond which the system has no access. Thus, information which falls behind the horizon before this time, although it is certainly lost to the system,  can still be recovered from the Hawking radiation.

  \end{itemize}

\subsection{The early time quantum extremal surfaces}

  \begin{itemize}
\item Now, we are interested in the behavior at early times to the past of the shock $x^+>t>0>x^-$.
  \item The central quantity is still given by the  generalized entropy defined by 
    \begin{eqnarray}
      S_{\rm gen}=\frac{A}{4G_N}+S_{\rm bulk}=\frac{\phi+\phi_0}{4G_N}+S_{\rm CFT}.
    \end{eqnarray}
    \item At early times the area term (dilaton field) corresponds to the original black hole with temperature $T_0$ with no gravitational backreaction.
   \item However, the bulk term $S_{\rm bulk}$, which is the entanglement entropy $S_{\rm CFT}$  of conformal matter in the region between the cutoff surface and the quantum extremal surface determined by the coordinates $(x^+,x^-)$, is now computed using the first formula (\ref{early}).

      \item This entanglement region is found by extremizing the generalized entropy over $(x^+,x^-)$. In other words, the  quantum extremal surfaces are given by the conditions
     \begin{eqnarray}
      \partial_{\pm}S_{\rm gen}=0\Rightarrow \partial_{\pm}\phi=-4G_N \partial_{\pm}S_{\rm CFT}.
     \end{eqnarray}
   \item The solution is given by a quantum extremal surface close to the horizon, viz \cite{Almheiri:2019psf}
     \begin{eqnarray}
       x^{\pm}\mp\frac{1}{\pi T_0}=\frac{k}{(\pi T_0)^2}f_{\pm}(\eta)~,~\eta=\frac{2\pi T_0 t}{1+\pi T_0 t}.\label{QESearly}
     \end{eqnarray}
     Here, we will not write down the functions $f_{\pm}(\eta)$ explicitly.

     Thus, the quantum extremal surface is close to the classical bifurcation surface of the original black hole horizon. It moves out towards the boundary in a spacelike, but nearly-null, direction.
     \item Recall that $t_{\infty}=\frac{1}{\pi T_1}+O(k)$ and from causality we must have $t_{\infty}\leq \hat{t}_{\infty}$ where $\hat{t}_{\infty}=\frac{1}{\pi T_0}$. The Poincare time $t=f(u)$ is in the range $t\leq t_{\infty}\leq \hat{t}_{\infty}$ where it reaches the value $t_{\infty}$ when the physical/boundary time $u$ goes to $\infty$.
   \item Early times correspond to $u$ of order $1$, i.e. $k u$ small and hence equation (\ref{diffeo}) takes the form
     \begin{eqnarray}
       \frac{1}{2}-\frac{t}{2t_{\infty}}=\exp(-2\pi T_1 u).
     \end{eqnarray}
     
      \item The regime of this approximation is defined by the range of $\eta$ given by 
      \begin{eqnarray}
        &&1-\eta=\frac{\hat{t}_{\infty}-t}{\hat{t}_{\infty}+t}\geq \frac{\hat{t}_{\infty}-t_{\infty}}{\hat{t}_{\infty}+t_{\infty}}=\frac{T_1-T_0}{T_1+T_0}-O(k).
     \end{eqnarray}
      In fact, since $T_1-T_0\gg k$, we have simply
      \begin{eqnarray}
        &&1-\eta \geq \frac{T_1-T_0}{T_1+T_0}.
      \end{eqnarray}
      
     \item Alternatively, $\eta$ must be in the range
      \begin{eqnarray}
        &&1-\eta=\frac{\hat{t}_{\infty}-t}{\hat{t}_{\infty}+t}\geq \frac{1}{2}-\frac{t}{2\hat{t}_{\infty}}\geq \frac{1}{2}-\frac{t}{2{t}_{\infty}}=\exp(-2\pi T_1 u)\nonumber\\
        &&u\geq -\frac{1}{2\pi T_1}\log(1-\eta).
     \end{eqnarray}
     \item  For very early times we have $\eta\longrightarrow 0$ and $u\longrightarrow 0$ or equivalently
      \begin{eqnarray}
       u= -\frac{1}{2\pi T_1}\log(1-\eta)\leq \frac{1}{2\pi T_1}\log\frac{T_1+T_0}{T_1-T_0}.
      \end{eqnarray}
      This corresponds to the regime of transient state of the black hole.
     \item The regime of the steady state of the black hole at early times corresponds to the regime when $\eta$ reaches its maximum value $1$ when the physical time $u$ is in the range 
   \begin{eqnarray}
       u\geq \frac{1}{2\pi T_1}\log\frac{T_1+T_0}{T_1-T_0}\simeq \frac{\beta_1}{2\pi}\log\frac{4E_1}{E_s}=t_{\rm HP}.
     \end{eqnarray}
In this equation, we have used $\pi^2 T^2=\frac{4\pi G_N}{\bar{\phi}_r}E$ and $(\pi T_1)^2=(\pi T_0)^2+\frac{4\pi G_N}{\bar{\phi}_r}E_S$. We have also made the approximation $T_1+T_0\simeq 2T_1$.

The time $t_{\rm HP}$ is the scrambling time which marks here the conclusion of the transient phase and the start of the Page curve behavior of the "young" black hole.

 \item Indeed, the corresponding entropy to the quantum extermal surface (\ref{QESearly}) is computed from (\ref{early}) and is given by
     \begin{eqnarray}
      S_{\rm gen}=...+\frac{c}{6}\log (u(1-\eta)\sinh(\pi T_1 u))+...
     \end{eqnarray}
     The contribution from the area term is negligible at early times.
 
   \item  Thus, at very early times we have a logarithmic increase $\log u$ due to the switching on of the coupling with the bath, i.e. to the addition of the entropy of the shock to the equilibrium entropy. Then, this is followed by a linear decrease $\log(1-\eta)=-2\pi T_1 u$ of the entropy due to the initial Hawking modes which did not have enough time to escape the black hole region and reach the bath. This process continues until $\eta$ reaches its maximum value $1$ at the scrambling time $t_{\rm HP}$ at which point the entropy reaches a minimum and the transient phase concludes.
   \item After this transient phase, the entropy starts to increase linearly as $\frac{c}{6}\log \sinh \pi T_1 u\sim\frac{c\pi T_1}{6}u$ due to the escape of Hawking modes into the bath. This is the true early time behavior of the entropy for a "young" black hole (although this is "young"  in the sense that we have a recent coupling of an "old" black hole to the bath and not to a recent formation of a black hole from gravitational collapse). This steady increase of the entropy due to the build up of entanglement of the Hawking radiation is the first leg of the unitary Page curve. 
    \end{itemize}

\section{The ``island'' conjecture}

 In most of this section we follow \cite{Almheiri:2019hni}. First, we start with a digression.
\subsection{The Randall-Sundrum model: A digression}

  We present here the Randall-Sundrum (RS) model \cite{Randall:1999vf,Randall:1999ee} which is a more succesfull extension of the   Arkani-Hamed-Dimopoulos-Dvali (VDD) model  \cite{Arkani-Hamed:1998jmv,Arkani-Hamed:1998sfv}.  These modes are inspired by string theory and they aim to solve among other things the hierarchy problem, i.e. why gravity is so weak compared to the other forces.

      The most important ingredients of these models are 1) extra dimensions and 2) Dp-branes.

      We consider one extra dimension labelled by $z$ given by the orbifold ${\bf S}^1/{\bf Z}_2$ and NOT by the circle ${\bf S}^1$. In other words, we impose the two compactification requirements:
\begin{itemize}
\item Periodicity $y\longrightarrow y+2y_c$, $y_c=\pi r_c$. The range of the extra dimension is $[0,2\pi r_c]$.

\item Orbifold symmetry $y\longrightarrow -y$. In other words, we identify $(x,y)=(x,-y)$ which allows us to take the range $[0,+\pi r_c]$. 
\end{itemize}

\begin{figure}[htbp]
   \begin{center}
     \includegraphics[width=10cm,angle=-0]{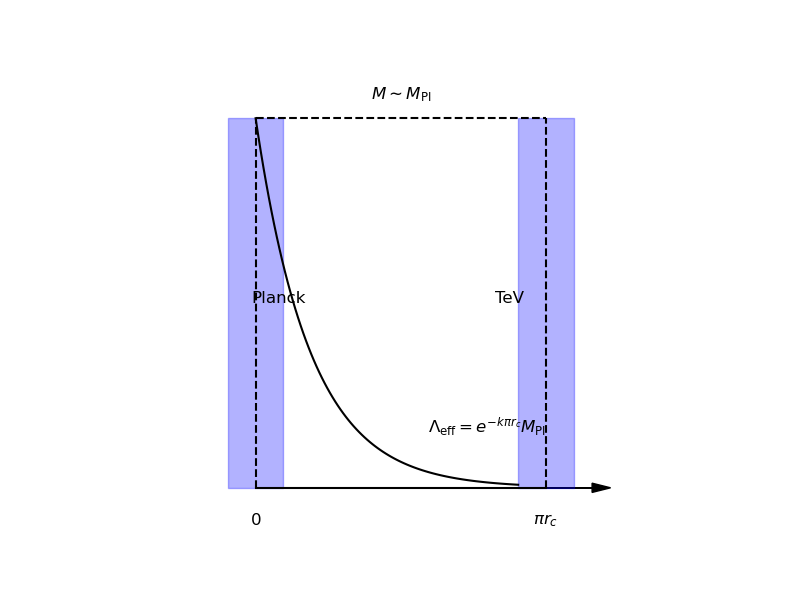}
\end{center}
\caption{The RS model.}\label{fig5}
  \end{figure}

Now, we place two four-dimensional 3-branes located at the two fixed points of the orbifold $y=y_c$ (the IR-brane where the standard model  and dark matter live) and $y=0$ (the UV-brane where quantum gravity lives).  We will also use $\varphi=y/r_c$. The five-dimensional spacetime (the bulk) is then bounded between these two branes. See figure (\ref{fig5}).

The action is given by 
\begin{eqnarray}
&&S=S_{\rm bulk}+S_{\rm IR}+S_{\rm UV}\nonumber\\
&&S_{\rm bulk}=\frac{1}{2}M_5^3\int d^4x \int_{-\pi}^{+\pi} d\varphi \sqrt{-G}(R-2\Lambda_{\rm bulk})\nonumber\\
&&S_{\rm IR}=\int d^4x\int_{-\pi}^{+\pi}d\varphi \sqrt{-g_{\rm IR}}(-V_{\rm IR}+{\cal L}_{\rm IR})\delta(\varphi-\pi)\nonumber\\
&&S_{\rm UV}=\int d^4x\int_{-\pi}^{+\pi}d\varphi \sqrt{-g_{\rm UV}}(-V_{\rm UV}+{\cal L}_{\rm UV})\delta(\varphi).
\end{eqnarray}
The dark matter fields on the IR-brane are denoted by $\phi$ and they can only interact gravitationlly  with the fields of the standard model. They are both contained in the Lagrangian density ${\cal L}_{\rm IR}$. This is the basic idea.

%\item Randall-Sundrum model 

The Einstein equations of motion derived from the above action can be solved by a a four-dimensional Poincare invariant background given by 

  \begin{eqnarray}
    ds^2=\exp(-2\sigma(\varphi))g_{\mu\nu}dx^{\mu}dx^{\nu}+dy^2.
    \end{eqnarray}
The warp factor $\exp(-2\sigma)$ is given in terms of the warping parameter $\sigma$ which is such that
\begin{eqnarray}
  \sigma=|y|\sqrt{\frac{-\Lambda_{\rm bulk}}{6}}.
  \end{eqnarray}
The vacuum energies on the branes must be related by 
\begin{eqnarray}
V_{\rm UV}=-V_{\rm IR}=6M_5^3 k^2~,~k^2=\frac{-\Lambda_{\rm bulk}}{6}.
\end{eqnarray}
The background solution is a five-dimensional anti-de Sitter spacetime ${\bf AdS}^5$ with radius $R=1/k$.

We substitute the solution in the action $S_{\rm bulk}$ and by assuming that the fields do not depend on the extra dimension $y$ we obtain the effective action
\begin{eqnarray}
  S_{\rm bulk}=\int d^4x \int_{-y_c}^{+y_c} dy\frac{1}{2}M_5^3 e^{-4k|y|}\sqrt{-g^{(4)}}e^{2k|y|}R^{(4)}=\frac{M_5^3}{2k}(1-e^{-2k|y_c|})\int d^4x \sqrt{-g^{(4)}}R^{(4)}.
  \end{eqnarray}
Thus, the four-dimensional Planck mass is given by 
\begin{eqnarray}
  M_{\rm Pl}^2=\frac{M_5^3}{k}(1-e^{-2k|y_c|}).
  \end{eqnarray}
The same considerations on the IR-brane gives the result that the effective gravitational coupling is suppressed by the warp factor, viz
\begin{eqnarray}
  \Lambda_{\rm eff}=\frac{1}{\sqrt{G}}=M_{\rm Pl}e^{-k|y_c|}=M_{\rm Pl}e^{-\pi kr_c}.
  \end{eqnarray}
In fact all VEV (vacuum expectation values) are suppressed or more precisely redshifted on the IR-brane.  This is the core idea behind brane world scenarios or brane cosmology.

The Planck mass $M_{\rm Pl}$ is obviously at the Planck scale which is of order $10^{19}$ GeV. The hierarchy problem can then be solved by choosing $M_5$, $k=-\Lambda_{\rm bulk}/6$ and $y_c$ appropriately.  But in this case (as opposed to ADD model) the size of the extra dimensions has no real impact on the ratio $M_{\rm Pl}/M_5$. 

However, we can choose $\Lambda_{\rm eff}\ll M_{\rm Pl}$ even for moderate choices of $kr_c$. For example, for $kr_c\sim 10$ the RS model can solve the hierarchy problem, i.e. $\Lambda_{\rm eff}$ is at the TeV scale with the choice $M_5\sim M_{\rm Pl}\sim k$.

 These models provide also the possibility that dark matter can be explained by the massive Kaluza-Klein gravitons obtained by compactification of the extra dimensions. For example, see \cite{Folgado:2019sgz,deGiorgi:2020qlg,Cai:2021nmk}.

The gravitational content of the Randall-Sundrum model is obtained by expanding the five-dimensional metric as follows
\begin{eqnarray}
  G_{MN}\longrightarrow G_{MN}+\kappa h_{MN}~,~\kappa=2/M_5^{3/2}.
  \end{eqnarray}
The field content of the theory is given by
\begin{itemize}
\item 1) A spin two tensor field (graviton).
\item 2) A spin one vector field (can be made to vanish).
\item 3) A spin zero scalar field (radion).
\end{itemize}
 These fields are five-dimensional fields. The graviton and the radion can also be made to decouple. Explicitly, we write
\begin{eqnarray}
  G_ {\mu\nu}=e^{-2k|y|-2\hat{u}}(\eta_{\mu\nu}+\kappa \hat{h}_{\mu\nu}) ~,~G_{55}=-(1+2\hat{u})^2~,~G_{\mu 5}=G_{5\mu}=0.
  \end{eqnarray}
We will not discuss the radion field which measures the width of the extra dimension.

By integrating the extra dimension we get four-dimensional fields. This is Kaluza-Klein reduction which is given for the metric by the field expansion 
\begin{eqnarray}
\hat{h}_{\mu\nu}(x,y)=\sum_{n=0}^{\infty}\frac{1}{\sqrt{r_c}}h_{\mu\nu}^{(n)}(x)\psi_n(\varphi).
\end{eqnarray}
The massless single 5D-graviton is transformed into a tower of massive 4D-gravitons (Kaluza-Klein modes). These are the particles of the dark matter.  The mode $n=0$ is precisely the  massless graviton of the theory of general relativity. The wave functions $\psi_n$ are determined in terms of the mass $m_n$ of the $n$-th graviton by the equation 
\begin{eqnarray}
\frac{d}{d\varphi}\bigg(e^{-4kr_c|\varphi|}\frac{d\psi_n}{d\varphi}\bigg)=-r_c^2m_n^2e^{-2kr_c|\varphi|}\psi_n.
\end{eqnarray}
The equation of motion of the $n-$th massive graviton is precisely the Pauli-Fierz equation of massive gravity given by 
\begin{eqnarray}
  (\eta_{\mu\nu}\partial^{\mu}\partial^{\nu}+m_n^2)h_{\mu\nu}^{(n)}(x)=0.
  \end{eqnarray}
The masses of the KK-graviton modes are given by
\begin{eqnarray}
m_n=kx_ne^{-\pi kr_c}~,~J_1(x_n)=0.
\end{eqnarray}
In other words, the masses $m_n$ (or more precisely $x_n$) are the roots of the Bessel function $J_1$.

This shows how dark matter candidates can be KK-gravitons in extra dimensions which is another very interesting idea worth pursuing.

\subsection{Holographic conformal matter}

    Again, we consider two-dimensional Jackiw-Teitelboim gravity coupled to a matter bulk theory given by a ${\rm CFT}_2$ theory. The JT gravity theory is also coupled to another copy of the same ${\rm CFT}_2$ representing the heat bath. The fields in the JT theory are the dilaton field $\phi$ and the metric $g_{ij}^{(2)}$ while the fields in the ${\rm CFT}_2$ theory are the matter fields $\chi$ and the metric $g_{ij}^{(2)}$. The action is given by
     \begin{eqnarray}
     S[g_{ij}^{(2)},\phi,\chi]=S_{\rm JT}[g_{ij}^{(2)},\phi]+S_{\rm CFT2}[g_{ij}^{(2)},\chi].\label{JT23}
    \end{eqnarray}
  We write the two-dimensional metric explicitly as
    \begin{eqnarray}
      ds^2=-\exp(2\rho(x))dx^-dx^+.
    \end{eqnarray}
  As we have seen, we can introduce new coordinates $w=w(x)$ in which the stress-energy-momentum tensor vanishes locally. The stress-energy-momentum tensor in the coordinates $x$ is determined by the conformal anomaly and it is given in terms of the diffeomorphism $w=w(x)$ by the relations
      \begin{eqnarray}
      T_{x^+x^+}=-\frac{c}{24\pi}\{w^+,x^+\}~,~ T_{x^-x^-}=-\frac{c}{24\pi}\{w^-,x^-\}.
      \end{eqnarray}
      These equations actually determine the diffeomorphism $w=w(x)$. By an additional Weyl transformation we can bring the above metric into the flat space form, viz $ds^2=-dw^-dw^+$. In fact, the stress-energy-momentum tensor vanishes identically in this flat metric. We have then the vacuum solution on flat space given by
      \begin{eqnarray}
      ds^2=-dw^-dw^+~,~T_{w^+w^+}=T_{w^-w^-}=0.
      \end{eqnarray} 
    Now we will assume that the above bulk ${\rm CFT}_2$ theory has a three-dimensional holographic gravity dual. In other words, the ${\rm CFT}_2$ theory lives on the boundary of some three-dimensional bulk theory. The three-dimensional metric is given by
       \begin{eqnarray}
    g_{ij}^{(3)}|_{\rm boundary}=\frac{1}{\epsilon^2}g_{ij}^{(2)}.\label{h1}
      \end{eqnarray} 
     The three-dimensional bulk geometry is ${\bf AdS}^3$. Indeed, the second copy of the ${\rm CFT}_2$, representing the heat bath, will live on the flat fixed boundary of a pure ${\bf AdS}^3$.

\begin{figure}[htbp]
   \begin{center}
 \includegraphics[width=17cm,angle=-0]{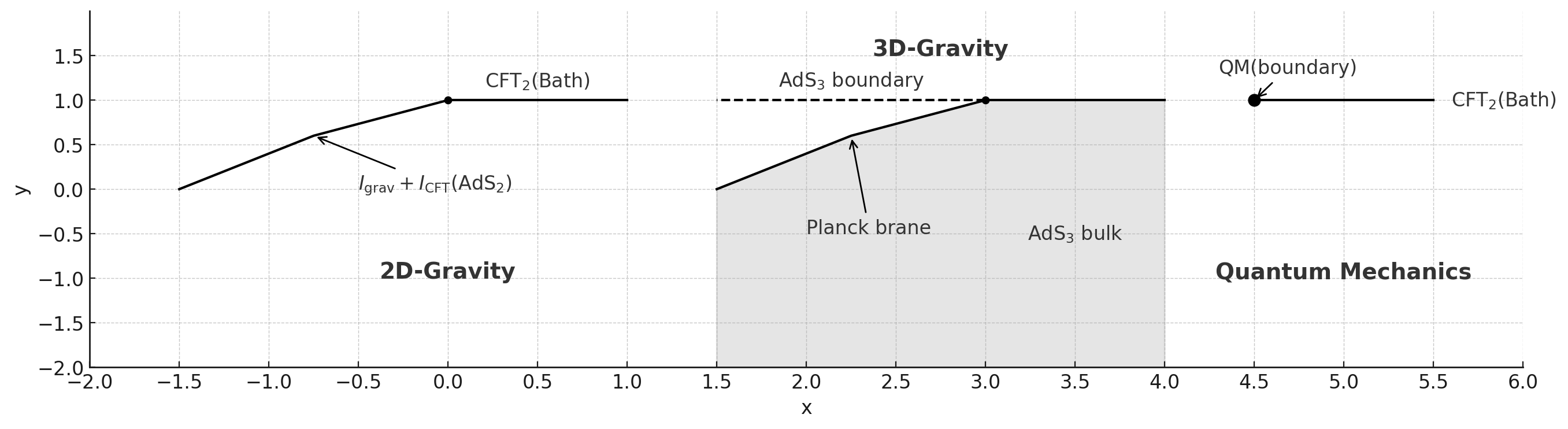}
\end{center}
\caption{Three equivalent description of ${\bf AdS}^2$ black hole.}\label{fig6}
  \end{figure}

     However, with respect to the first copy of the ${\rm CFT}_2$ representing matter fields, there are two differences with the usual ${\rm AdS}/{\rm CFT}$ correspondence. First, the corresponding three-dimensional bulk theory does not have a fixed boundary metric since $g_{ij}^{(2)}$ is a dynamical field. Second, there is another scalar field (the dilaton $\phi$) propagating on this two-dimensional boundary in addition to the matter scalar fields $\chi$. The boundary itself in this case is therefore a dynamical space  since we are integrating over the metric $g_{ij}^{(2)}$ (and the dilaton field $\phi$ which determines the metric). This dynamical space is analogous to the Planck brane of the Randall-Sundrum model \cite{Randall:1999vf,Randall:1999ee}. Hence, the three-dimensional bulk is locally ${\bf AdS}^3$ with a dynamical boundary where the two-dimensional theory with action (\ref{JT23}) lives. The bulk metric near the Planck brane is given explicitly by the ${\bf AdS}^3$ metric
   \begin{eqnarray}
ds^2=\frac{-dw^+dw^-+dz_w^2}{z_w^2}.
   \end{eqnarray}
 The Planck brane is located at
   \begin{eqnarray}
z_w=\epsilon e^{-\rho(x)}\sqrt{\frac{dw^+}{dx^+}\frac{dw^-}{dx^-}}.\label{h2}
   \end{eqnarray}
 The evaporation of a two-dimensional ${\bf AdS}^2$ black hole (into a heat bath) can then be described by three different but equivalent systems (where $\sigma_y=(y^+-y^-)/2$):
   \begin{itemize}
   \item {\bf $2$D-Gravity:} A two-dimensional Jackiw-Teitelboim/conformal-field-theory model of gravity-matter interaction living in $\sigma_y<0$ coupled to a two-dimensional conformal field theory living in $\sigma_y>0$.
   \item {\bf $3$D-Gravity:} A three-dimensional theory of gravity in ${\bf AdS}^3$ spacetime with a dynamical boundary (Planck brane) in the region $\sigma_y<0$ and a fixed boundary in the region $\sigma_y>0$.
   \item {\bf Quantum Mechanics:} The dAFF/Yang-Mills conformal/matrix quantum mechanics (the holographic dual of ${\bf AdS}^2$ or noncommutative ${\bf AdS}^2_{\theta}$ space) living at the point $\sigma_y=0$ coupled to  a two-dimensional conformal field theory living in $\sigma_y>0$.
   \end{itemize}
   See figure (\ref{fig6}).

  \subsection{The island and entanglement wedges at late times}

     We have already shown how to obtain the correct Page curve of an evaporating black hole by computing the quantum extremal surfaces which extremize the generalized entropy of the black hole by following the prescription of \cite{Engelhardt:2014gca}.

     By assuming that the state of the black hole and the Hawking radiation is pure we can immediately conclude that the Hawking radiation is also characterized by the correct Page curve. However, this does not solve the information loss paradox.

     A real solution of the information loss paradox amounts to the direct calculation of the Page curve of the Hawking radiation by computing its quantum extremal surfaces and showing that they coincide with the quantum extremal surfaces of the black hole. This direct calculation will involve  in a crucial way the so-called ``quantum extremal islands'' which are located deep inside the black hole interior behind the event horizon.

     The necessity of these islands in the correct calculation of the von Neumann entropy of the radiation has already being discussed but their construction will be made explicit now by means of holography. Indeed, by assuming that the bulk matter is holographic we can escalate the problem from two-dimensional gravity to three-dimensional gravity where the physical meaning of these islands is much more transparent.

     In fact, these islands provide a concrete realization of the ER=EPR proposal \cite{Maldacena:2013xja}. In other words, the entanglement between the interior modes of the black hole within the island and the modes of the Hawking radiation is precisely equivalent to the geometric connection between the island and the asymptotic region of the black hole through the extra dimension provided by the higher dimensional gravity theory.

     We are therefore interested in computing the quantum extremal surfaces of the Hawking radiation holographically, i.e.  in the three-dimensional gravity theory. This calculation is equivalent to leading order to the Ryu-Takayanagi formula for extremizing areas.

     Let us start by outlining the holographic calculation, i.e. the calculation from the perspective of the three-dimensional gravity of the  generalized entropy in one interval corresponding to the quantum extremal surfaces of the black hole system. First, recall that from the perspective of the two-dimensional gravity, the generalized entropy of an interval ${I}_y$ (the accessible region) extending from the boundary to the point $y$ in the bulk (which is the boundary of the accessible region in this case) is given by
      \begin{eqnarray}
         S[{\bf Bl-Ho}]\equiv S_{\rm gen}(y)=\frac{\phi(y)}{4G_N^{(2)}}+S_{\rm 2d~bulk}[I_y].
      \end{eqnarray}
    The bulk entropy $S_{\rm 2d~bulk}[I_y]$ is the von Neumann entanglement or fine grained-entropy of the bulk conformal matter fields $\chi$, dilaton field $\phi$ and the bulk two-dimensional metric $g_{ij}^{(2)}$. This entropy is dominated by the matter fields $\chi$ in the limit of a large number of degrees of freedom of the ${\rm CFT}2$.

    As we have seen, the generalized entropy $  S_{\rm gen}(y)$ is minimized in the spatial direction but maximized in the temporal direction. In other words, the obtained surface is in fact an extremal surface found by extremizing the generalized entropy over the point $y$.
      
      The Ryu-Takayanagi formula instructs us to extend the point $y$ to the one-dimensional minimal area surface $\Sigma_y$ in the bulk which bounds the accessible region in ${\bf AdS}^3$. Here, in the context of three-dimensional gravity, $\Sigma_y$ is an interval. The generalized entropy $S_{\rm gen}(y)$ is then given by the Ryu-Takayanagi formula as follows
      \begin{eqnarray}
        S_{\rm gen}(y)=\frac{\phi(y)}{4G_N^{(2)}}+\frac{{\rm Area}(\Sigma_y)}{3G_N^{(3)}}.
      \end{eqnarray}
      At late times we consider the spatial slice $\Sigma_{\rm late}$  which is an interval between the point $0<\sigma_y\equiv \sigma_0\ll 1$ (in the heat bath) and the point $y^+\equiv y_e^+$ (at the quantum extremal surface). As we have shown, the quantum extremal surface lies immediately behind the horizon. The quantum extremal surface at time $u$ corresponds to the intersection with the horizon of the future directed light ray which was emitted from the boundary at the earlier time $y^-=u - t_{\rm HP}$ where  $t_{\rm HP}$ is the scrambling time.
      
    The spatial slice $\Sigma_{\rm late}$ corresponds in the three-dimensional geometry to the bulk region bounded by the minimal surface area $\Sigma_y$ and $\Sigma_{\rm late}$.
      
\begin{figure}[htbp]
   \begin{center}
     \includegraphics[width=18cm,angle=-0]{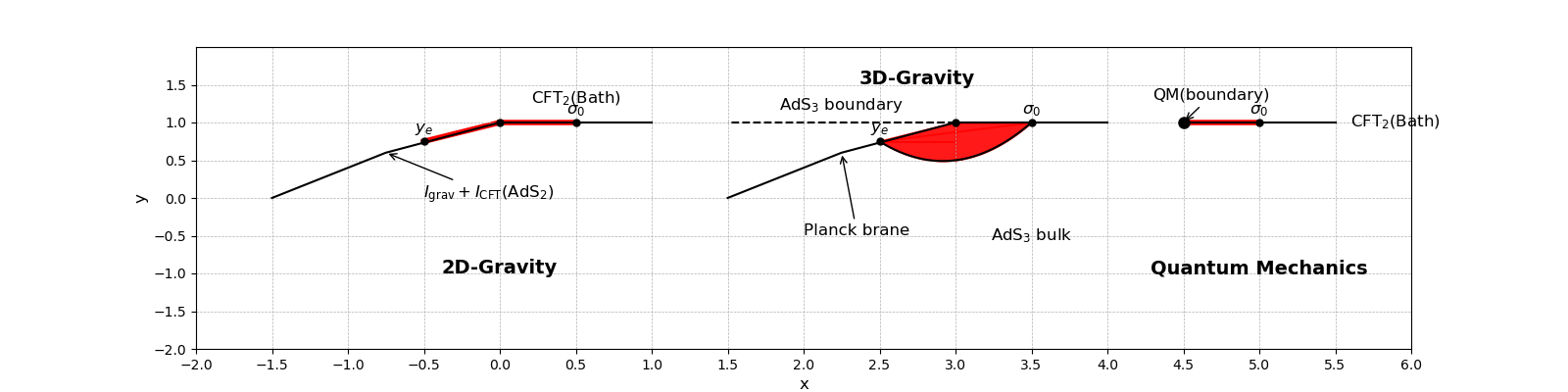}
\end{center}
\caption{Entanglement wedges in the three theories for the black hole.} \label{fig7p1}
\end{figure}

The entanglement wedge of the black hole at late times is therefore given in the two-dimensional picture by the causal domain of the spacelike slice connecting the two points $(u,\sigma_0 )$ and $(y_e^+, y_e^-)$ while in the three-dimensional picture it is given by the bulk region bounded by the minimal surface area $\Sigma_y$ and the spatial slice $\Sigma_{\rm late}$.  See figure (\ref{fig7p1}).
     
For Hawking radiation, at late times, we consider the disconnected spatial slice $\Sigma_{\rm late}^{'}$  which corresponds to the union of two disconnected intervals. The first interval, in the heat bath, starts at the point $0<\sigma_y\equiv \sigma_0\ll 1$ and goes to infinity and is the complement of the interval considered in the black hole case. The second interval, which corresponds to the island, is found deep inside the black hole beyond the point $y^+\equiv y_e^+$ (where the quantum extremal surface is located).

Clearly, the region in the three-dimensional bulk corresponding to the radiation is complement to the region  corresponding to the black hole. Thus, from the three-dimensional perspective the above two intervals correspond to a simply connected bulk region. Indeed, the minimization of the entanglement entropy according to the Ryu-Takayanagi formula will show that the  accessible spatial slice $\Sigma_{\rm late}^{'}$ of the radiation will connect the exterior region (Hawking radiation) to the interior region (island). 

Hence, the entanglement wedge of the radiation contains the interior region that was not covered by the entanglement wedge of the black hole  and thus the quantum extremal surface is found to be the same.

We use now the fact that the interval $I_y\equiv [y_e^+,\sigma_0]$ (black hole) is complementary to the interval $I_y^{'}\equiv [0,y_e^+]\cup [\sigma_0,\infty[$ (radiation+island). In other words, the corresponding von Neumann entropies are the same since these two intervals correspond to the entanglement wedges  $\Sigma_{\rm late}$ and $\Sigma_{\rm late}^{'}$ of the black hole and the radiation+island respectively.

    From the three-dimensional perspective it is much easier to see that the entanglement entropy of the radiation+island is equal to the entanglement entropy of the black hole as they correspond to two regions which are complementary.   These two regions (the black hole and the radiation+island regions) are  thus found in a pure state. 
  
  The entanglement entropy of the radiations is then computed as follows:
  \begin{eqnarray}
    &&S_{\rm 2d~bulk}[I_y]=S_{\rm 2d~bulk}[I_y^{'}]\Rightarrow\nonumber\\
      && S_{\rm gen}(y)=\frac{\phi(y)}{4G_N^{(2)}}+S_{\rm 2d~bulk}[I_y]=\frac{\phi(y)}{4G_N^{(2)}}+S_{\rm 2d~bulk}[I_y^{'}]\Rightarrow\nonumber\\
        &&S[{\bf RAD}]\equiv S_{\rm gen}(y)=\frac{\phi(y)}{4G_N^{(2)}}+S_{\rm 2d~bulk}[I_y^{'}]\Rightarrow\nonumber\\
            &&S[{\bf RAD}]=\frac{{\rm Area}(\partial{\cal I})}{4G_N^{(2)}}+S[{\bf rad}\cup{\cal I}].     
  \end{eqnarray}
 In general, the von Neumann entropy of the Hawking radiation involves an island ${\cal I}$ behind the horizon and is given by
   \begin{eqnarray}
    S[{\bf RAD}]={\rm min}\bigg\{{\rm ext}\bigg[\frac{{\rm Area}{\partial{\cal I}}}{4G_N}+S[{\bf rad}\cup{\cal I}]\bigg]\bigg\}.     
   \end{eqnarray}
   The extremization is done over the choices of islands ${\cal I}$ followed by a minimization over all these extrema. We insist, following  \cite{Almheiri:2019hni}, that ${\bf RAD}$ represents the quantum state of Hawking radiation (which is not known) whereas ${\bf rad}$ represents the state of the radiation in the semi-classical description. This rule gives us then the entropy of the exact quantum state (not the quantum state itself) by using only semi-classical physics.
 
\begin{figure}[htbp]
   \begin{center}
      \includegraphics[width=17cm,angle=-0]{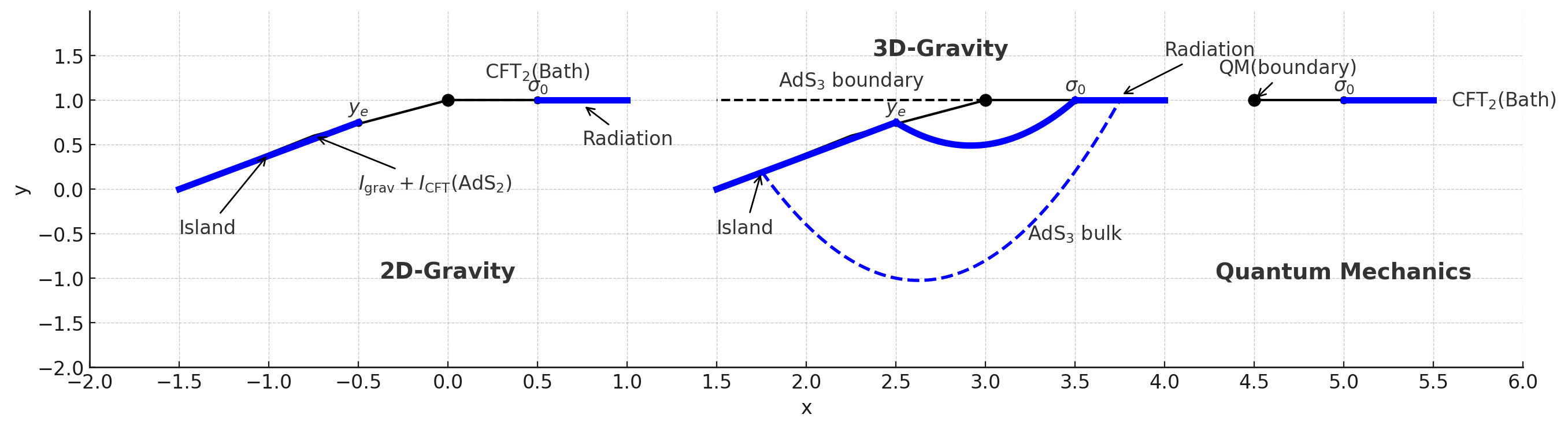}
\end{center}
\caption{Entanglement wedges in the three theories for the radiation+island.} \label{fig7p2}
  \end{figure}

The entanglement entropy of the radiations at late times is then given explicitly by the result (\ref{Sent}), i.e. it is dominated by the area term of the quantum extremal surface and hence it is given by the Bekenstein-Hawking entropy $ S_{B-H}$ at the corresponding temperature. Indeed, equation (\ref{Sent}) can be put in the form
   
   \begin{eqnarray}
     &&S[{\bf RAD}]= S[{\bf Bl-Ho}]=S_{\rm gen}-S_0= S_{B-H}(T_1(y^-))=\frac{A(y^-)}{4 G_N}\nonumber\\
     &&A(y^-)\equiv \phi(y^-)=\bar{\phi}_r 2\pi T_1(y^-)~,~T_1(y^-)=T_1e^{-\frac{ky^-}{2}}.
%A=\phi_0+\phi=\phi_0+\bar{\phi}_r 2\pi T_1e^{-\frac{ky^-}{2}}\Rightarrow S-S_0=\frac{\bar{\phi}_r}{4G_N}2\pi T_1e^{-\frac{ky^-}{2}}.\label{Sent}
   \end{eqnarray}
    The scrambling time $t_{\rm HP}$ is given in terms of this entropy $S_{\rm gen}-S_0\equiv S[{\bf RAD}]= S[{\bf Bl-Ho}]$ by the relation
      \begin{eqnarray}
                      t_{\rm HP}=u-y^-=\frac{\beta}{2\pi}\log \frac{16}{c}(S_{\rm gen}-S_0)~,~\beta=\frac{1}{T_1(y^-)}.
      \end{eqnarray}
 See figure (\ref{fig7p2}).

  \subsection{The entanglement wedges at early times and Page curve}

  \begin{itemize}
  \item Thus, the entropy of the black hole at late times corresponds to a non-trivial quantum extremal surface and is dominated by the area term. This entropy is found to be a linearly decreasing function of time which starts at the Page time from the Bekenstein-Hawking entropy $ S_{B-H}|_{T=T_1}$ associated with the perturbed temperature $T_1$ and then approaches zero at infinite time. We have then 
    \begin{eqnarray}
     S[{\bf Bl-Ho}]_{\rm late~times}=S_{\rm gen}-S_0= S_{B-H}(T_1(y^-)).
   \end{eqnarray}
  \item As we have seen, at early times, the quantum extremal surface of the black hole is found to be the vanishing surface.
    \item Thus, the entanglement entropy of the black hole is dominated at early times by the bulk term (computed for the black hole region) and is a linearly increasing function of time starting at zero time from the Bekenstein-Hawking entropy $ S_{B-H}|_{T=T_0}$ associated with the unperturbed temperature $T_0$. Explicitly, we have
       \begin{eqnarray}
                    S[{\bf Bl-Ho}]_{\rm early~times}=S_{\rm gen}-S_0=\frac{\bar{\phi}_r}{4G_N}2\pi T_0+k \frac{\bar{\phi}_r}{4G_N} 2\pi T_1 u.
       \end{eqnarray}
     \item The correct Page curve is thus reproduced for the black hole.
        \item The entanglement entropy of the radiation is controlled by the island which is intimately connected to the quantum extremal surface.
       \item       A non-vanishing island appears in the spacetime after a scrambling time. The bulk entropy involves therefore both the radiation and the island and is always small since the island degrees of freedom purify the Hawking modes in the radiation region. In other words, the generalized entropy of the radiation is controlled by the area term (boundary of the island) and thus starts large from the  Bekenstein-Hawking entropy of the black hole and decreases to zero as the area decreases with time.
       \item However, the non-vanishing quantum extremal surface corresponds to the global minimum of the entropy only at late times.
         \item Hence, the entanglement entropy of the radiation at late times is dominated by the area term and is a decreasing function of time which starts from the  Bekenstein-Hawking entropy $ S_{B-H}|_{T=T_1}$ associated with the perturbed temperature $T_1$ and decreases to zero as the area decreases with time.  We have then the purity condition
    \begin{eqnarray}
     &&S[{\bf RAD}]_{\rm late~times}= S[{\bf Bl-Ho}]_{\rm late~times}=S_{B-H}(T_1(y^-)).
   \end{eqnarray}
     \item There is no islands at very early times since the quantum extremal surface is vanishing. Hence, the entanglement entropy of the radiation at these very early times is solely given by the bulk term (computed for the radiation region) which is an increasing function of time.
     \item In fact, if we simply compute the entropy of the radiation without the island we will always find an increasing function of time (representing the continued accumulation in the radiation region of the Hawking modes which are entangled with their interior partners).
       \item However, the vanishing quantum extremal surface corresponds to the global minimum of the entropy only at early times.
\item Hence, the entanglement entropy of the radiation at early times is dominated by the bulk term and is an increasing function of time which starts from $0$ and increases to $2S_{\rm B-H}(T_1)$ as time increases to infinity. Indeed, we have
 
  \begin{eqnarray}
    S[{\bf RAD}]_{\rm early~times}&=&S[{\bf Bl-Ho}]_{\rm early~times}-\frac{\bar{\phi}_r}{4G_N}2\pi T_0\nonumber\\
    &=&S_{\rm gen}-S_0-\frac{\bar{\phi}_r}{4G_N}2\pi T_0\nonumber\\
    &=&k \frac{\bar{\phi}_r}{4G_N} 2\pi T_1 y^-\nonumber\\
    &=&2S_{\rm B-H}(T_1)(1-e^{-\frac{ky^-}{2}})\nonumber\\
    &=&2\big(S_{\rm B-H}(T_1)-S_{\rm B-H}(T_1(y^-))\big).
  \end{eqnarray}
\item Purity requires that we have 
  \begin{eqnarray}
    S[{\bf RAD}]_{\rm early~times}+\frac{\bar{\phi}_r}{4G_N}2\pi T_0=S[{\bf Bl-Ho}]_{\rm early~times}.
  \end{eqnarray}
\item We reproduce therefore the same Page curve for the Hawking radiation.
\item Page time is the time at which the late time behavior equals the early time behavior. At this time a phase transition occurs from the vanishing quantum extremal surface (no island) to the non-vanishing quantum extremal surface (island). The Page time is thus determined as follows:
  
\begin{eqnarray}
  &&    S_{B-H}(T_1(y^-))=2S_{\rm B-H}(T_1)(1-e^{-\frac{ky^-}{2}})+\frac{\bar{\phi}_r}{4G_N}2\pi T_0\Rightarrow\nonumber\\
  &&e^{-\frac{ky^-}{2}}=2(1-e^{-\frac{ky^-}{2}})+\frac{S_{\rm B-H}(T_0)}{S_{\rm B-H}(T_1)}\Rightarrow\nonumber\\
  &&y^-=\frac{2\ln\frac{3}{2}}{k}~,~\frac{S_{\rm B-H}(T_0)}{S_{\rm B-H}(T_1)}\ll 1.
  \end{eqnarray}
\item The entanglement wedges of the black hole and the radiation before and after the Page time are fundamentally distinguished from each other by the presence of the island in the black hole interior which becomes, at late times after the Page time, causally connected to the Hawking radiation at infinity.
  \end{itemize}
%  \begin{figure}[htbp]
%   \begin{center}
%     \includegraphics[width=10cm,angle=-0]{entropy3p1.png}
%      \includegraphics[width=10cm,angle=-0]{entropy3p2.png}
%\end{center}
%\caption{The vanishing surface (before Page time) and the quantum extremal surface or QES (after Page time).}
%  \end{figure}

\subsection{The ${\bf AdS}^2$ eternal black hole: The two-intervals entropy of radiation}

  We consider here an eternal ${\bf AdS}^2$ black hole, in Jackiw-Teitelboim gravity with matter given by
    a two-dimensional ${\rm CFT}$ with central charge $c$, in thermal equilibrium. This involves two entangled black holes, interacting with two copies of a heat bath given by the same ${\rm CFT}$, in a thermofield double state represented by the Hartle-Hawking state.

   This system is much simpler than the case of an evaporating ${\bf AdS}^2$ black hole since there is no gravitational backreaction.

\begin{figure}[htbp]
    \begin{center}
      \includegraphics[width=15cm,angle=-0]{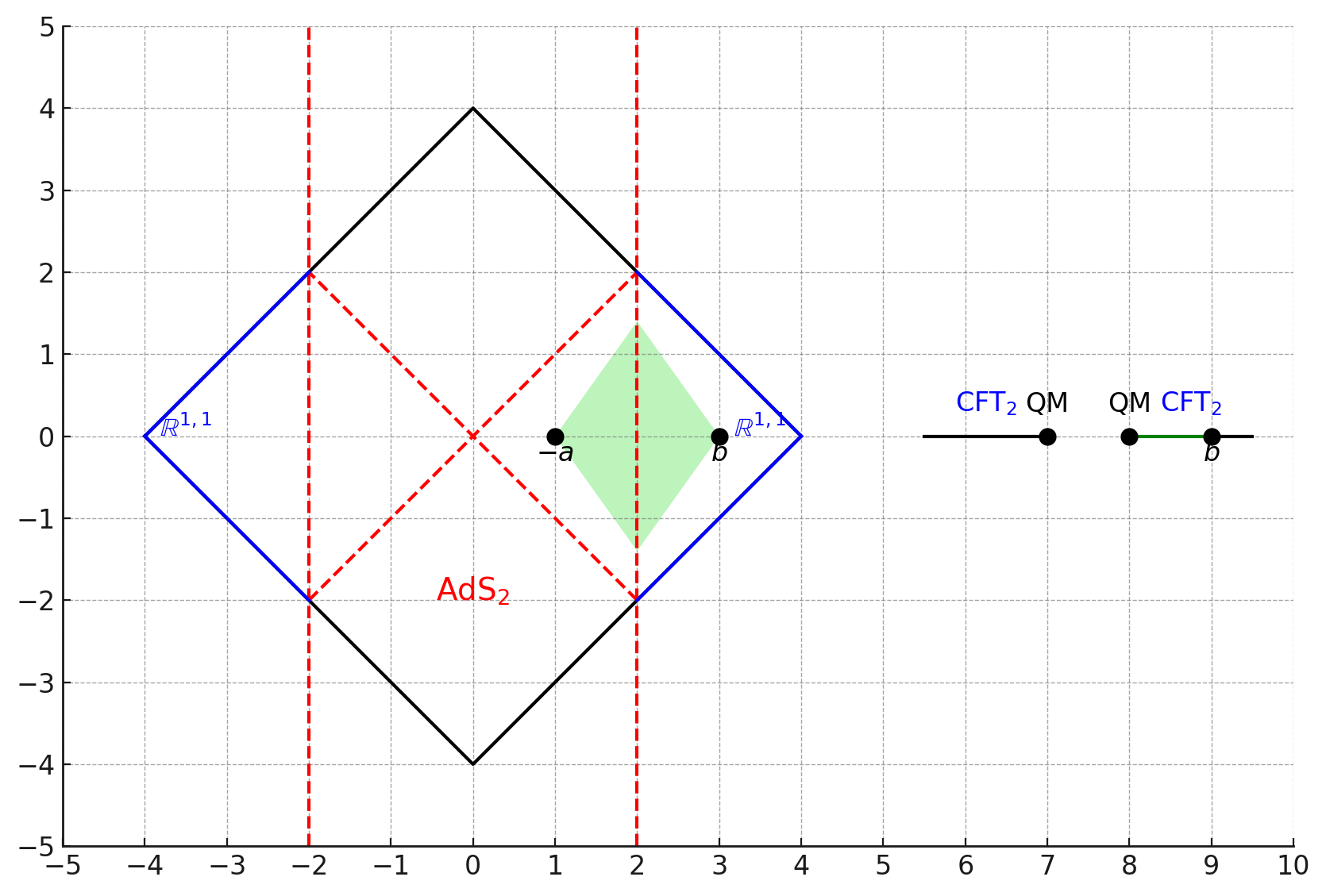}
\end{center}
\caption{A simplified version of the information paradox: An eternal ${\bf AdS}^2$ black hole in thermal equilibrium.}\label{fig8}
   \end{figure}

   This system is also dual to two copies of the same conformal quantum mechanics residing at the two boundaries. See figure (\ref{fig8}).

  We work in the coordinates $y^{\pm}$ in which the right black hole has the metric and dilaton profiles given by
    \begin{eqnarray}
      ds^2=-\frac{4\pi^2}{\beta^2}\frac{dy^+dy^-}{\sinh^2\frac{\pi}{\beta}(y^+-y^-)}~,~\phi=\phi_0-\frac{2\pi \phi_r}{\beta}\frac{1}{\tanh \frac{\pi}{\beta}(y^+-y^-)}.
    \end{eqnarray}
    Here, $\beta$ is the inverse temperature.
    
  This system provides a new version of the information paradox for a black hole in contact with a heat bath in the Hartle-Hawking state. See  \cite{Mathur:2014dia} and \cite{Almheiri:2019qdq,Almheiri:2019yqk}. In this version of the problem, the entropy of the combined system of black hole and bath initially grows but eventually stabilizes (at twice the  Bekenstein-Hawking value of the entropy of one of the boundaries) due to the presence of islands. Indeed, the energy absorbed by the black hole from the bath is re-emitted by the black hole into the bath without any correlation, i.e. as thermal radiation and thus the von Neumann entropy increases with time indefinitely until it exceeds twice the Bekenstein-Hawking entropy (corresponding to the two boundaries). This non-unitary behavior is impossible which is the paradox and is only cured by the presence of the island which modifies the quantum extremal surface dominating the generalized entropy at late times.

    However, in this case the islands can extend outside the black hole horizon which may conflict with causality if not for the so-called quantum focusing conjecture which ensures that the entanglement wedge remains within the causal wedge \cite{Bousso:2015mna}.

    The entropy of the black hole is computed in a single interval $[0,b]_R$ in the quantum mechanical description. Clearly, the point $b_R$ is in the right conformal quantum mechanics. This interval corresponds in the gravitational description to the region $[-a,b]_R$. The generalized entropy is a function of $a$  and is given by  \cite{Almheiri:2019qdq,Almheiri:2019yqk}
      \begin{eqnarray}
        S_{\rm gen}(a)=\frac{2\pi \phi_r}{\beta}\frac{1}{\tanh \frac{2\pi a}{\beta}}+\frac{c}{6}\log\frac{\sinh^2\frac{\pi(a+b)}{\beta}}{\sinh\frac{2\pi a}{\beta}}.\label{2In0}
        \end{eqnarray}
      The location $a$ of the quantum extremal surface  is determined from the extremum value, viz
      \begin{eqnarray}
        \partial_aS_{\rm gen}(a)=0\Rightarrow \frac{\sinh \frac{\pi(a-b)}{\beta}}{\sinh \frac{\pi (a+b)}{\beta}}=\frac{12\pi \phi_r}{c\beta}\frac{1}{\sinh\frac{2\pi a}{\beta}}.\label{2In}
        \end{eqnarray}
      This location $a$ of the quantum extremal surface, in the limit $\phi_r/c\beta\gg 1$,  is such that we have the entropy
      \begin{eqnarray}
       S_{\rm BH}=S_0+\frac{c}{12}\exp(\frac{2\pi}{\beta}(a-b))~,~\frac{\phi_r}{c\beta}\gg 1.
      \end{eqnarray}
      The Bekenstein-Hawking entropy $S_{\rm BH}$ is given by the value of the generalized entropy $S_{\rm gen}$ evaluated at the location $a$ of the quantum extremal surface and it is given by
      \begin{eqnarray}
       S_{\rm BH}=S_0+\frac{2\pi \phi_r}{\beta}.\label{2In1}
      \end{eqnarray}
      Thus, $a-b$ is the scrambling time in this context. Clearly, the value of $a$ corresponds to a point outside the horizon, i.e. the quantum extremal surface (island) is outside the horizon.

      The entropy of the radiation (collected on the right) will involve an island which is contained in the entanglement wedge of the region complementary to $[0,b]_R$. This complementary region is the two-intervals given by the union $[-\infty,0]_L\cup [b,+\infty]_R$. This two-intervals corresponds, in the gravitational description, to the region $[-\infty,\infty]_L\cup [-\infty,-a]_R\cup [b,+\infty]_R$.  In other words, the island corresponds to the right bulk region $[-\infty,-a]_R$.

      By construction, the entropy of the complementary region $[-\infty,0]_L\cup [b,+\infty]_R$ is equal to the entropy of the single interval  $[0,b]_R$.

      A more realistic two-intervals for the calculation of the entropy of radiation starts with the union $[-\infty,-b]_L\cup [b,+\infty]$ in the quantum mechanical description.  This two-intervals corresponds, in the gravitational region, to the region $R=[-\infty,-b]_L\cup [b,+\infty]$ which is also the entanglement wedge in this case.

      This region  $R=[-\infty,-b]_L\cup [b,+\infty]$  lies on the $t=0$ slice in the bath region. This is the case which will give an information paradox for the eternal ${\bf AdS}^2$ black hole. The two intervals $[-\infty,-b]_L$ and $ [b,+\infty]_R$ can be made to lie on the same non-zero time slice $t$ by means of the isometries under time translations in the two black holes.

      The entropy in this case is the entropy of two intervals in the thermofield double state which is found to be given by \cite{Almheiri:2019qdq,Almheiri:2019yqk}
       \begin{eqnarray}
S=\frac{c}{3}\log \big(\frac{\pi}{\beta}\cosh\frac{2\pi t}{\beta}\big)=\frac{2\pi c}{3\beta}t+..~,~t\gg \beta.
       \end{eqnarray}
       This is the no-island result which provides a new version of the information loss paradox. See figure (\ref{fig8p2}).

       However,  the complement interval, to the above considered two-intervals, is $[-b,0]_L\cup[0,b]_R$ with an entanglement wedge given by the region $[-b_L,b_R]$. Thus, this no-island result corresponds really to a single interval $[-b_L,b_R]$. 

       We need to include the island $I=[a,-\infty]\cup [-\infty,-a]$ in the calculation which will turn the problem into a genuine two-intervals calculation. The single interval and the two-intervals cases are shown in figure (\ref{fig8p2}). The value of $a$ is obtained from extremizing the generalized entropy for the two-intervals case given by the island rule
       \begin{eqnarray}
    S_{\rm gen}=\frac{{\rm Area}(\partial I)}{4G_N}+S_{\rm matter}(R\cup I).
       \end{eqnarray}
       Luckily, the extremization of this entropy,  which depends on the cross-ratio of the four endpoints $-b_L$, $a_L$, $b_R$, $-a_R$ of the two intervals $[-b,a]_L$ and $[-a,b]_R$,  is not required since we already know the answer. In fact, it is expected that the quantum extremal surfaces coming from the two-intervals calculation should agree with the single integral calculation. In other words, we expect that the location $a$ of the quantum extremal surface  to be given by the same formula found in the single interval case, i.e. by (\ref{2In}).

       Indeed, in a replicated $n$ geometry (discussed in the next section), we can perform an operator product expansion (OPE) when the two endpoints $-a$ and $b$ are close to each other to find that the generalized entropy is dominated by the identity operator in the OPE.

       This gives the result that the generalized entropy for the two-intervals case equals twice the generalized entropy of the single interval case, given by equation (\ref{2In0}), evaluated at the quantum extremal surface. This means in particular that the location $a$  of the quantum extremal surface is indeed given by the formula (\ref{2In}). This  means also that at late times $t\gg\beta$, the generalized entropy of the radiation, identified with the generalized entropy of the two-intervals, is constant in time given by twice the Bekenstein-Hawking value $S_{\rm BH}$ in (\ref{2In1}), viz
       
      \begin{eqnarray}
       S_{\bf RAD}=2S_{\rm BH}=2(S_0+\frac{2\pi \phi_r}{\beta})~,~.
      \end{eqnarray}
See figure (\ref{fig8p2}).

  \begin{figure}[htbp]
    \begin{center}
     \includegraphics[width=15cm,angle=-0]{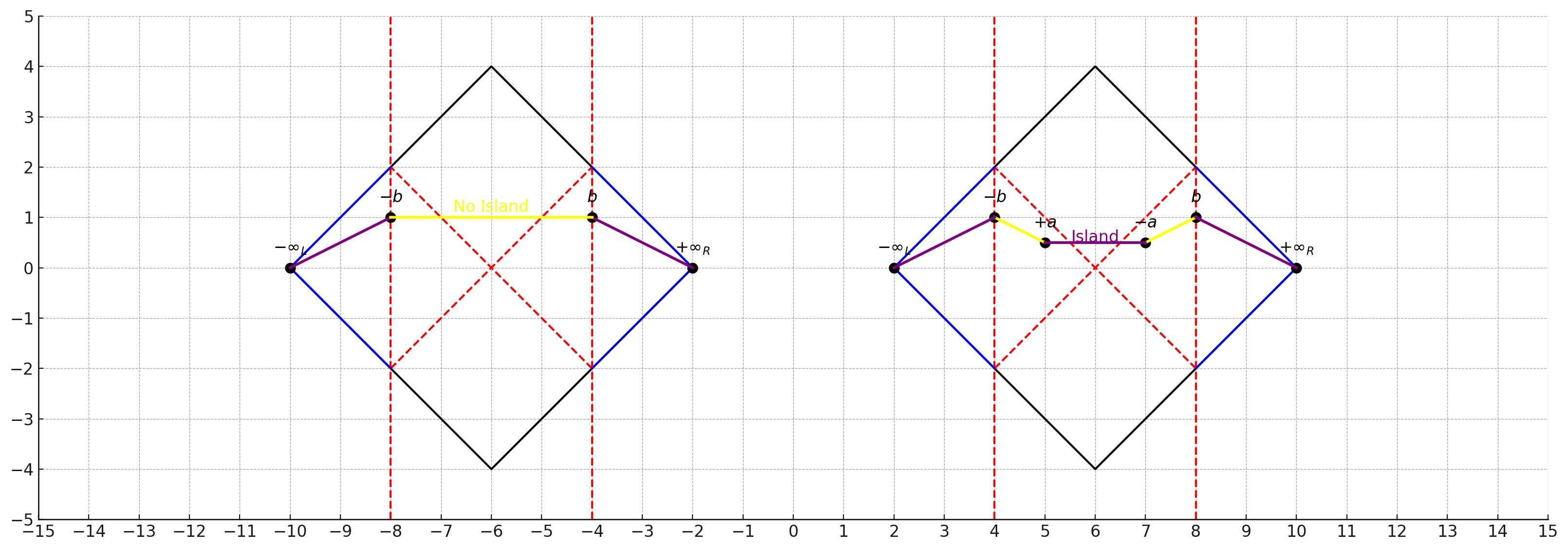}  
\end{center}
\caption{The one-interval (no island) and the two-intervals (island) cases for an eternal black hole in thermal equilibrium.}\label{fig8p2}
   \end{figure}

\begin{figure}[htbp]
    \begin{center}
     \includegraphics[width=15cm,angle=-0]{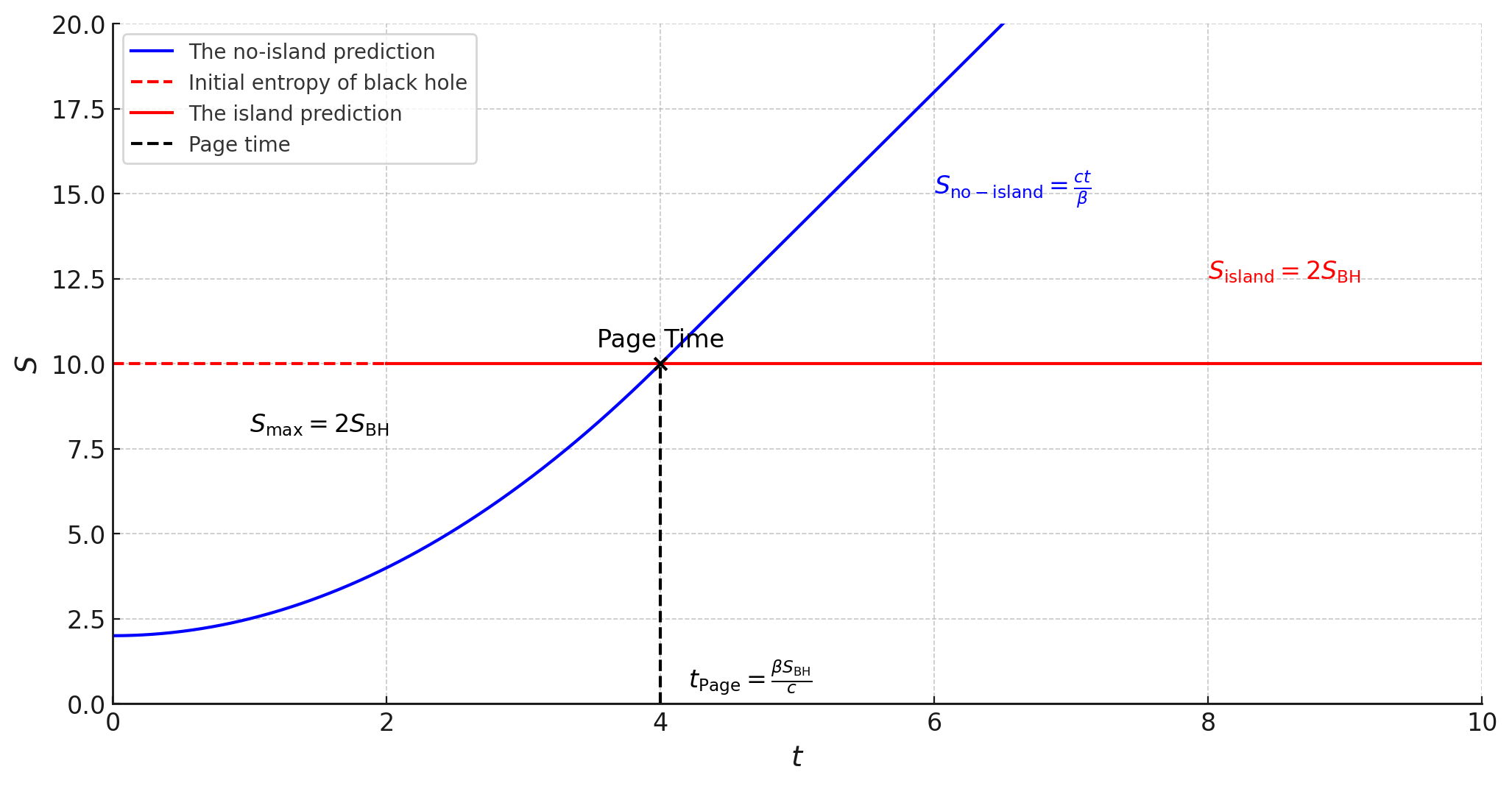}
\end{center}
\caption{The Page curve of an eternal black hole in thermal equilibrium.}\label{fig8p2}
   \end{figure}

\subsection{The replica wormhole picture}

  \begin{itemize}
  \item The replica trick involves computing the entropy by considering $n$ copies of the system, each in the same quantum state, and then cyclically gluing these copies together along the region of interest, i.e. along the interval whose entropy is being calculated, creating thus a new manifold with a specific structure where the field configurations on the different copies (replicas) are connected.
     \item In fact, the quantum extremal surface prescription can be derived from a gravitational theory by means of the replica trick \cite{Faulkner:2013ana,Lewkowycz:2013nqa,Dong:2016hjy,Dong:2017xht}.

   \item We are interested here in the construction outlined in \cite{Almheiri:2019qdq} which shows that the off-shell replicated action in the limit $n\longrightarrow 1$ is precisely the generalized fine-grained von Neumann entropy. See also  \cite{Penington:2019kki}.

  \item First, we note that the von Neumann entanglement entropy $S$ can be computed in terms of the so-called Renyi entropy $S_n$ by means of the  relations
    \begin{eqnarray}
      S&=&-Tr\rho\log\rho=S_n~,~n\longrightarrow 1\nonumber\\
      S_n&=&\frac{1}{1-n}\log(Tr\rho^n)=-\frac{\partial}{\partial n}(Tr\rho^n).
      \end{eqnarray}
    Here, $\rho$ is the density matrix of the radiation or the black hole. Thus, we need to compute $Tr\rho^n$ by pasting together $n$ copies (replicas) of the original system, which are glued cyclically along the intervals whose entropy is being computed, and then extend the resulting function to non-integer values of $n$ through analytic continuation \cite{Callan:1994py,Casini:2009sr}.
     \item This can be connected to the action as follows
    \begin{eqnarray}
     Tr\rho^n=\frac{Z_n}{Z_1^n}\Rightarrow S=(1-n\frac{\partial}{\partial n})\ln Z_n~,~n\longrightarrow 1.
      \end{eqnarray}
    Clearly, $Z_n$ is the partition function of the $n$-replicated system.

  \item The replica trick involves manifolds $\tilde{\cal M}_n$ with fixed geometry in the non-gravitational region while in the gravitational region any geometry which satisfies the boundary conditions is allowed.
    \item Thus, the partition function of the $n$-replicated system is given by $Z_n=Z_n(\tilde{\cal M}_n)$. This corresponds to an effective action for the geometry which combines gravitational and quantum field contributions. The gravitational part is in fact evaluated at a saddle point resulting in a classical metric with quantum matter contributions.
    \item By imposing an extra replica symmetry the manifold $\tilde{\cal M}_n$ reduces to the single manifold ${\cal M}_n=\tilde{\cal M}_n/Z_n$ which has conical singularities and twist fields. %This simplifies the problem because one can study a single copy and then extend the results to the full replicated manifold.
      Furthermore, if the saddle point respects the replica symmetry, the geometry and fields are periodic under the cyclic permutation of the replicas.
    \item However, the single manifold ${\cal M}_n=\tilde{\cal M}_n/Z_n$  involves, as we have just mentioned, conical singularities and twist fields.
  \item The conical singularities are enforced, in the gravitational region, by means of codimension-two cosmic branes with tension $4G_N T_n=1-1/n$. In two dimensions these branes are just points while in four dimensions they are strings. We must also have twist operators, inserted at the positions of these branes, for the $n$ copies of the matter theory. The positions of these branes (which are the fixed points of the $Z_n$ action) are determined using Einstein's equations.
  \item There are additional twist fields inserted at the endpoints of intervals in the non-gravitational region. These twist fields, which encode the entanglement structure of the matter fields, create branch cuts where the matter quantum fields are required to match across different replicas in order to enforce  the cyclic identification of the replica trick.

  \item The original problem on the replicated manifold $\tilde{\cal M}_n$ is rewritten as a problem on the single manifold ${\cal M}_n$ where $n$ copies of the matter field theory are living.
    
    \item The gravitational part of the action is essentially changed by the addition of the cosmic branes action, viz
     \begin{eqnarray}
   \frac{1}{n}S_{\rm grav}[\tilde{\cal M}_n]=S_{\rm grav}[{\cal M}_n]+T_n\int_{\Sigma_{d-2}} \sqrt{g}.
     \end{eqnarray}
    \item The $n$ copies of the quantum field theory living on ${\cal M}_n$ are connected through the branch cuts or twist operators. These cuts enforce boundary conditions such that the field on the $i$-th replica is connected to the field on the $(i+1)$-th replica across the cut which means that the matter degrees of freedom on different replicas are not independent but are correlated through these boundary conditions.
\item The partition function on the replicated manifold  $Z_n$ is not simply the product of the partition functions of $n$ independent copies of the system but includes additional terms that account for the correlations induced by the twist operators. For instance, the partition function can be written as $Z_n=\langle {\cal T}_1...{\cal T}_k\rangle$, i.e. it is the expectation value of the product of twist operators reflecting the correlations between the replicas.

 \item    The total effective action for the system is a sum of the above gravitational action and the action of the quantum fields on the replicated manifold $\tilde{\cal M}_n$ which is equal to the action of $n$ copies of the quantum fields on the manifold ${\cal M}_n$, viz
 \begin{eqnarray}
   \frac{1}{n}S_{n}^{\rm tot}&=&\frac{1}{n}S_{\rm grav}[\tilde{\cal M}_n]+\frac{1}{n}S_{\rm matt}[\tilde{\cal M}_n]\nonumber\\
   &=&S_{\rm grav}[{\cal M}_n]+T_n\int_{\Sigma_{d-2}} \sqrt{g}+\frac{1}{n}\sum_{i=1}^nS_{\rm matt}^i[{\cal M}_n].
     \end{eqnarray}
     \item For $n\simeq 1$, the action is then expanded perturbatively, starting from the original solution ${\cal M}_1=\tilde{\cal M}_1$, as follows
\begin{eqnarray}
   \frac{S_{n}^{\rm tot}}{n}=S_1^{\rm tot}+{\delta}(\frac{S_n^{\rm tot}}{n})~,~n\longrightarrow 1.
     \end{eqnarray}
The perturbative correction includes contributions from the tension of the cosmic branes and from the twist fields inserted at the positions of these branes both of which are evaluated on ${\cal M}_1$ since these two effects are already of order $n-1$. We have then
\begin{eqnarray}
   \frac{S_{n}^{\rm tot}}{n}=S_1^{\rm tot}+(n-1)S_{\rm gener}(w_i).
\end{eqnarray}
$S_{\rm gener}(w_i)$ is the generalized von Neumann entropy evaluated at the positions of the cosmic branes given by
\begin{eqnarray}
   S_{\rm gener}=\frac{{\rm Area}}{4G_N}+S_{\rm matter}.
\end{eqnarray}
This combines the area term (for holographic theories) and quantum corrections from matter fields. The extremization of the generalized entropy functional (coming from the extremization of the action functional) determines the correct entropy.
    \item The cosmic branes (conical singularities) and twist fields together ensure that the correct boundary conditions and singularities are imposed on the replicated geometry allowing therefore for the computation of the Renyi entropy $S_n$ and, in the limit $n\longrightarrow \infty$, of the von Neumann entropy. This procedure leads to the quantum extremal surface prescription, showing that the off-shell action evaluated near $n\simeq 1$ yields the generalized entropy, thus connecting the replica trick with holographic entropy calculations.

  \item In the case when the interior of the black hole is fully connected to the $n$ replicas we get the so-called "replica wormholes", which leads to the island rule, whereas in the case when the interior is fully disconnected from the replicas we get the usual result of Hawking. 
    \item The "replica wormholes" are new gravitational saddles of the gravitational path integral that appear when calculating the entropy of Hawking radiation in a gravitational setting \cite{Almheiri:2019qdq,Penington:2019kki}. These wormholes connect different replicas leading to non-trivial contributions to the path integral.

    \item The "replica wormholes" are the physical principle underlying the island rule  which in turn leads to a unitary Page curve. Hence, certain regions (islands) inside the black hole contribute to the entanglement entropy of the radiation which implies that parts of the black hole interior are encoded in the radiation resolving many aspects of the black hole information paradox.
    \item In \cite{Almheiri:2019qdq}, the replica trick is applied to the calculation of the entanglement entropy for  the eternal ${\bf AdS}^2$ black hole. The entanglement entropy was calculated for both the single interval case (black hole) and the two-intervals case (Hawking radiation) where "replica wormholes" are constructed explicitly and the correct quantum extremal surfaces are obtained.
\item In particular, when replica wormholes are included in the two-intervals case, the saddle point analysis shows that islands, which  refers to those regions of spacetime that are disconnected from the asymptotic boundary, become relevant, i.e. they contribute to the entanglement entropy of the radiation.  These islands appear as regions inside the black hole horizon that are part of the entanglement wedge of the radiation. The generalized entropy in their presence becomes 
  \begin{eqnarray}
    S_{\rm gen}=\frac{{\rm Area}(\partial I)}{4G_N}+S_{\rm matter}(R\cup I).
  \end{eqnarray}
  In this equation $\partial I$ is the boundary of the island $I$ and $R$ is the radiation region. Thus, the inclusion of islands means that information about the interior of the black hole can be encoded in the radiation observed at infinity.
  
\item This shows explicitly how to obtain a unitary Page curve from a gravitational path integral by including "replica wormholes" in the saddle point calculation.  This also shows the validity of the notion of entanglement wedge reconstruction in AdS/CFT correspondence which  means that the entire bulk geometry, including regions inside the event horizon, can be reconstructed from the boundary state, implying that the information about the interior is accessible from the radiation.
\end{itemize}

  \subsection{Replica wormhole versus Hawking saddle }
  \begin{figure}[htbp]
   \begin{center}
      \includegraphics[width=8cm,angle=-0]{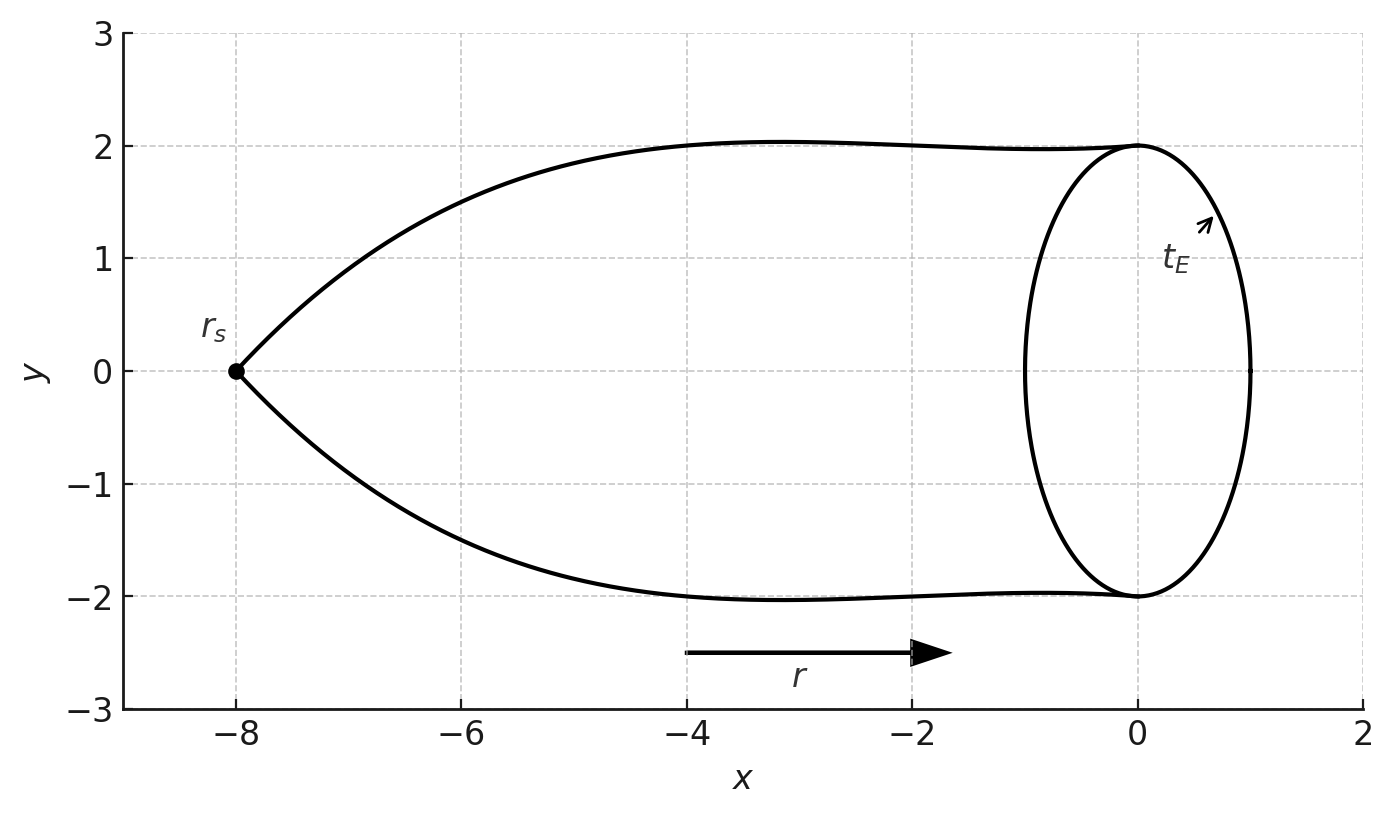}
\end{center}
\caption{Euclidean black hole: The cigar configuration.}\label{cigar}
  \end{figure}
  \begin{itemize}
      \item As an explicit example, we consider here a replicated geometry with $n=2$ which is relevant for the calculation of the purity of the state of the black hole. This schematic, but very illuminating, discussion is taken from \cite{Almheiri:2020cfm}.
      \item An evaporating black hole can be viewed as having two future regions, the future region of the outside universe, and the future of the interior where the singularity lies.
       % \item We will consider the limit in which the black hole has completely evaporated.
  \item An Euclidean black hole (for example the Schwarzschild metric after Wick rotation) has the geometry of a cigar where the tip is the horizon $r=r_s$ and far away, for $r\gg r_s$, the Euclidean time is a circle of circumference $\beta=4\pi r_s$ which is the inverse temperature of the partition function calculating the black hole Euclidean path integral. The periodicity in time is required in order to avoid a conical singularity at $r=r_s$.  Thus, an Euclidean black hole has no interior as shown in figure (\ref{cigar}).
  \item The initial state of the black hole is a pure state $|\Psi(t_0)\rangle$. The final state $|\Psi(t)\rangle=U(t,t_0)|\Psi(t_0)\rangle$  is obtained by evolving $|\Psi(t_0)\rangle$ using the gravitational path integral. The corresponding density matrix is given by $\rho(t)=|\Psi(t)\rangle\langle \Psi(t)| $ with matrix elements $\rho_{ij}(t)=\langle i|\rho(t)|j\rangle=\langle i|\Psi(t)\rangle \langle\Psi(t)|j\rangle$ computed by means of the same gravitational path integral. See figure (\ref{fig10}). Similarly, we compute $Tr\rho=\sum_i\rho_{ii}(t)$ by connecting the exterior regions, i.e. by identifying the final and initial states $|i\rangle$ and $|j\rangle$ and then summing over them which amounts to gluing the interiors together as required by entanglement.  See figure (\ref{fig10}).

    Here, $|i\rangle$ are taken to be states of radiation since we are assuming that the black hole has evaporated completely and we are interested in the entanglement entropy of radiation.
  \item Thus, in order to compute {\bf ${\rho}_{ij}$} we start at the exterior either from the index $i$ or from the index $j$ then follow the entanglement until we reach either the interior corresponding to the ket $|\Psi(t)\rangle$ or the interior corresponding to the bras $\langle \Psi(t)|$. In $Tr\rho$ the two indices $i$ and $j$ are identified and the two interiors are naturally glued together.
     \begin{figure}[htbp]
   \begin{center}
     \includegraphics[width=14cm,angle=-0]{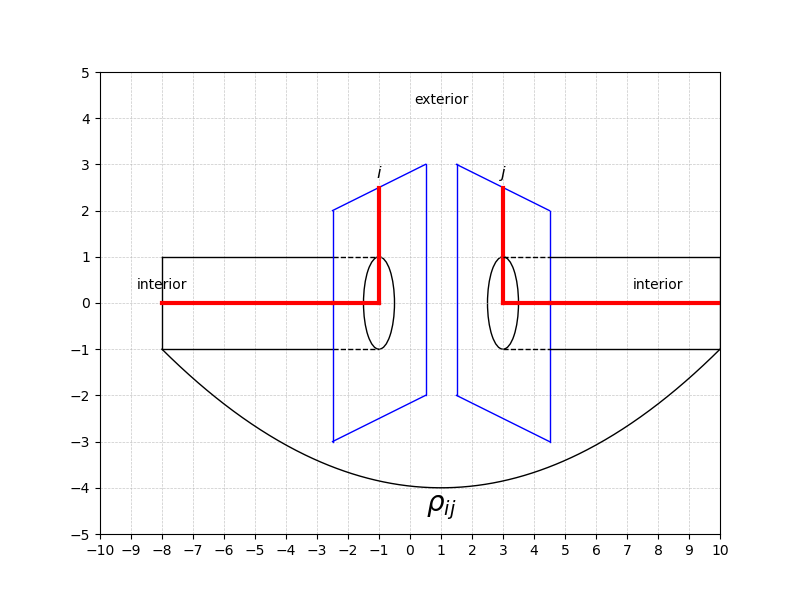}
      \includegraphics[width=14cm,angle=-0]{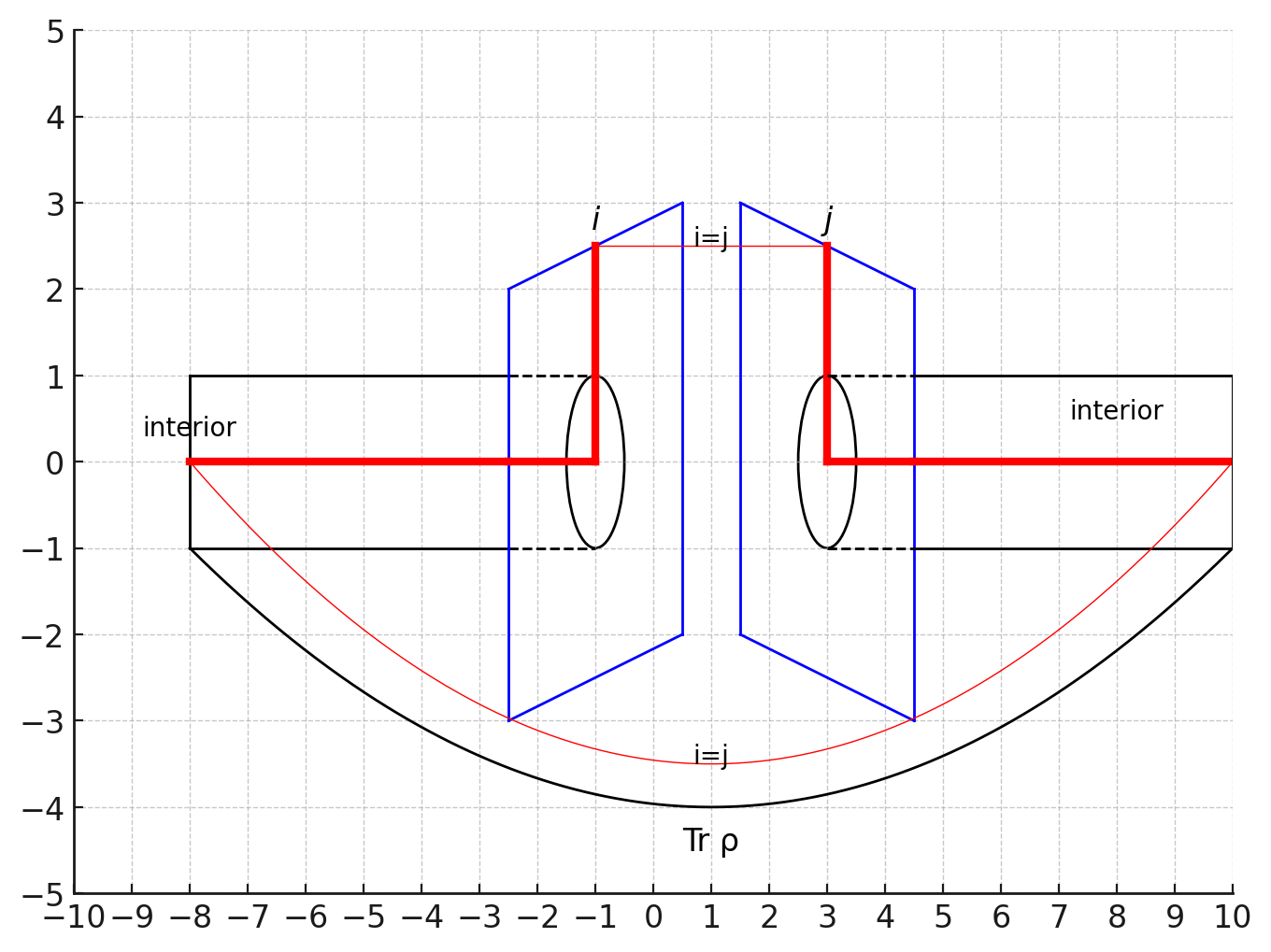}
\end{center}
\caption{The matrix element $\rho_{ij}$ and the trace $Tr\rho$ in the gravitational path integral. The red line represents entanglement.}\label{fig10}
  \end{figure}
  \item We want really to compute the purity of the black hole  state defined by
    \begin{eqnarray}
      Tr\rho^2(t)=\sum_{i,j}\rho_{ij}(t)\rho_{ji}(t).
      \end{eqnarray}
    We have two possibilities. Either $\rho$ is a pure state density matrix for which $Tr\rho^2=(Tr\rho)^2$ (zero entropy) or $\rho$ is a mixed state density matrix for which $Tr\rho^2\ll (Tr\rho)^2$ (large entropy).
  \item As before, we compute $Tr\rho^2(t)$ by connecting the exterior regions and then gluing together the interior regions.   But in a gravitational path integral we are required to sum over all possible topologies. In this case, the two topologically distinct possibilities are:
    \begin{enumerate}   
    \item {\bf Hawking Saddle (Mixed State):} See figure (\ref{hawking}). This is the standard saddle point obtained originally by Hawking.  In this case the interiors corresponding to each of the two factors  $\rho_{ij}(t)$ and  $\rho_{ij}(t)$ are separately glued together, i.e. we have two interiors. We can start from the exterior at the index $i$ of the first factor $\rho_{ij}(t)$ and follow the entanglement through the interior until we reach the exterior at the index $j$ of the first factor. Clearly, the index $i/j$ of the first factor $\rho_{ij}(t)$ is identified with the index $i/j$ of the second factor $\rho_{ji}(t)$. Hence, when we reach the index $j$ of the first factor, we would have really reached the index $j$ of the second factor, and then by following the entanglement through the other interior we will reach the exterior at the index $i$ of the second factor which is identified with the index $i$ of the first factor. This whole path completes one closed loop for which obviously $Tr\rho^2\ne (Tr\rho)^2$.  
    \item {\bf Replica Wormhole (Pure State):}  See figure (\ref{replica}). This is actually the dominating saddle point. In this case the interior corresponding to the ket $|\Psi(t)\rangle$ in the first factor $\rho_{ij}(t)$ is glued together with the interior corresponding to the bras $\langle\Psi(t)| $ in the second factor $\rho_{ji}(t)$.  Similarly, the interior corresponding to the bras $\langle \Psi(t)|$ in the first factor $\rho_{ij}(t)$ is glued together with the interior corresponding to the ket $|\Psi(t)\rangle$ in the second factor $\rho_{ji}(t)$. Thus, in this case we have a single interior. Again, the index $i/j$ of the first factor $\rho_{ij}(t)$ is identified with the index $i/j$ of the second factor $\rho_{ji}(t)$.

      We can then start from the exterior at the index $i$ of the first factor $\rho_{ij}(t)$ and follow the entanglement through the glued interior until we reach the exterior at the index $i$ of the second factor  $\rho_{ji}(t)$. Since the indices $i$ in the two factors are identified this path by itself corresponds to a closed loop.  Similarly, we can start from the exterior at the index $j$ of the first factor $\rho_{ij}(t)$ and follow the entanglement through the glued interior until we reach the exterior at the index $j$ of the second factor  $\rho_{ji}(t)$. Since the indices  $j$ in the two factors are identified this path corresponds also to a  closed loop. Then, we have obviously $Tr\rho^2= (Tr\rho)^2$  in this case.      
    \end{enumerate}
    \item The Hawking saddle is thus characterized by a very large entropy (mixed state) whereas the replica wormhole is characterized by zero entropy (pure state).  Hence, the replica wormhole dominates over the Hawking saddle in the path integral. 
\item In summary, in the case of the Hawking saddle, we have the identifications $i/j\longrightarrow i/j$ and the entanglement relations $i/j\longrightarrow j/i$. But, in the case of the replica wormhole, we have the identifications $i/j\longrightarrow i/j$ as well as the entanglement relations $i/j\longrightarrow i/j$. 
  
    \item Thus, in the Euclidean formulation, the Hawking saddle will correspond to two disconnected copies of the cigar geometry while the replica wormhole will correspond to two Euclidean black holes joined through the interior.
 
  \end{itemize}

   \begin{figure}[htbp]
   \begin{center}
     \includegraphics[width=15cm,angle=-0]{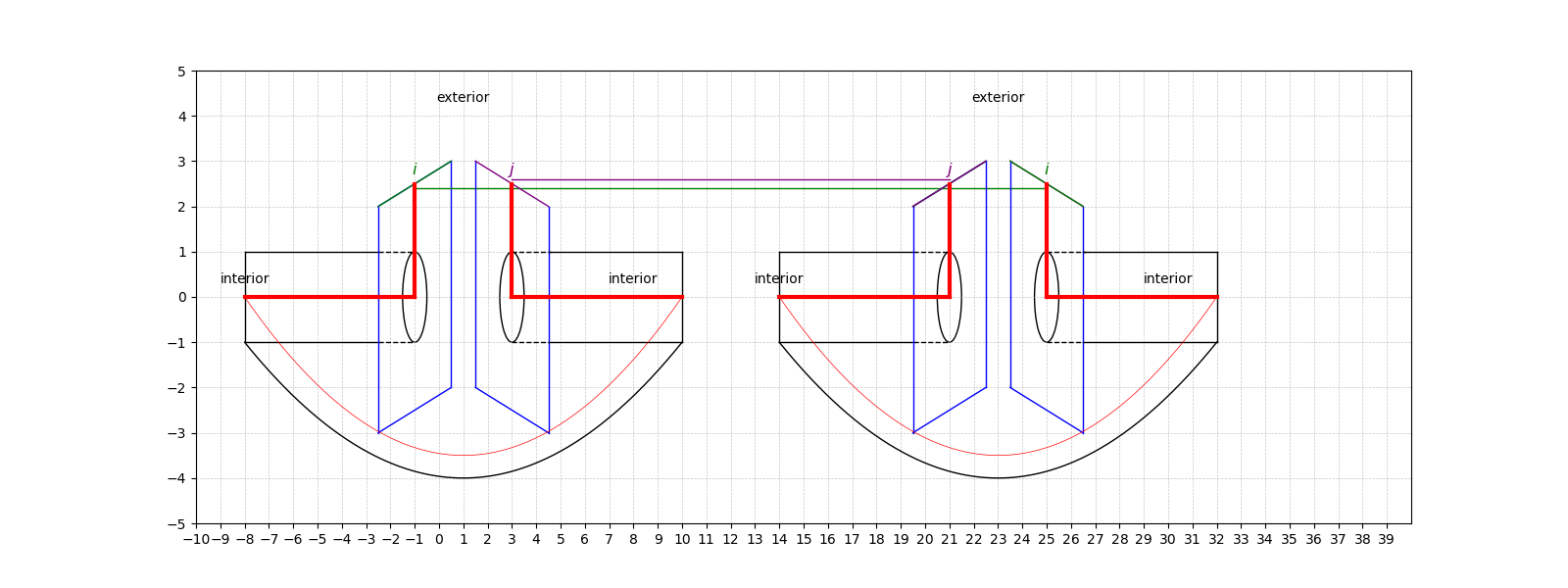}
\end{center}
\caption{The Hawking saddle point (first configuration).}\label{hawking}
   \end{figure}

   \begin{figure}[htbp]
     \begin{center}
        \includegraphics[width=15cm,angle=-0]{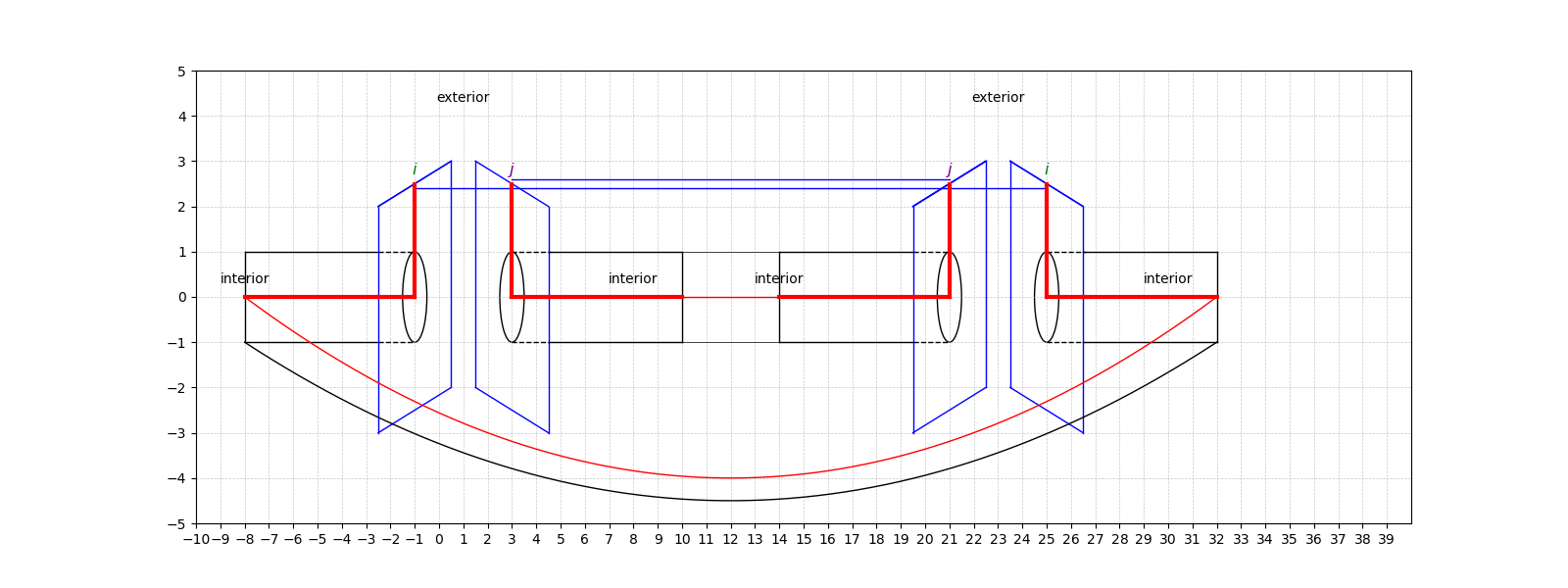}
     \includegraphics[width=15cm,angle=-0]{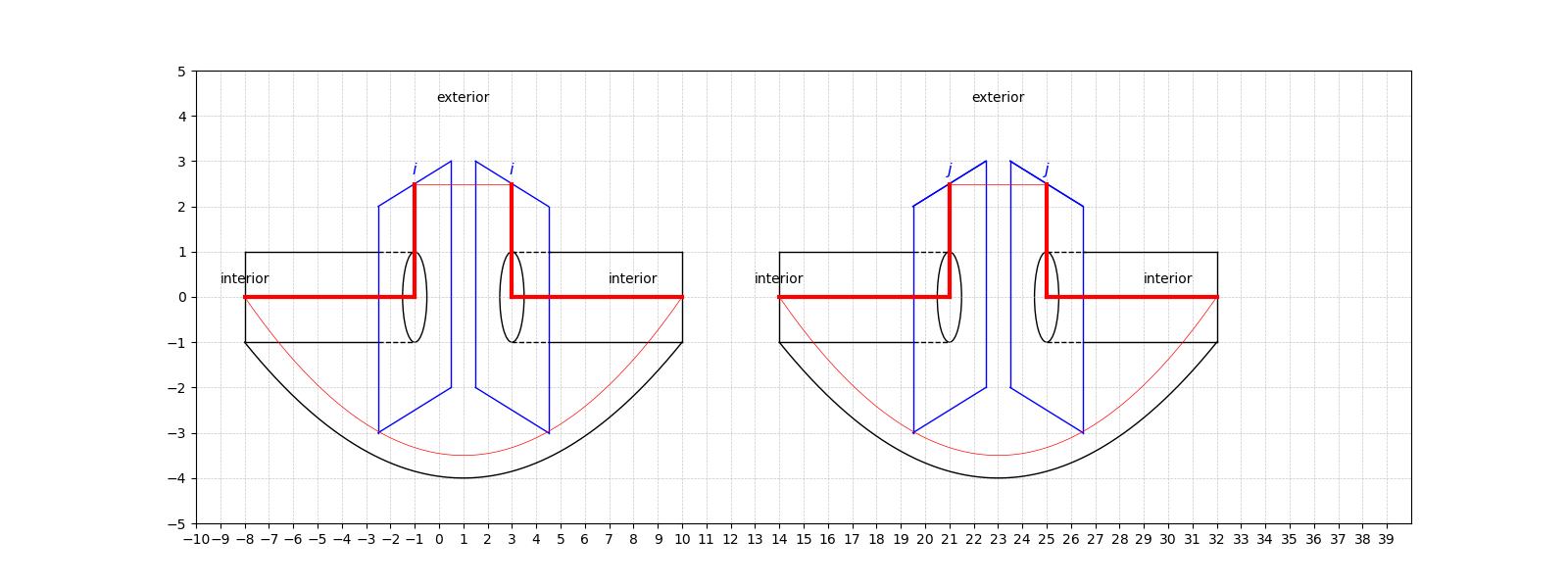}
\end{center}
\caption{The replica wormhole saddle point (second configuration) which is also given by the third configuration.}\label{replica}
   \end{figure}
   
   \begin{figure}[htbp]
   \begin{center}
     \includegraphics[width=10cm,angle=-0]{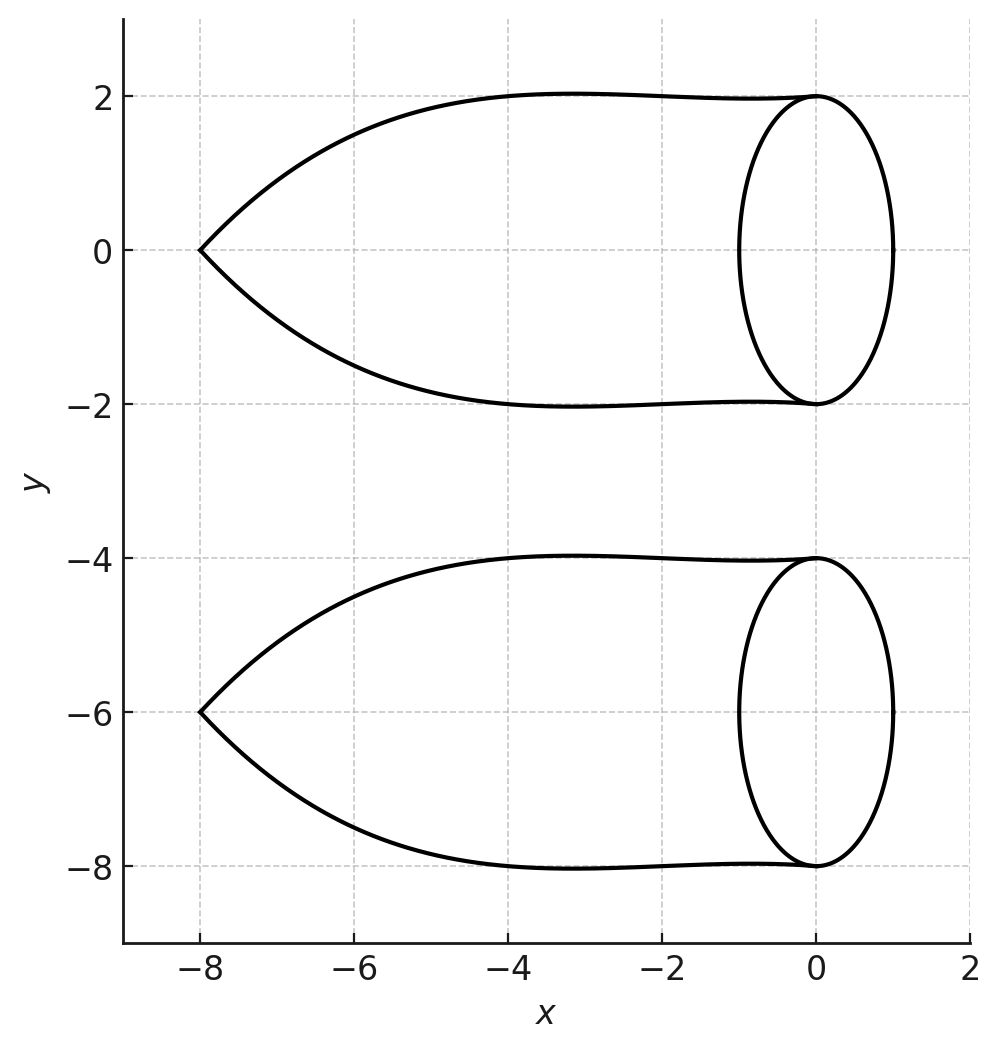}
       \includegraphics[width=10cm,angle=-0]{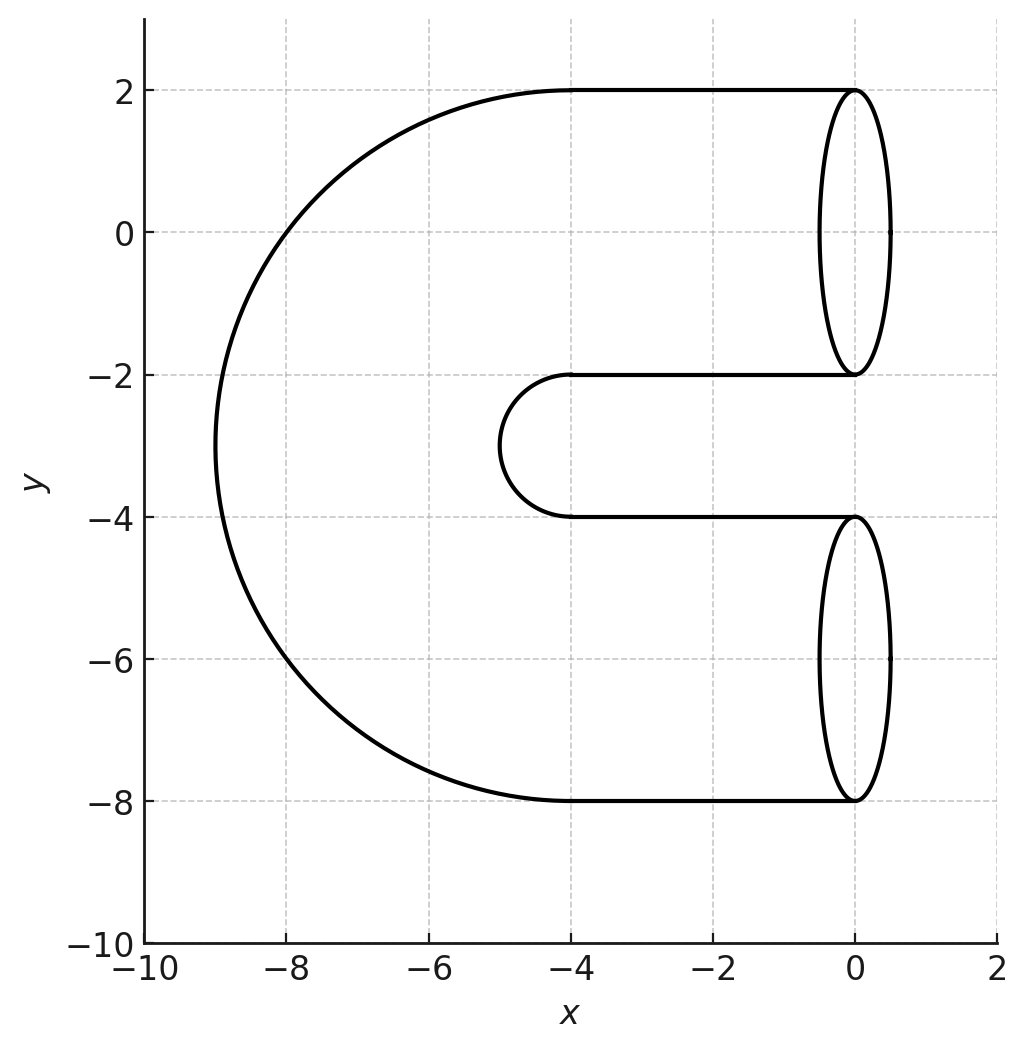}
\end{center}
\caption{The Hawking saddle point (cigar configurations) and the replica wormhole saddle point (two connected cigars). }
  \end{figure}

%\section{Exercise}

%\begin{frame}
 % \frametitle{Exercise}
  
 % \begin{itemize}
 % \item Show equation (\ref{Sent}) by following similar steps to those which led to equations (\ref{cond1}) and (\ref{cond2}).
 % \item By using equation (\ref{h1}) derive (\ref{h1}). Show that the extrinsic curvature of the two-dimensional boundary is given in terms of the stress-energy-momentum  tensor of the ${\bf CFT}_2$.
    
 % \end{itemize}
 % \end{frame}

\section{Summary}

  JT gravity, quantum extremal surfaces, and the island conjecture provide a robust framework to resolve the black hole information paradox. The distinction between von Neumann entropy and Bekenstein-Hawking entropy highlights the differences in fine-grained and coarse-grained descriptions of entropy, crucial for understanding information preservation in black hole dynamics. The replica trick is an essential tool in computing entropy, facilitating the understanding of complex entanglement structures in quantum gravity.

\end{document}